\documentclass{aa}  

\usepackage{graphicx}
\usepackage{txfonts}

\usepackage{hyperref}
\hypersetup{
    colorlinks=true,
    linkcolor=red,
    filecolor=red,      
    urlcolor=red,
    }
\hypersetup{colorlinks,citecolor=blue}
\usepackage{graphicx}
\usepackage{natbib}
\usepackage{threeparttable}
\usepackage{float}
\begin{document}

\title{The imprint of AGN-driven outflows on the CGM: the case of Ly$\alpha$ nebulae around high-z quasars}

\author{
  Silvia C. Rueda-Vargas\inst{1}
  \and Vincenzo Mainieri\inst{1}
  \and Giulia Tozzi\inst{2}
  \and Darshan Kakkad\inst{3}
  \and Fabrizio Arrigoni-Battaia\inst{4}
  \and Tiago Costa\inst{5}
  \and Emanuele Paolo Farina\inst{6,7}
  \and Chris M. Harrison\inst{5}
}

\institute{
  European Southern Observatory, Karl-Schwarzschild-Strasse 2, 85748 Garching bei M\"unchen, Germany\\
  \email{silviacarolina.ruedavargas@eso.org}
  \and Max-Planck-Institut f\"ur extraterrestrische Physik, Gie{\ss}enbachstra{\ss}e 1, 85748 Garching bei M\"unchen, Germany
  \and Centre for Astrophysics Research, University of Hertfordshire, Hatfield AL10 9AB, UK
  \and Max-Planck-Institut f\"ur Astrophysik, Karl-Schwarzschild-Stra{\ss}e 1, 85748 Garching bei M\"unchen, Germany
  \and School of Mathematics, Statistics and Physics, Newcastle University, Newcastle upon Tyne, NE1 7RU, UK
  \and International Gemini Observatory/NSF NOIRLab, 670 N A’ohoku Place, Hilo, Hawai'i 96720, USA
  \and INAF -- Osservatorio di Astrofisica e Scienza dello Spazio di Bologna, via Gobetti 93/3, I-40129, Bologna, Italy
}

\date{}

\abstract{Some cosmological hydrodynamical simulations predict that outflows driven by active galactic nuclei (AGN) play a key role in powering the Ly$\alpha$ nebulae observed around high-redshift quasars. In such simulations, AGN feedback seeded as powerful outflows leads to extended and luminous nebulae whose morphology and surface-brightness profiles accurately reproduce the observations, while suppressing AGN feedback leads to compact and faint nebulae. This link might arise from outflows opening up a channel for Ly$\alpha$ photons to escape from the galactic nucleus to the circumgalactic medium (CGM). The main aim of this paper is to test this theoretical prediction using observations, by comparing the physical properties of outflows and Ly$\alpha$ nebulae. We analyze integral-field unit data obtained with VLT/ERIS and GEMINI/GNIRS to trace the ionized gas in the interstellar medium (ISM) of a sample of six quasars at $z\sim2-3$, using the [\ion{O}{iii}] emission line. We detect powerful outflows in all the quasars of our sample, with velocities $>1500~\mathrm{km~s^{-1}}$ and kinetic energies $ \gtrsim 2\times10^{43}~\mathrm{erg~s^{-1}}$. Four of our quasars are spatially resolved and show signs of extended [\ion{O}{iii}] emission out to distances $>2$ kpc from the central supermassive black hole. When excluding one outlier, we find a positive monotonic correlation between the outflow power and the Ly$\alpha$ nebulae size ($\rho=0.89$, $p=0.03$) and luminosity ($\rho=0.6$, $p=0.28$). Additionally, we find evidence of spatial alignment between the ionization cone and the inner and brightest regions of the Ly$\alpha$ nebula. Our results provide tentative evidence in support of the theoretical prediction that AGN-driven outflows at ISM scales open a low-optical-depth path for central Ly$\alpha$ photons to reach the CGM and create extended nebulae.}


\maketitle
%

\section{Introduction}

The influence of supermassive black hole activity on the evolution of the host galaxy has been extensively studied in the last decades \citep[e.g.][]{kormendy2013coevolution}. It is widely accepted now that the black hole (BH) and host galaxy properties are closely correlated, as evidenced by the scaling relations between the BH mass and the velocity dispersion of the bulge ($M_{\mathrm{BH}}-\sigma$) \citep{ferrarese2000fundamental, gebhardt2000relationship} and the BH and galaxy mass ($M_{\mathrm{BH}}-M_*$) \citep{haring2004black}. These correlations are believed to be shaped during phases of galaxy evolution in which the supermassive BH is actively accreting and interacting with its surrounding environment through feedback mechanisms. Such phases are known as AGN. AGN feedback injects energy and momentum into the host galaxy via multiple channels, including radiation and outflows, operating on a broad range of spatial scales: from the accretion disk (< 1 pc from the supermassive BH) to the ISM (10 pc $-$ 10 kpc), CGM (10 $-$ 100 kpc), and the intragroup or intracluster medium (IGrM/ICM, 100 kpc $-$ 1 Mpc) \citep{harrison2024observational}. Among these channels,  outflows are generally considered the dominant mechanism for transferring significant amounts of energy and momentum into the host galaxy \citep{springel2005black, di2005energy}. However, their effect on the surrounding gas is still a matter of debate. On the one hand, outflows might suppress star formation (SF) either by heating the gas and preventing it from cooling or by removing such gas from the host galaxy. These mechanisms are referred to as preventive and ejective feedback, respectively \citep[e.g.,][]{cano2012observational, cresci2015blowin, carniani2016fast}. On the other hand, outflows might trigger SF by compressing the gas in the ISM or due to their large content of dense molecular gas \citep[e.g.,][]{zubovas2013agn, maiolino2017star, gallagher2019widespread}.

AGN-driven outflows may originate as highly ionized, high-velocity winds, often reaching relativistic speeds ($\gtrsim 0.1 c$), launched from the inner accretion disk at sub-pc scales \citep[e.g.][]{king2003black, gibson2009catalog, hidalgo2020survey, chartas2021multiphase, vietri2025extremely}. When propagating outward, these fast winds interact with the ISM, driving shocks (a forward shock into the ISM and a reverse shock into the wind) which create a large-scale expanding bubble of hot gas \citep{faucher2012physics, costa2020powering}. The subsequent propagation of this bubble is typically modeled in two regimes: the energy-driven mode, where the shocked gas retains its thermal energy, or the momentum-driven mode, where rapid cooling of the shocked gas dominates \citep{zubovas2012clearing, costa2014feedback, marconcini2025unveiling}. The material accelerated and displaced by the internal pressure of this bubble forms the large-scale, kpc-scale outflow observed in the ISM.

Outflows are observed in multiple different phases through different observational tracers \citep{harrison2024observational}. The hot ionized phase ($T_{\mathrm{gas}}\approx 10^{5}-10^{8}$ K and $n_{\mathrm{gas}}\approx10^{6}-10^{8} \mathrm{cm^{-3}}$) is typically observed at nuclear scales through highly ionized absorption lines in the X-ray spectrum of the AGN such as \ion{O}{viii}, \ion{Fe}{xxv} and \ion{Fe}{xxvi} \citep[e.g.][]{reeves2003massive, longinotti2013rise, chartas2021multiphase}. The warm ionized phase ($T_{\mathrm{gas}}\approx 10^{3}-10^{5}$ K and $n_{\mathrm{gas}}\approx10^{2}-10^{5}\mathrm{cm^{-3}}$) is traced using emission lines in the optical, UV or near-infrared such as [\ion{O}{iii}] or H$\alpha$ \citep[e.g.,][]{harrison2014kiloparsec, woo2016prevalence, kakkad2016tracing, kakkad2020super, kakkad2023dissecting, scholtz2020kashz, tozzi2021connecting, tozzi2024super}. This phase is one of the most studied, and thanks to the new capabilities of JWST, it is now accessible for quasars at $z \gtrsim 4$. The molecular phase ($T_{\mathrm{gas}}\approx 10^{1}-10^{3}$ K and $n_{\mathrm{gas}}>10^{3}\mathrm{cm^{-3}}$), in contrast, is mostly detected at $z<1$, as a consequence of the long integration times needed to spatially resolve it. This phase is detected, for example, by using rotational CO lines such as CO(1-0) or CO(2-1) in the sub-milimiter, rovibrational $\mathrm{H}_2$ lines in the near-infrared, or (OH) transitions in the far-infrared \citep[E.G.,][]{feruglio2010quasar, sturm2011massive, rupke2013breaking}. The neutral atomic phase ($T_{\mathrm{gas}}\approx 10^{2}-10^{3}$ K and $n_{\mathrm{gas}}\approx10^{1}-10^{2} \mathrm{cm^{-3}}$) is commonly traced in radio wavelengths using the \ion{H}{i} absorption line, the sodium doublet absorption NaID or low ionization UV absorption lines such as \ion{Mg}{ii} \citep{morganti2016another, cazzoli2016neutral}. In this work, we investigate the warm ionized phase of AGN-driven outflows by analyzing the [\ion{O}{iii}] $\lambda 4959, 5007$ doublet. Although the cold molecular phase is thought to carry most of the mass and the hot phase most of the energy \citep{speranza2024multiphase, fluetsch2021properties, harrison2024observational}, the warm ionized phase provides a crucial tracer of the momentum and energy coupling that drives feedback into the galaxy environment. This phase is predicted to correspond to the swept-up ISM gas, originating from the cooling boundary layer of the nuclear AGN-wind bubble \citep{torrey2020impact, richings2021unravelling}. The [\ion{O}{iii}] $\lambda4959, 5007$ doublet is particularly well suited for our study, as it arises from a forbidden transition and can therefore only be produced in low-density environments such as the ISM or the AGN Narrow-Line Region (NLR). In contrast, while permitted emission lines such as H$\alpha$ or H$\beta$ also exhibit outflowing components, they can feature broad-line emission ($\mathrm{FWHM>1500~km~s^{-1}}$) originating from the inner AGN broad-line region (BLR) or from star-forming regions \citep{schreiber2019kmos3d}.

On scales beyond the ISM, integral field spectrographs (IFSs), such as VLT/MUSE \citep{bacon2010muse} and Keck/KCWI \citep{mclean2012ground}, have made it possible to reach unprecedentedly low levels of surface brightness (SB), enabling the detection in emission of the cold and diffuse gas in the IGrM/ICM \citep[e.g.,][]{tornotti2025high} and CGM around bright quasars. Using mainly the Ly$\alpha$ emission line, it has been possible to detect the nebulae surrounding these quasars. The Ly$\alpha$ nebulae are now routinely detected around high-redshift quasars, and are characterized for having a wide range of morphologies, extents from tens to hundreds of kiloparsec and luminosities $L_{\mathrm{Ly\alpha}} > 10^{44}~\mathrm{erg~s^{-1}}$ \citep[e.g.,][]{borisova2016ubiquitous, arrigoni2019qso, cai2019evolution, vayner2023cold, lobos2025qso}. The emission of Ly$\alpha$ photons is attributed to recombination radiation from photoionized gas by the AGN or star formation (SF) \citep[e.g.,][]{cantalupo2014cosmic, geach2009chandra, kollmeier2009lyalpha}, gravitational cooling of collisionally excited hydrogen \citep[e.g.,][]{haiman2000lyalpha, rosdahl2012extended}, or resonant scattering of the photons from the BLR of the AGN \citep[e.g.,][]{dijkstra2008polarization}. 

Cosmological, hydrodynamical simulations have investigated the physical mechanism driving the extended Ly$\alpha$ nebulae observed in the CGM of high-redshift quasars. \citet{costa2022agn} explores three possible scenarios for the origin of the Ly$\alpha$ photons: i) a mix of recombination and gravitational cooling, ii) a mix of recombination, gravitational cooling, and BLR scattering, or iii) pure BLR scattering. Additionally, \citet{costa2022agn} further studies the impact of AGN feedback by varying the quasar luminosity, and therefore the feedback power. In the absence of quasar feedback, the predicted Ly$\alpha$ escape fractions are extremely low ($f_{\mathrm{esc}} \sim 0.1\%$, $17\%$, and $0\%$ for recombination, cooling, and BLR scattering, respectively), resulting in nebulae that are significantly fainter and more compact than observed by \citet{farina2019requiem}. In contrast, when the quasar luminosity is increased to $3 \times 10^{47} \mathrm{erg,s^{-1}}$ and $5 \times 10^{47} \mathrm{erg,s^{-1}}$, the escape fractions rise substantially ($f_{\mathrm{esc}} \sim 20\%$, $30\%$, and $73\%$ at $3 \times 10^{47} \mathrm{erg,s^{-1}}$; $39\%$, $30\%$, and $87\%$ at $5 \times 10^{47} \mathrm{erg,s^{-1}}$). Under these conditions, the simulated nebulae reach luminosities within the observed range and reproduce the SB profiles reported by \citet{farina2019requiem}, characterized by a central plateau that steepens within $\sim 10$ kpc and drops rapidly beyond $\sim 30$ kpc. However, the impact of outflows is not always positive for the detection of extended Ly$\alpha$ emission: when they become too powerful, the central optical depths drop to the point that the Ly$\alpha$ photons escape directly from the nucleus rather than scatter through the halo, leading to a shrinking of the nebula.

The results of \citet{costa2022agn} suggest a potential connection between AGN-driven outflows and the properties of Ly$\alpha$ nebulae around high-redshift quasars. Such a link may arise through two main channels: (i) an enhanced production of ionizing photons, which boosts both recombination and BLR emissivities, and increased $f_{\mathrm{esc}}$ when the feedback power increases, and/or (ii) a substantial reduction in the optical depth of neutral hydrogen and dust in the host galaxy due to the effect of AGN-driven outflows, facilitating the escape of Ly$\alpha$ photons from the galactic nucleus \citep{costa2022agn, arrigoni2019qso, arrigoni2018inspiraling}. In this paper, we aim to test these theoretical predictions with observations and assess whether, and to what extent, the properties of AGN-driven outflows are linked to those of the Ly$\alpha$ nebulae surrounding high-redshift quasars. We do this by analyzing the [\ion{O}{iii}]$\lambda 4959, 5007$ emission at ISM scales of a sample of six quasars at $z\sim2-3$ where a Ly$\alpha$ nebula has been previously detected using Keck/KCWI or VLT/MUSE. This paper is split as follows: in Sect. \ref{sect:sample} we describe our sample, its characteristics, and the observing strategy for VLT/ERIS and GEMINI/GNIRS, and in Sect. \ref{sect:data_reduction} we describe the data reduction process for both data sets. In Sect. \ref{sect:data_analysis} we describe in detail the analysis procedure consisting of the H$\beta$+[\ion{O}{iii}] spectral fitting and the Ly$\alpha$ nebulae detection and properties. In Sect. \ref{sect:results} we present the results on the ionized outflows and their properties, and in Sect. \ref{sect:discussion} we compare the properties of the derived outflows with those of the previously detected Ly$\alpha$ nebulae. Throughout this work, we adopt a flat $\Lambda$CDM cosmology with $H_{0}=67.4~\mathrm{km~s^{-1}~Mpc^{-1}}$, $\Omega_{\mathrm{m}}=0.315$, and $\Omega_{\Lambda}=0.685$.


\section{Sample} \label{sect:sample}

\subsection{Sample selection}

Our sample was extracted from two previous surveys analyzing the properties of extended Ly$\alpha$ nebulae around quasars at cosmic noon. The first one, presented in \cite{cai2019evolution}, uses Keck/KCWI \citep{morrissey2018keck} to target 16 ultraluminous QSOs at $z=2.1-2.3$. The second, presented in \cite{arrigoni2019qso} as the QSO MUSEUM I, consists of 61 QSOs at $z=3.03-3.46$ observed with VLT/MUSE \citep{bacon2010muse}. From both surveys, we discarded targets with limited visibility at the telescope location (VLT or GEMINI) during the corresponding observing period, and those where the [\ion{O}{iii}]$\lambda 5007$ line profile was affected by atmospheric features. We selected a subsample of targets representative of the wide observed range of Ly$\alpha$ nebulae extensions and velocity dispersions.

\begin{figure}[t]
    \centering
    \includegraphics[width=1.1\linewidth]{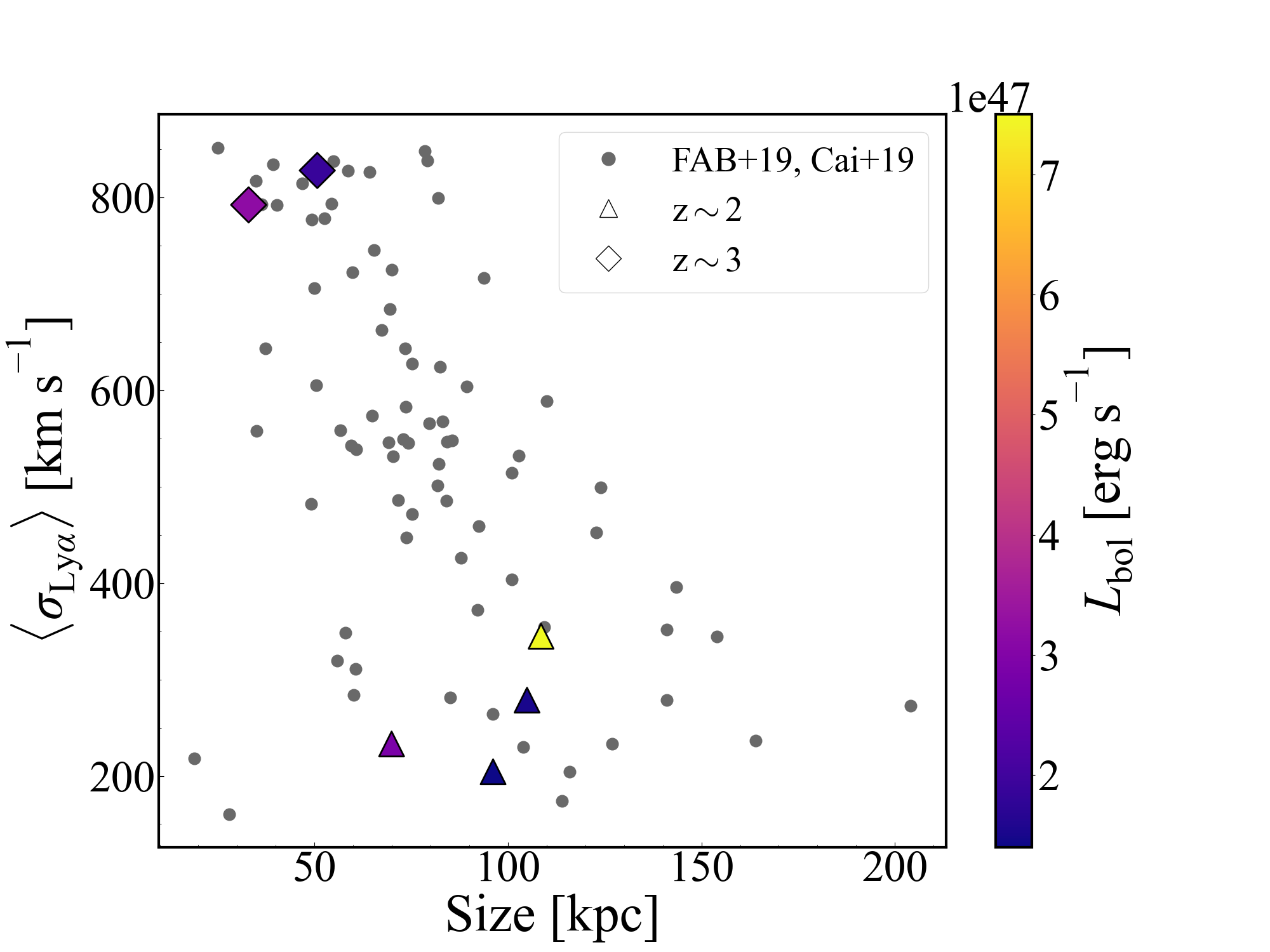}
    \caption{Distribution of the Ly$\alpha$ nebulae around the targets in our sample in the velocity dispersion–size plane. The gray dots represent the parent sample from which our sources were selected, namely \citet{arrigoni2019qso} (FAB+19) and \citet{cai2019evolution}. Triangular and diamond markers indicate the targets selected for our study at $z\sim2$ and $z\sim3$, respectively. The points are color-coded by the bolometric luminosity $L_{\mathrm{bol}}$.}
    \label{fig:lya_distribution}
\end{figure}

Our final sample consists of six Type I QSOs, spanning nebula sizes in the range of $50 < \mathrm{size}<150~\mathrm{kpc}$ and velocity dispersion in the range of $150<\sigma < 850~\mathrm{km~s^{-1}}$. Fig. \ref{fig:lya_distribution} shows the location of our sample in the Ly$\alpha$ nebulae velocity dispersion-size plane, obtained from the parent sample analyzed in \citet{arrigoni2019qso} and \citet{cai2019evolution}. We re-estimate these nebulae properties in Section \ref{sect:lya_props} using our own definition. Our targets lie in the redshift range $z\sim2.3-3.3$, the peak of SF and accretion on massive black holes. This range is even more relevant in this context due to the significant evolution in the SB of the nebulae reported between z$\sim$2 and z$\sim$3,
while there is little evolution between z$\sim$3 and z$\sim$6 \citep{farina2019requiem, arrigoni2019qso, fossati2021muse}. Table \ref{tab:sample_props} records the main properties of the QSOs in our sample. 

\begin{table*}[t]
    \renewcommand{\arraystretch}{1.2}
    \centering
    \caption[]{Main properties of the 6 observed Type I QSOs in our sample.}
    \label{tab:sample_props}
    \begin{tabular}{c c c c c c c c c}
        \hline \hline
        Target ID & R.A. & DEC & $z$ & $L_{\mathrm{bol}}$ & Instr. & Seeing \ion{O}{iii}&Instr. & Seeing Ly$\alpha$\\
        & (J2000) & (J2000) & & [$10^{47}~\mathrm{erg~s^{-1}}$] & \ion{O}{iii} & [arcsec] & Ly$\alpha$ & [arcsec] \\
        \hline
        Q0050+0051 & 00:50:21.22 & +00:51:35.0 & 2.241 & 1.4 & ERIS & 0.54 & KCWI & 0.9\\
        Q0052+0140 & 00:52:33.67 & +01:40:40.8 & 2.297 & 2.9 & ERIS & 0.51 & KCWI & 0.8\\
        Q2121+0052 & 21:21:59.04 & +00:52:24.1 & 2.373 & 1.5 & ERIS & 0.60 & KCWI & 0.8\\
        Q2123-0050 & 21:23:29.46 & -00:50:52.9 & 2.282 & 7.5 & ERIS & 0.94 & KCWI & 1.0\\
        SDSSJ2319$-$1040 & 23:19:34.80 & -10:40:36.0 & 3.162 & 1.9 & ERIS & 0.95 & MUSE & 1.4\\
        SDSSJ1427$-$0029 & 14:27:55.85 & -00:29:51.1 & 3.359 & 3.2 & GNIRS & 0.8 & MUSE & 0.9\\
        \hline
    \end{tabular}

    \vspace{4mm}
    \tablefoot{
        From left to right, the columns report the target’s name,
        right ascension and declination, spectroscopic redshift, bolometric luminosity,
        the instrument used to trace the [\ion{O}{iii}]$\lambda$5007 line, the seeing of such observations, the instrument used to trace the extended Ly$\alpha$ emission surrounding the quasars, and the seeing of such observations. The redshift $z$ is estimated using the [\ion{O}{iii}]$\lambda 5007$ line. The bolometric luminosity $L_{\mathrm{bol}}$ is computed using the monochromatic luminosity $L_{\lambda}(1350\text{\AA})$, and the bolometric correction factor given in \citet{shen2011catalog} and adapted from \citet{richards2006spectral}:
        $L_{\rm bol} = 3.81 \times \lambda L_{\lambda}(1350\text{\AA})$. The obtained values can have up to 0.3\,dex of intrinsic uncertainty.
    }
\end{table*}

\subsection{Observations}

\subsubsection{VLT/ERIS}

The ERIS \citep{davies2023enhanced} observations of the objects in our sample were acquired in service-mode under the ESO program 113.26AT (PI: Mainieri) at the VLT 8.2m telescope UT4 in noAO mode. The observations were performed with the SPIFFIER IFS with a pixel scale of 250mas/pixel, and a field of view (FoV) of $8\arcsec\times8\arcsec$, which is optimal to map the ISM regions of quasars at cosmic noon. To cover the redshifted [\ion{O}{iii}]$\lambda5007$ emission line, we used the $\mathrm{H_{low}}$ (1.45$-$1.87 $\mu$m, spectral resolution of R$\sim5200$) and $\mathrm{K_{low}}$ (1.93$-$2.48 $\mu$m, R$\sim5600$) grating for the targets at $z\sim2$ and $z\sim3$, respectively. Individual exposures of 3600 s were obtained and repeated three times for the $z\sim2$ objects (a total of 3 h per target) and four times for the $z\sim3$ objects (a total of 4 h per target). Dithering within the ERIS FoV was applied to enable accurate sky subtraction without loss of on-source time. 

\subsubsection{GEMINI/GNIRS}

The GNIRS observations were conducted as part of the program GN-2024A-Q-302 (PI: Farina) at Gemini North. The instrument was configured in the LR-IFU mode, which has a FoV of $3.15\arcsec\times4.80\arcsec$. We used the Short Blue camera, which provides a pixel scale of $0.15\arcsec$/pixel and a grating of 32 l/mm, for a nominal spectral resolution of R$\sim$1700. To cover the redshifted [\ion{O}{iii}]$\lambda5007$ emission line from the quasar, the observations were taken in the K-band. The data was collected at two position angles: PA=0 and PA=90, each comprising 2$\times$ABA' sequences (where B indicates an empty sky location and A and A' are two slightly dithered exposures on the source). Each exposure lasted 200 s, for a total time of 1600 s on target.

\section{Data reduction} \label{sect:data_reduction}

\subsection{VLT/ERIS} \label{sect:eris_reduction}

We performed the data reduction of the targets observed with VLT/ERIS using the ESO ERIS-SPIFFIER Pipeline (v. 1.7.0), which consists of eight consecutive tasks. The first step, carried out by the \texttt{eris\_ifu\_dark} task, calculates a master dark frame and a bad pixel mask using five dark exposures taken during the observation night. Next, the task \texttt{eris\_ifu\_detlin} uses a set of one hundred and fifty flat frames taken with increasing exposure time to create a map of detector pixels with non-linear response to input light. As the raw detector image is bent horizontally, \texttt{eris\_ifu\_distortion} performs a North-South data reduction by deriving a set of 2D polynomials using calibration frames from slit masks and arc lamps. These polynomials warp each slitlet to a rectified 64-pixel-wide image, aligning edges and straightening spectral traces.

Following, \texttt{eris\_ifu\_flat} uses ten flat frames to determine the master flat and corrects for detector responsiveness. The wavelength calibration is performed by the task \texttt{eris\_ifu\_wavecal}, which uses three arc lamps from Neon, Argon, and Kripton to compare to a reference line list. The next step in the data reduction process uses a standard star observed right before or after the science observation. However, as this was not available, for all the targets in our sample we used the observation of the archival standard star Feige 110 in the same band, with the same pixel scale and grating, closest to the night of our science observation. We use the observation of the standard star with the task \texttt{eris\_ifu\_stdstar} to calculate the instrument response function and the flux calibration. The data cubes of the multiple science frames in an observation are created by the task \texttt{eris\_ifu\_jitter}. This task combines and applies the previously mentioned corrections to each of the raw science frames. Additionally, it performs the sky subtraction according to the Ric Davies method \citep{davies2007method} using the closest in time off-source sky frame. 

A correction of the Differential Atmospheric Refraction is applied at this stage to account for the effect of the stratified density structure of Earth's atmosphere, which displaces a source by an amount dependent on the wavelength. The final step in the data reduction process corresponds to the combination of the data cubes obtained by \texttt{eris\_ifu\_jitter} for the multiple frames within a single observation, and for different observations of the same target. This step is carried out by the task \texttt{eris\_ifu\_combine\_hdrl}. As the target is dithered inside the FoV, its position in each frame is different; therefore, the combination of data cubes is done based on the offset of the source in each frame, relative to its position in the first frame of the first observation. The offsets in the \textit{x} and \textit{y} directions are calculated by subtracting the positions of the flux centroids of two consecutive data cubes, estimated using a 2D-Gaussian fit. 

We processed the final combined data cube of each target to correct for telluric absorption and performed the flux calibration using archival observations of standard stars observed at similar airmass and with the same setup as the science targets.

The end product corresponds to a flux and wavelength-calibrated, sky-corrected combined cube, with a pixel scale of 125 mas/pixel. The rest-frame optical wavelength range varies slightly for each target, but it is generally between  $4600-5600\AA$.

\subsection{GEMINI/GNIRS}

We performed the data reduction of the targets observed with GEMINI/GNIRS within the Gemini IRAF environment \citep[][]{GeminiIRAF}, adding custom made routines to improve the local sky subtraction and the alignment of the different frames. We carried out the single cube combination with the pyfu package\footnote{https://github.com/jehturner/pyfu}. The telluric absorption correction and relative flux calibration were performed using A0V stars observed at a similar airmass and in the same configuration as the science targets.

\section{Data analysis} \label{sect:data_analysis}

\subsection{Spectra fitting} \label{sect:spec_fitting}

The main aim of this work is to use the [\ion{O}{iii}]$\lambda 5007$ emission line to trace and characterize the presence of AGN-driven outflows in the ISM. This procedure requires spatially resolving the kinematics of the gas at such scales, which implies performing a spaxel-by-spaxel fit of the IFU observations to identify the [\ion{O}{iii}]$\lambda 5007$ line properties across the FoV. This process was carried out in two main steps, following the approach from previous studies tracing AGN-driven outflows \citep{kakkad2020super, tozzi2021connecting}. First, we model the BLR emission using an integrated spectrum extracted from the nuclear region of the quasar. This BLR emission is unresolved in our observations; therefore, we do not expect any change in its kinematics across the FoV (see Section \ref{sect:mod_BLR}). Then, we use this BLR template to perform the spaxel-by-spaxel modeling and spatially resolve the ionized emission at ISM scales (see Section \ref{sect:spax_fit}).

\subsubsection{Modeling the unresolved emission} \label{sect:mod_BLR}

\begin{figure*}[ht]
    \centering
    \begin{tabular}{cc}
        \includegraphics[width=0.5\textwidth]{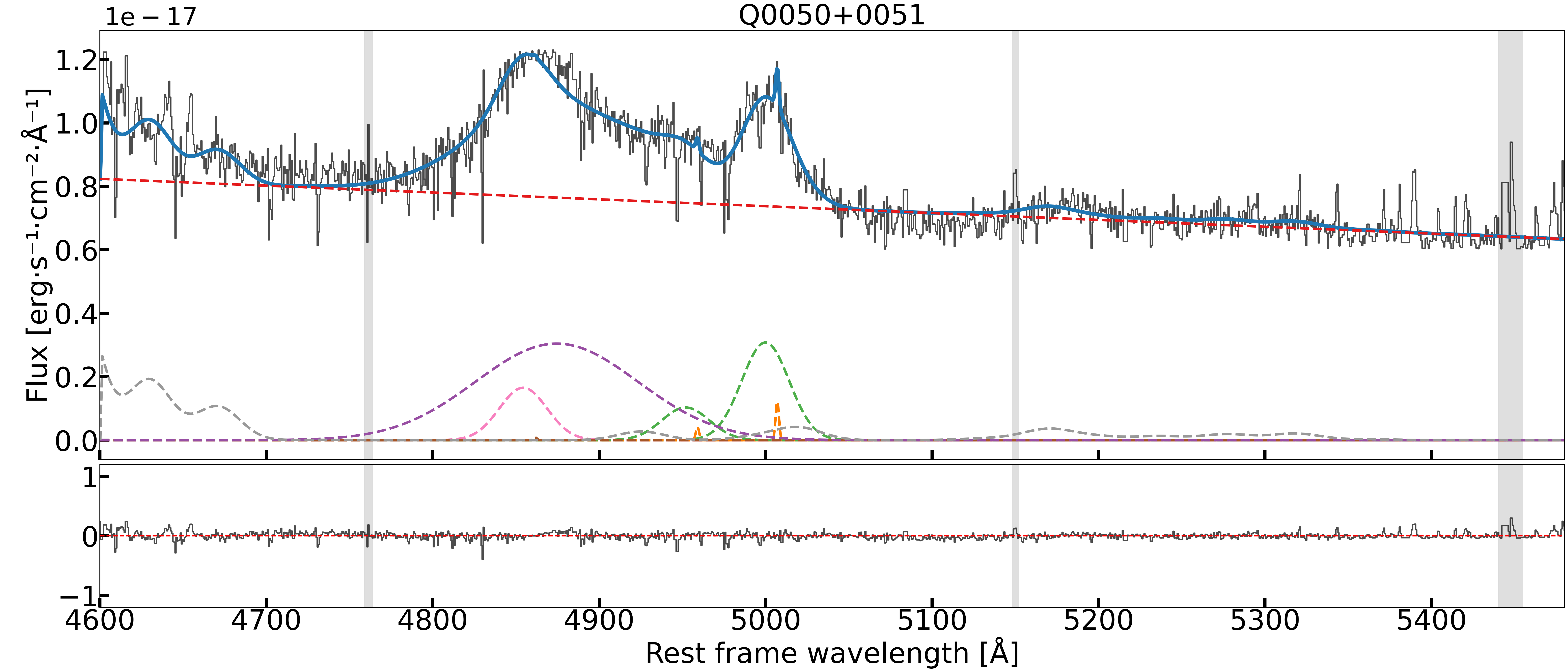} &
        \includegraphics[width=0.5\textwidth]{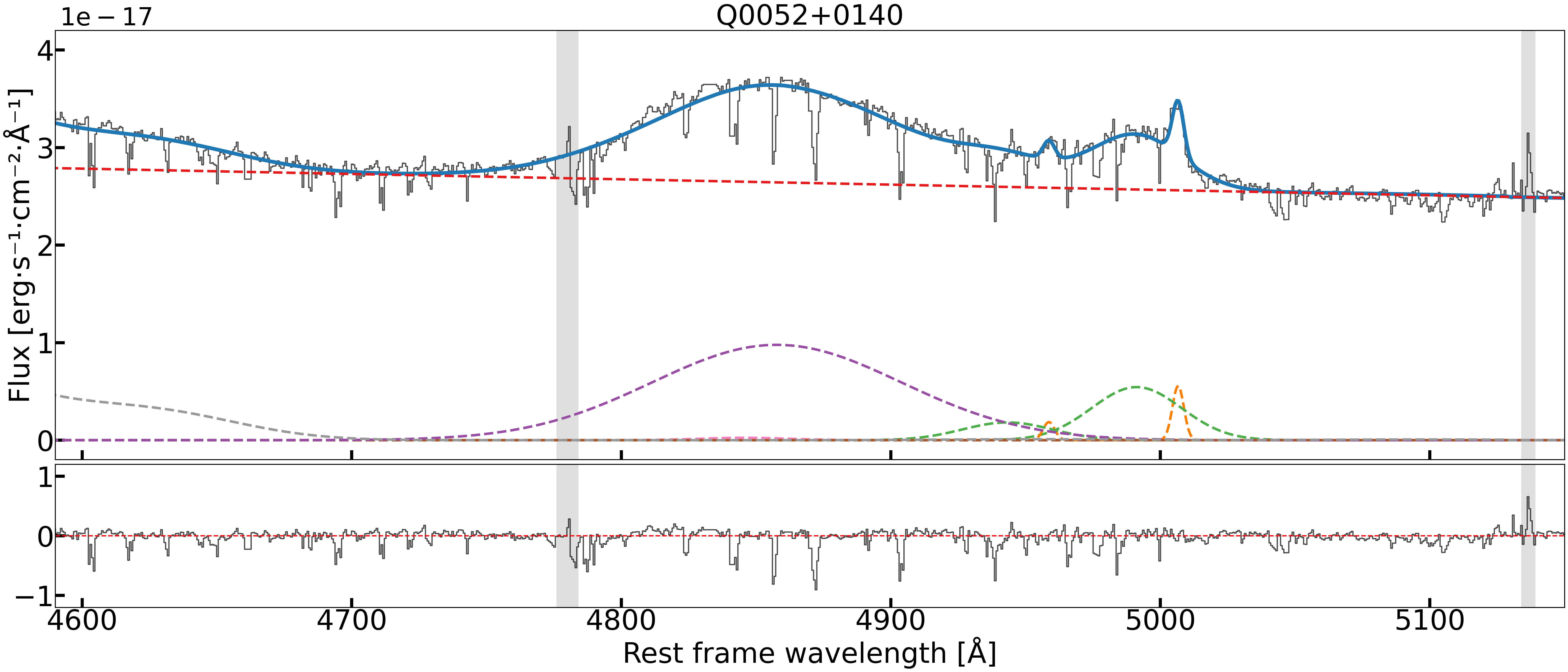} \\
        \includegraphics[width=0.5\textwidth]{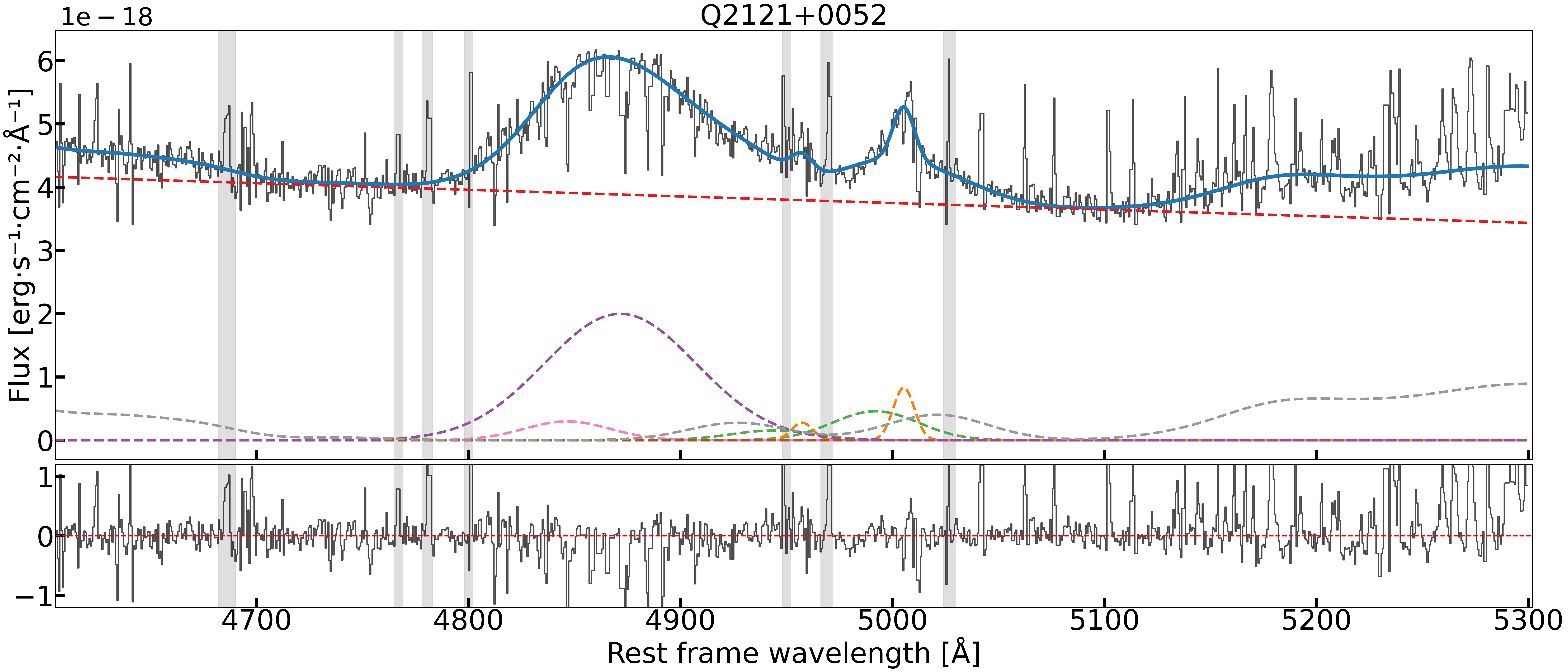} &
        \includegraphics[width=0.5\textwidth]{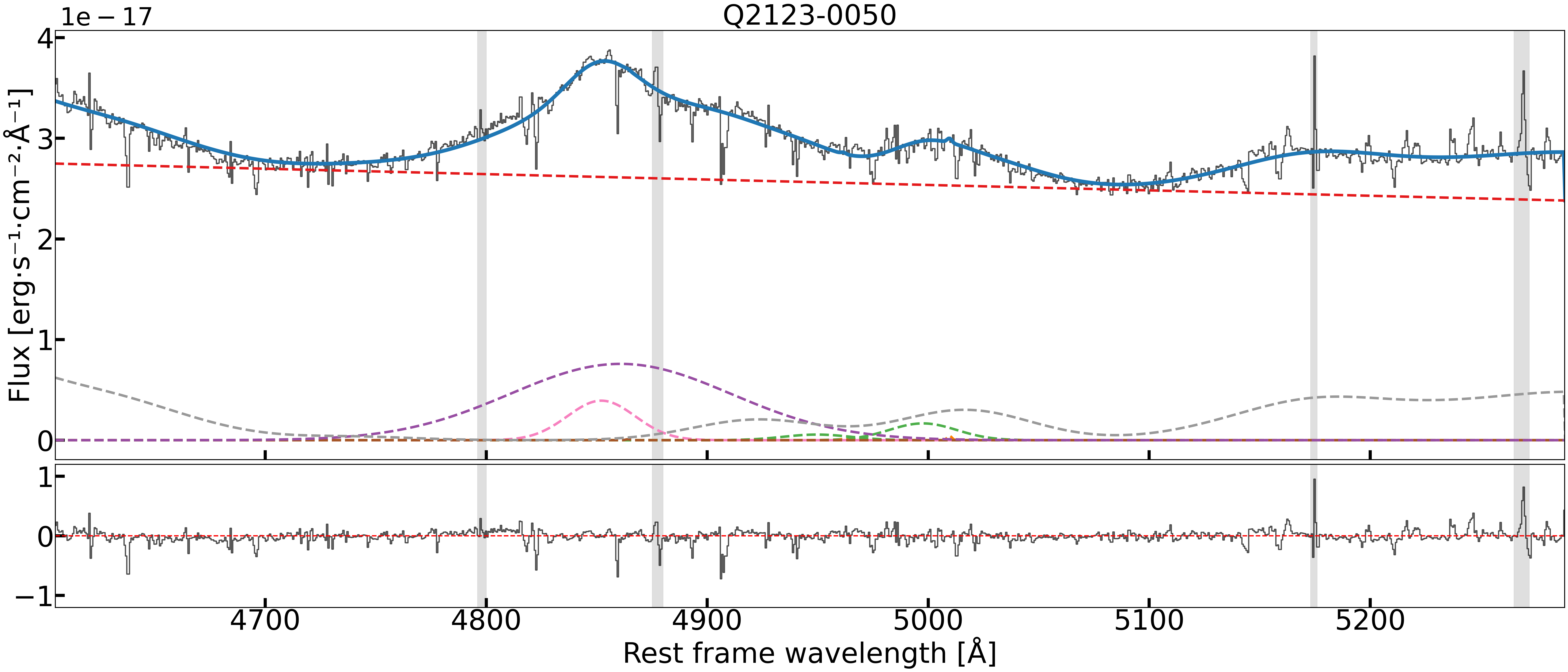} \\
        \includegraphics[width=0.5\textwidth]{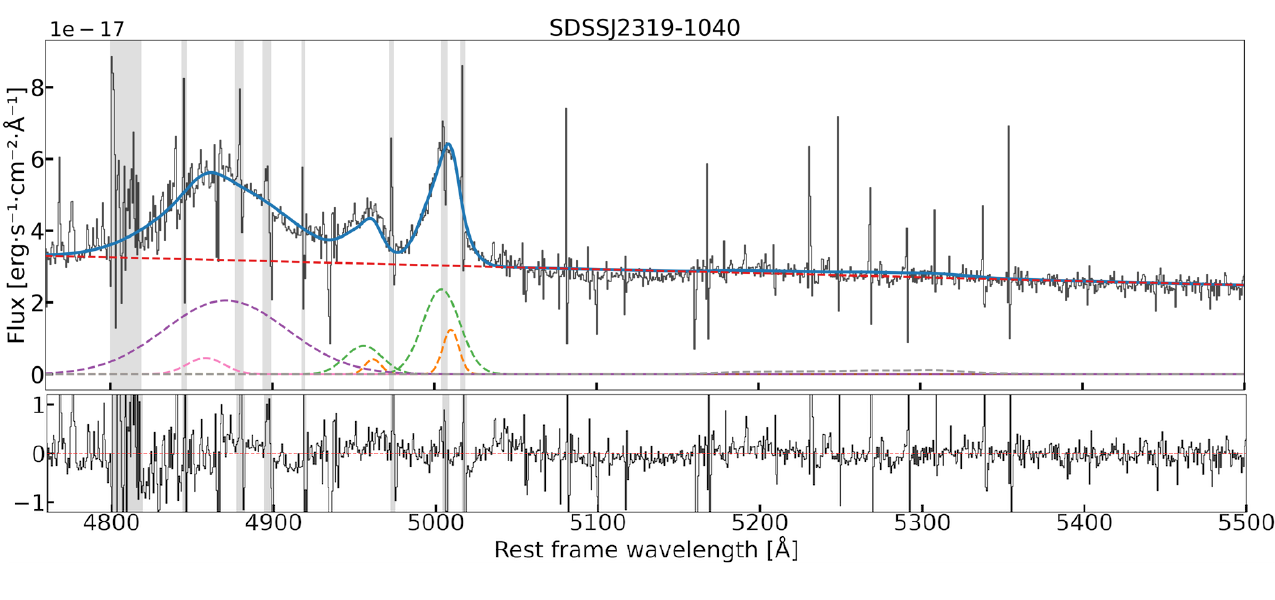} &
        \includegraphics[width=0.5\textwidth]{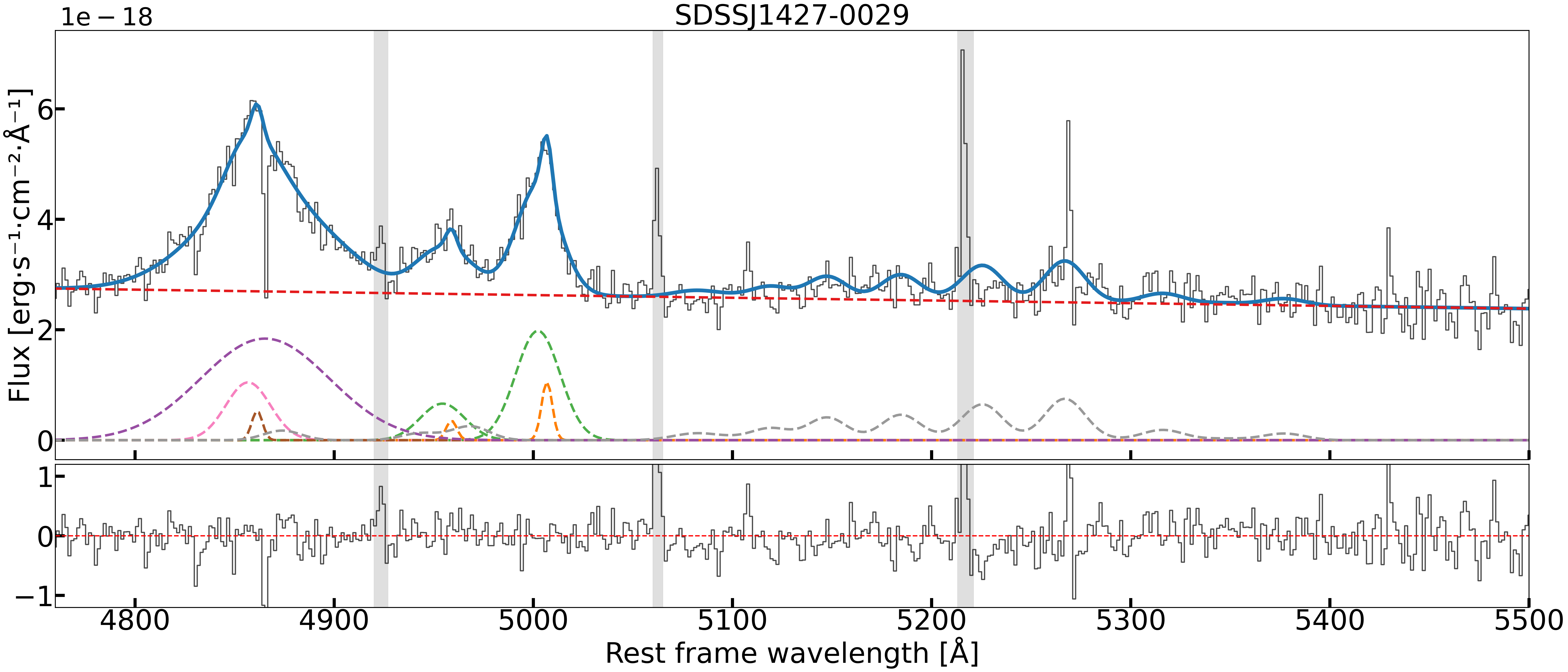} \\ 
    \end{tabular}
    \caption{Top: Fit model for an integrated spectrum with an aperture of one pixel around the quasar center.  The orange lines represent the [\ion{O}{iii}]$\lambda4959,5007$ NLR component. The green lines represent the outflowing component in the [\ion{O}{iii}]$\lambda4959,5007$ line. The H$\beta$ line is modeled with a broad component for the BLR shown in purple, a narrow component for the NLR shown in brown, and an additional broad component for the outflowing gas in pink. The FeII emission is shown in grey and the linear continuum in red. The total profile is shown as a solid blue line. Bottom: Residuals of the total fit. The gray-shaded bands in both panels correspond to remnant sky emission after applying the sigma-clipping algorithm mentioned in Sect. \ref{sect:mod_BLR}.}
    \label{fig:org_fit}
\end{figure*} 

We start by modeling the emission from an integrated spectrum extracted from a circular aperture with a radius of one pixel, centered around the quasar centroid. On this spectrum, we restricted the wavelength range to exclude noisy channels at the edges. The spectrum was centered on the systemic velocity given by the peak of the [\ion{O}{iii}]$\lambda 5007$ line. As the sky–subtraction procedure described in Section \ref{sect:eris_reduction} is not optimal, we applied a sigma–clipping algorithm, rejecting flux values above $6\sigma$ and below $2\sigma$. The clipped values were replaced using a piecewise linear interpolation method. The resulting \textit{H}-band spectrum contains the doublet [\ion{O}{iii}]$\lambda 4959, 5007$, a strong H$\beta$ line, a contribution from the quasar continuum, and FeII emission (e.g., see Fig. \ref{fig:org_fit}). Each of the spectral components is modeled simultaneously using the \texttt{lmfit} package in Python, as explained below.

\begin{itemize}

    \item The continuum was modeled using a linear function (red line in Fig. \ref{fig:org_fit}). 
    
    \item Each of the components of the doublet [\ion{O}{iii}]$\lambda 4959, 5007$ was modeled with two Gaussians: a narrow component (orange line in Fig. \ref{fig:org_fit}), and a broad component to reproduce the extended line wing(s), which point to the presence of outflows (green line in Fig. \ref{fig:org_fit} ). For the narrow component of [\ion{O}{iii}]$\lambda 5007$, the prior on the centroid was set to the vacuum wavelength of $5008.239\AA$, with $\mathrm{FWHM\leq150~km~s^{-1}}$. For the broad component, the centroid was restricted to the range [$4900-5100~\AA$], and the prior on the FWHM was set to be three times higher than that of the narrow component. The centroid of the narrow and broad components of [\ion{O}{iii}]$\lambda 4959$ was set to have a fixed offset of $47.94\AA$ from that of the narrow and broad components of [\ion{O}{iii}]$\lambda 5007$, respectively. This value is estimated from the vacuum wavelength of [\ion{O}{iii}]$\lambda 4959$ ($4960.295\AA$). The FWHM of both components of [\ion{O}{iii}]$\lambda 4959$ was fixed to be equal to that of both components of [\ion{O}{iii}]$\lambda 5007$, as they are expected to originate from the same gas. 
    
    The ratio of the amplitudes of [\ion{O}{iii}]$\lambda5007$ to [\ion{O}{iii}]$\lambda 4959$ was fixed to 2.99 for both the narrow and broad components, as demonstrated by atomic physics \citep{storey2000theoretical, dimitrijevic2007flux}.

    \item The H$\beta$ line was modeled using three Gaussian components: a very broad one ($\mathrm{FWHM>1500~km~s^{-1}}$), centered at the vacuum wavelength of H$\beta$ ($\lambda 4862.721\AA$) (purple line in Fig. \ref{fig:org_fit}), and an additional narrow (brown line in Fig. \ref{fig:org_fit}), and broad component (pink line in Fig. \ref{fig:org_fit}). The centroid and FWHM of these last two narrow and broad components of H$\beta$ were tied to those of [\ion{O}{iii}]$\lambda 5007$, respectively.
    
    \item The FeII emission was modeled using the semi-analytic templates of \citet{kovavcevic2010analysis} (grey line in Fig. \ref{fig:org_fit}). We fit each of the 21 templates and chose the one that generated the fit with the minimum $\chi^2$. 
   
\end{itemize}

After modeling and fitting the integrated spectrum with the multiple components described above, we determined the velocity centroid and FWHM of the BLR contribution to H$\beta$, together with the relative amplitudes of the components of the adopted Fe II template. These parameters were kept fixed during the spaxel–by–spaxel fitting, as we do not expect them to change across the FoV. This assumption is based on the spatially unresolved nature of the BLR (H$\beta$+Fe II), which implies that its emission scales in amplitude like the instrumental PSF across the FoV.

\begin{figure*}[h]
    \centering
    \includegraphics[width=1\linewidth]{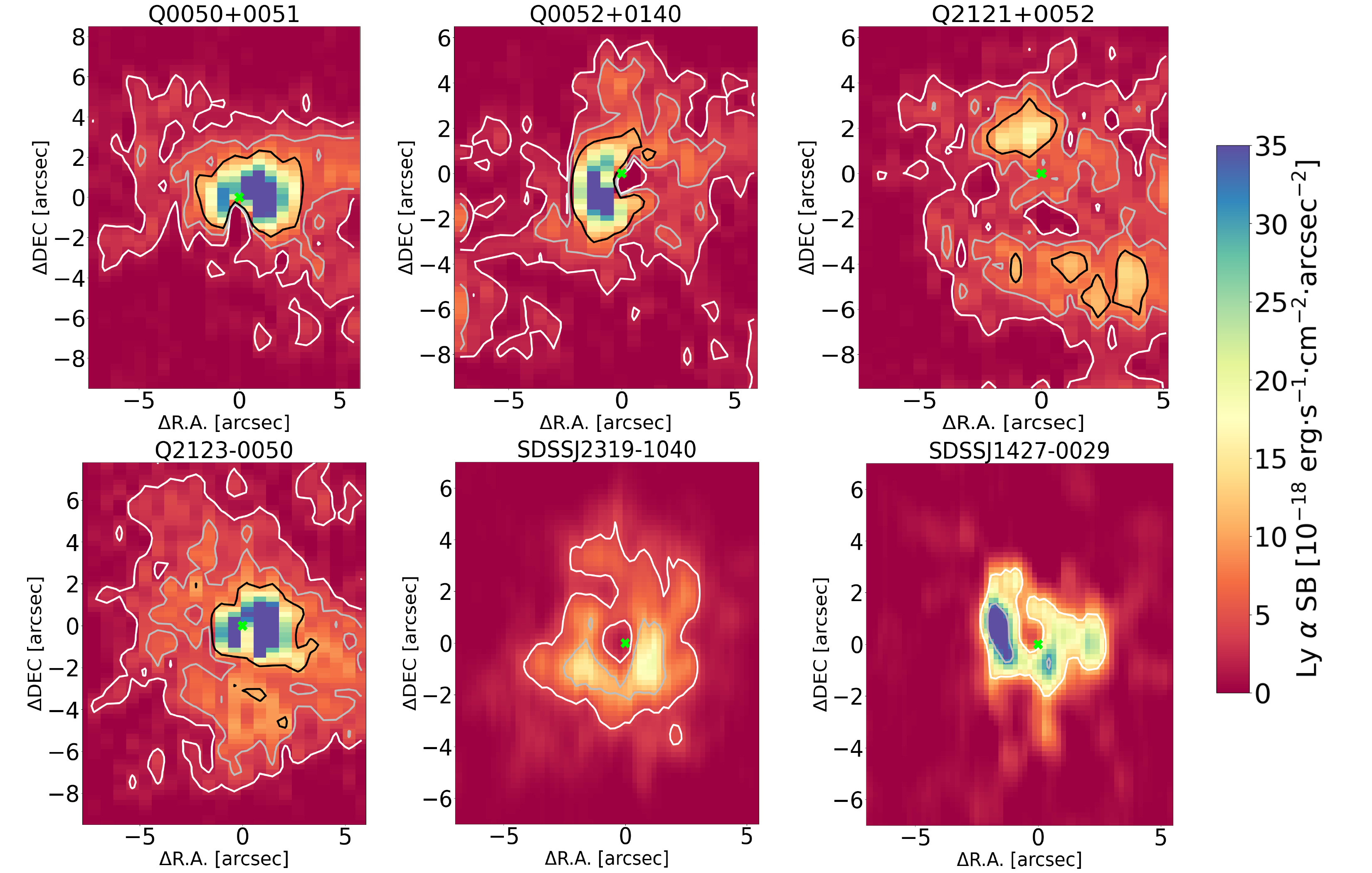}
    \caption{Optimally extracted SB maps of the Ly$\alpha$ nebulae surrounding the quasars (see Sect.~\ref{sect:lya_props}). The white, grey and black contours represent $2\sigma$, $5\sigma$ and $10\sigma$, respectively. The values of $\sigma$ are reported in Table~\ref{tab:Lya_props}. The green cross in the center represents the position of the quasar, determined by estimating the photocentroid of a channel map in a wavelength where the Ly$\alpha$ line is not present. North is up and East is left.}
    \label{fig:Lya-SB_maps}
\end{figure*}

\subsubsection{Spaxel-by-spaxel fitting} \label{sect:spax_fit}

We modeled the emission in each spaxel of the datacube using the same spectral components as for the integrated spectrum (Sect. \ref{sect:mod_BLR}), fixing the BLR velocity centroid and dispersion to the values derived previously and allowing only its amplitude to vary. Because the S/N varies across the FoV, applying this model uniformly may lead to overfitting in low S/N regions. To account for this, we evaluated each spaxel individually to determine the number of Gaussian components required to reproduce the [\ion{O}{iii}]$\lambda4959,5007$ doublet and H$\beta$ emission. For this purpose, we tested two different models: Model A, identical to that used for the integrated spectrum modeling the BLR emission (Sect. \ref{sect:mod_BLR}), with two Gaussian components for each [\ion{O}{iii}] feature and three for H$\beta$. And Model B, a simplified version with one Gaussian component per [\ion{O}{iii}] feature and two for H$\beta$: a very broad component ($\mathrm{FWHM>1500~km~s^{-1}}$) tracing the emission from the BLR and a narrow component tracing the emission from the NLR.

For both models, we subtracted the contribution of all unresolved components (continuum + H$\beta$ + [\ion{Fe}{ii}]) and that of the [\ion{O}{iii}]$\lambda4959$ line from the observed spectrum. This procedure isolates the [\ion{O}{iii}]$\lambda5007$ emission and removes blending with adjacent components. The residual spectrum obtained after subtracting Model A was fitted with two Gaussians, while that obtained after subtracting Model B was fitted with a single Gaussian.

We then computed the Bayesian Information Criterion (BIC) for both versions of the fit of the subtracted spectrum.  We chose to fit the spaxel with Model A when the inclusion of additional Gaussian components provided a statistically significant improvement (i.e., when the BIC difference between the two models exceeded a target-specific threshold). Otherwise, Model B was adopted to avoid overfitting. The resulting modeled cube was visually inspected, and further manual adjustments were applied where necessary.

The properties of the [\ion{O}{iii}]$\lambda5007$ emission line ([\ion{O}{iii}] hereafter) used in the subsequent analysis were derived from the preferred fit of the subtracted spectrum.

\subsection{Ly$\alpha$ nebulae properties} \label{sect:lya_props}

The extended Ly$\alpha$ nebulae around the quasars in our sample were studied in a previous work by \cite{cai2019evolution} for targets at z$\sim2$. These observations with KCWI have a FoV of $16.8\arcsec\times20\arcsec$ and seeing in the range of $0.8\arcsec-1.0\arcsec$. For targets at redshift z$\sim3$, the nebulae were studied by \cite{arrigoni2019qso} using MUSE. These observations have a FoV constrained to the central $55\arcsec\times55\arcsec$, with seeing of $1.07\arcsec$. Table \ref{tab:sample_props} summarizes the instrument used to detect the extended Ly$\alpha$ emission around our quasars. From these studies, we used the extracted SB maps and re-analyzed them to homogeneously define the main properties of the Ly$\alpha$ nebulae across our sample. The Ly$\alpha$ SB maps are shown in Fig. \ref{fig:Lya-SB_maps}, smoothed with a Gaussian filter of $\mathrm{FWHM}=1"$. We overlay the contours corresponding to S/N=2,5 and 10, i.e., each contour encloses regions where the SB lies above the 2, 5, and 10$\sigma$ limit, respectively. $\sigma$ corresponds to the noise level estimated in the previous studies and tabulated in Table \ref{tab:Lya_props}.

We characterize the Ly$\alpha$ nebulae using the S/N>2 isophote. For each source, we measure the area enclosed by this contour and define $D_{\mathrm{Ly\alpha}}$ as the diameter of a circle with the same area, which we adopt as our size estimator. Size uncertainties are dominated by calibration errors, which we assume as $5\%$ \cite{bacon2010muse}. Therefore, we constrain the Ly$\alpha$ nebulae size by measuring the S/N>1.9 and S/N>2.1 isophotes, which we quote as the upper and lower bounds, respectively.

The Ly$\alpha$ luminosity, $L_{\mathrm{Ly\alpha}}$, is obtained by integrating the SB within the S/N>2 contour. In some cases, this isophote extends beyond the FoV, therefore, both sizes and luminosities correspond to lower limits. The luminosity uncertainty is estimated as
$\Delta L_{\mathrm{Ly\alpha}}=\sigma_{\mathrm{bkg}}\cdot A_{\mathrm{pix}} \cdot N_{\mathrm{pix}}$,
where $\sigma_{\mathrm{bkg}}$ is the standard deviation of the SB in source–free regions, $A_{\mathrm{pix}}$ is the pixel area, and $N_{\mathrm{pix}}$ is the number of pixels enclosed by the S/N>2 contour. The resulting measurements are listed in Table \ref{tab:Lya_props}.

\begin{table}[!t]
    \renewcommand{\arraystretch}{1.4}
    \centering
    \caption[]{Properties of the Ly$\alpha$ nebulae around the QSOs in our sample.}
    \label{tab:Lya_props}
    \begin{tabular}{c c c c}
        \hline \hline
        Target ID & $2\sigma$\tablefootmark{$\star$} & $D_{\mathrm{Ly\alpha}}$ & $L_\mathrm{{Ly\alpha}}$  \\
        &&[kpc] & [$10^{43}~\mathrm{erg~s^{-1}}$] \\
        \hline
        Q0050+0051 & 1.8 & $> 96.16^{+0.09}_{-0.10}$ & $> 2.94 \pm 0.01$  \\
        Q0052+0140 & 1.7 & $> 69.93^{+0.90}_{-0.32}$ & $> 1.35 \pm 0.01$ \\
        Q2121+0052 & 1.8 & $> 104.93^{+0.01}_{-0.14}$ & $> 2.51 \pm 0.02 $ \\
        Q2123$-$0050 & 2.0 & $> 108.56^{+0.35}_{-0.12}$ & $> 5.42 \pm 0.01$\\
        SDSSJ2319$-$1040 & 4.1 & $50.72^{+0.14}_{-0.12}$ & $2.32 \pm 0.10$ \\
        SDSSJ1427$-$0029 & 12.4 & $32.98^{+0.05}_{-0.06}$ & $3.22 \pm 0.18$\\
        \hline
    \end{tabular}

    \vspace{4mm}
    \tablefoot{
        From left to right: Target ID, SB uncertainty, Ly$\alpha$ nebulae size, and luminosity determined using the S/N>2 isophote. \\
        \tablefoottext{$\star$}{In units of $10^{-18}~\mathrm{erg~s^{-1}cm^{-2}arcsec^{-2}}$.}
    }
\end{table}

\begin{figure*}[ht]
\centering

\setlength{\tabcolsep}{2pt}
\renewcommand{\arraystretch}{0.0}

\begin{minipage}[c]{0.84\textwidth}  
  \centering
  \begin{tabular}{@{}ccc@{}}
    \includegraphics[width=0.33\textwidth]{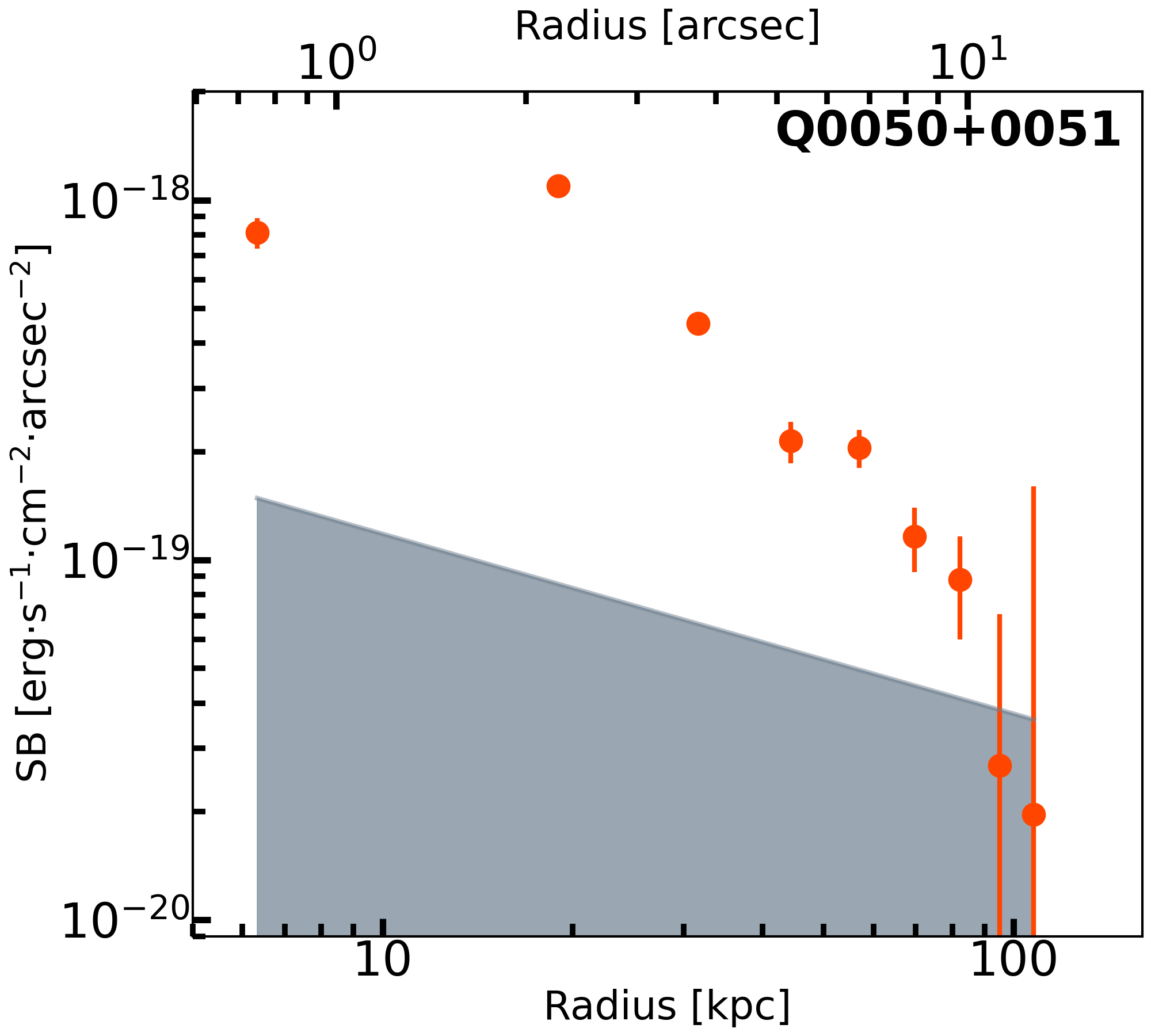} &
    \includegraphics[width=0.33\textwidth]{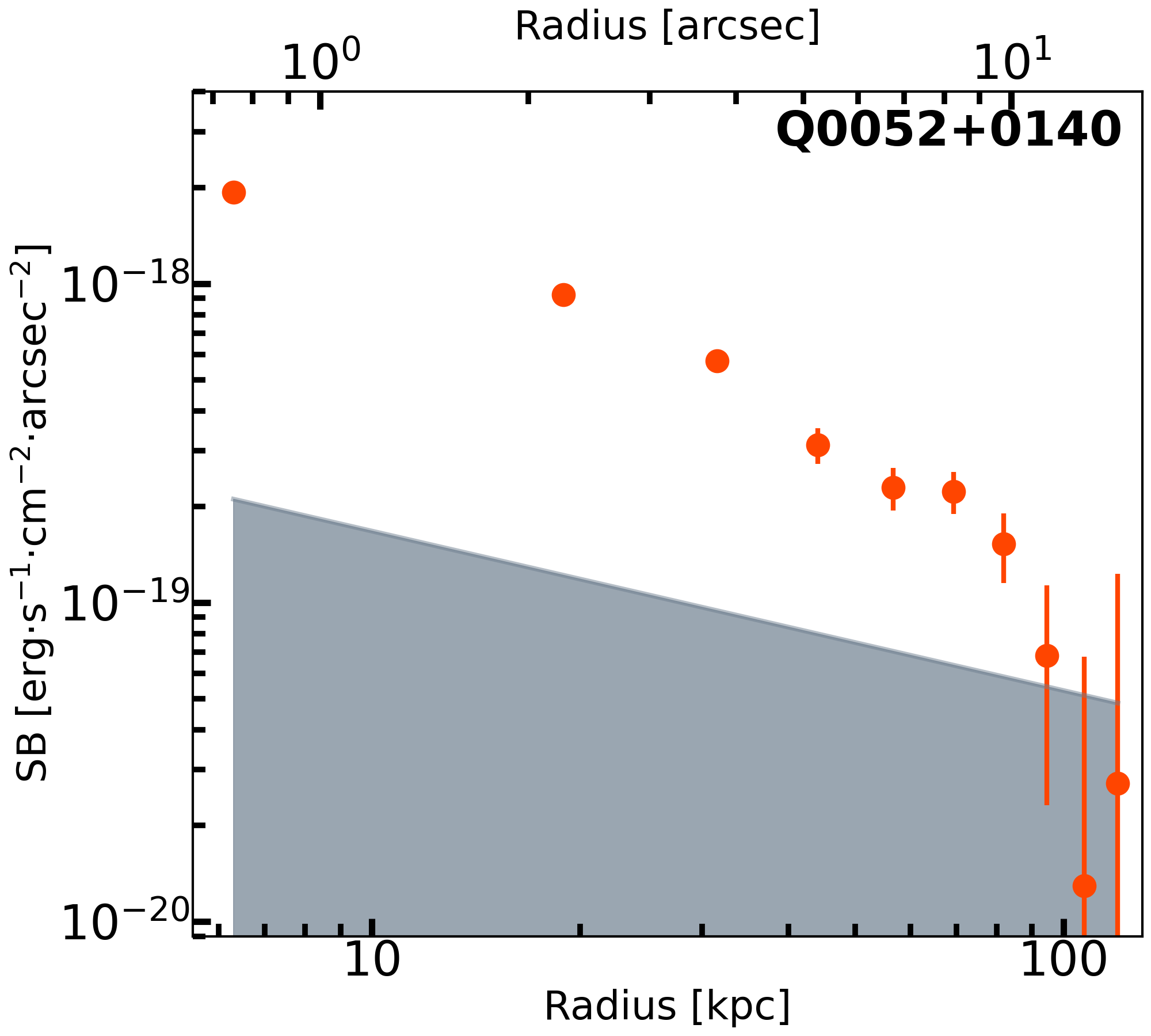} &
    \includegraphics[width=0.33\textwidth]{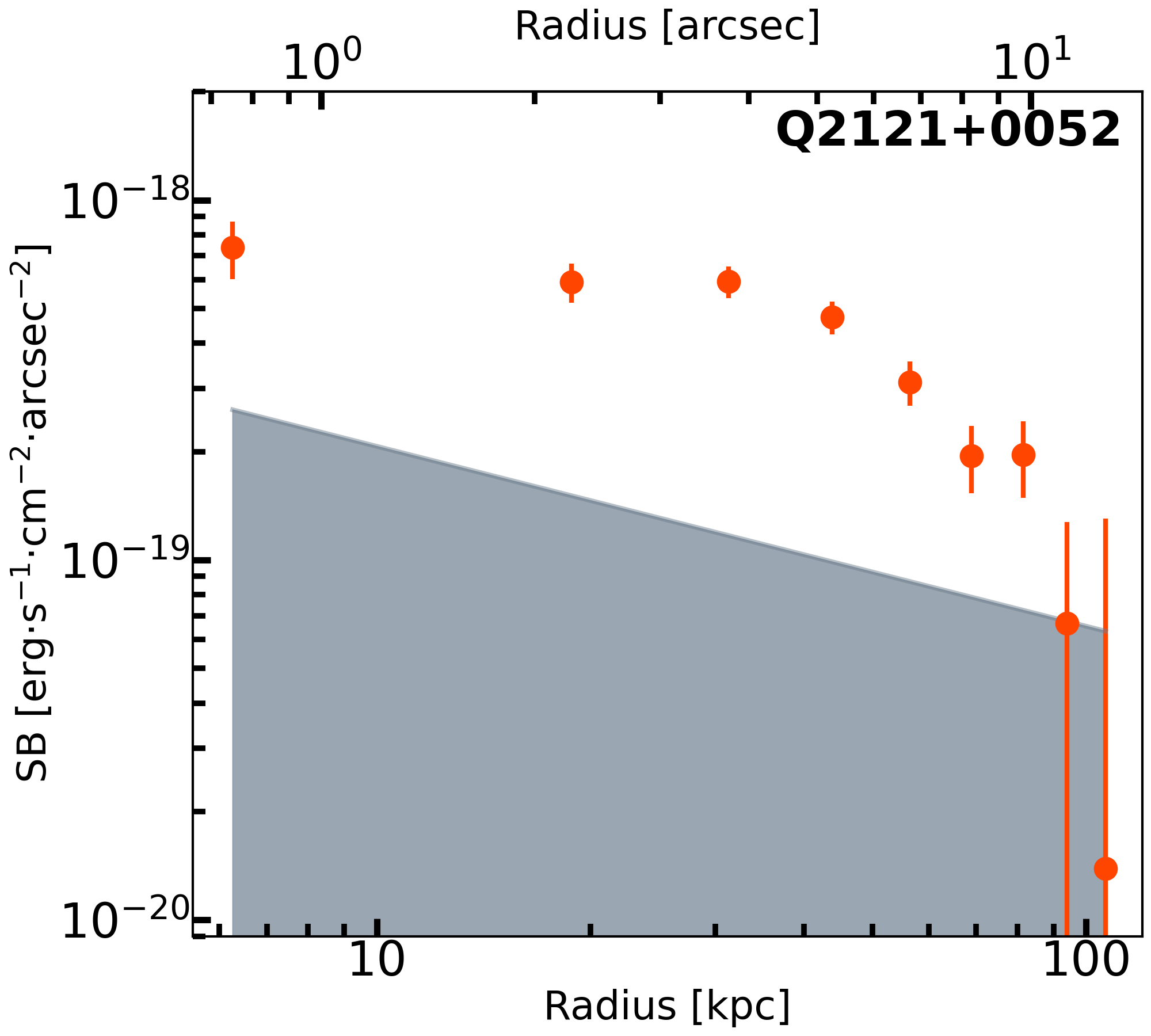} \\
    \includegraphics[width=0.33\textwidth]{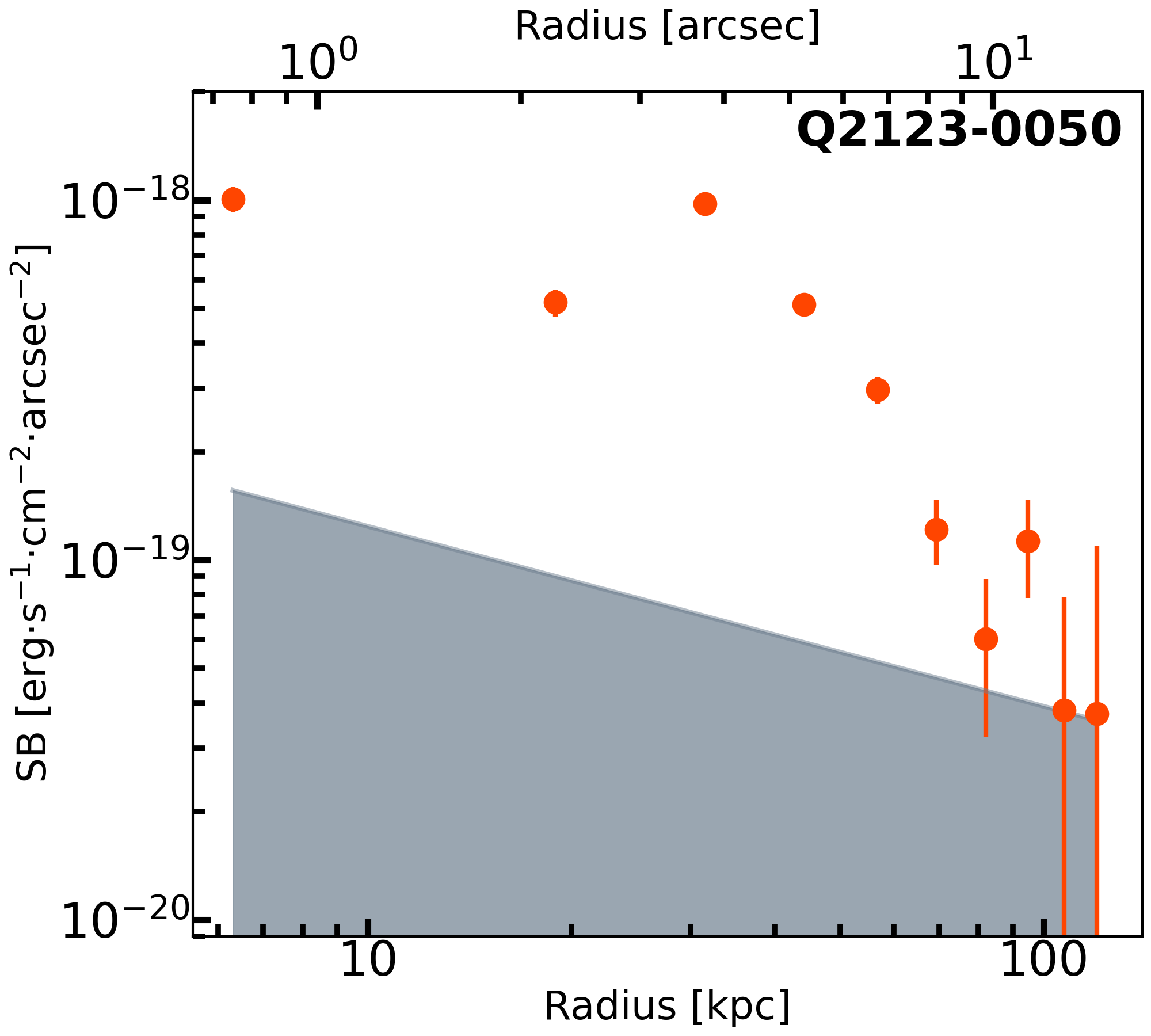} &
    \includegraphics[width=0.33\textwidth]{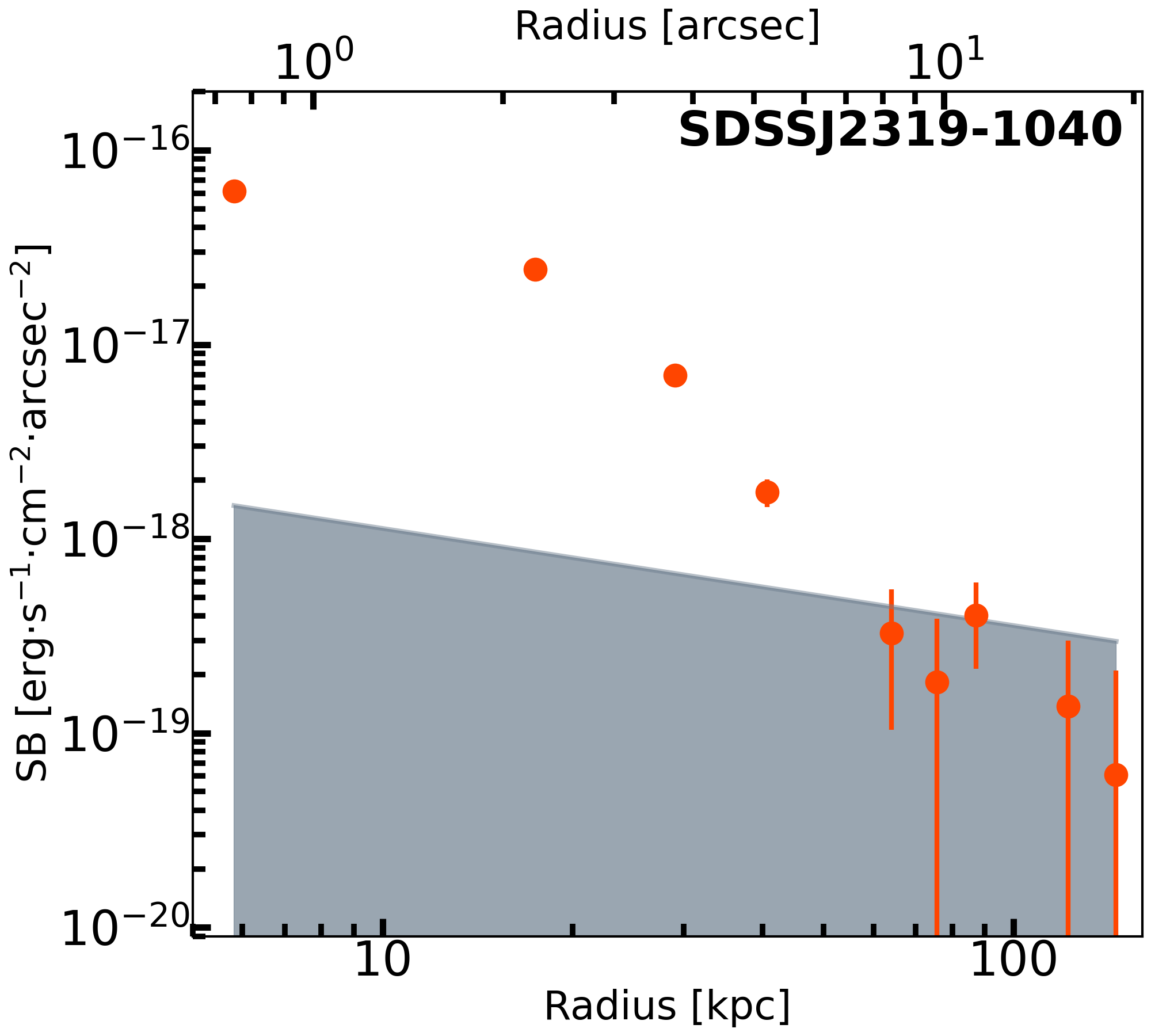} &
    \includegraphics[width=0.33\textwidth]{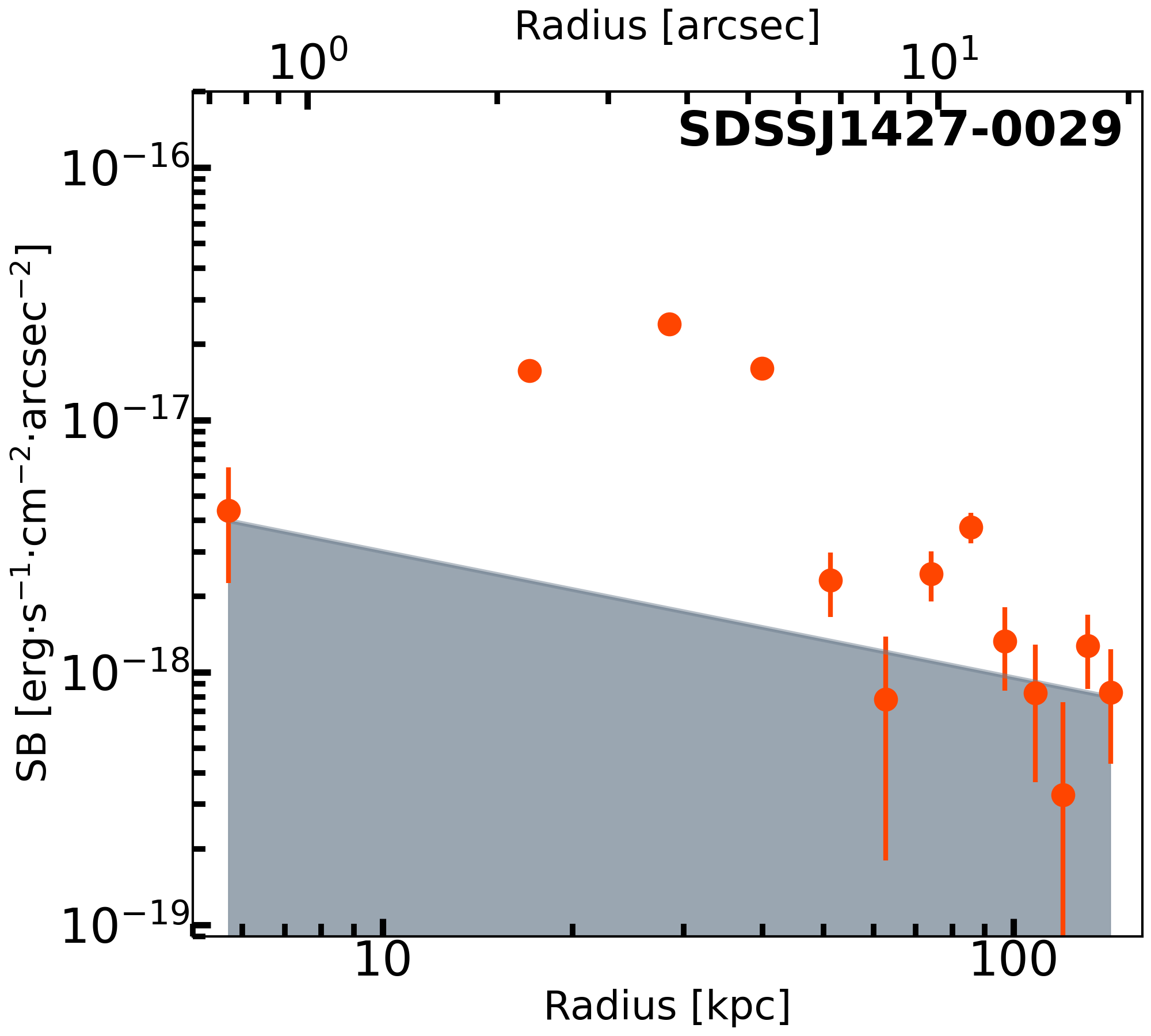} \\
  \end{tabular}
\end{minipage}%
\caption{Ly$\alpha$ SB radial profiles. The red points show the average SB per radial annulus, while the grey line denotes the $2\sigma$ detection limit estimated for each annulus. }
\label{fig:SB_lya_prof_p}
\end{figure*}

Following the approach by \cite{arrigoni2019qso}, we used $30\AA$ narrow-band images to derive the SB radial profiles of the nebulae, shown in Fig. \ref{fig:SB_lya_prof_p}. To do this, we created a grid of radial distances of every pixel from the quasar center, considering the rectangular pixel scale. Based on this, we defined bins (annuli) of the same radial distance and estimated the mean SB for each bin. The grey line in Fig. \ref{fig:SB_lya_prof_p} shows $\mathrm{SB_{annulus}^{limit}}$, corresponding to the $2\sigma$ detection limit estimated for each annulus. This value was estimated as $\mathrm{SB_{annulus}^{limit}}=2\mathrm{SB_{arcsec^2}^{lim}}/\sqrt{\mathrm{A_{annulus}}}$, where $\mathrm{A_{annulus}}$ corresponds to the area of the annulus, and $\mathrm{SB_{arcsec^2}^{lim}}$ corresponds to the SB limit per $\mathrm{arcsec^2}$. The latter was estimated as $\mathrm{SB_{arcsec^2}^{lim}} = \mathrm{SB_{pix}}\sqrt{A_{\mathrm{pix}}}$, where $\mathrm{A_{pix}}$ corresponds to the pixel area and $\mathrm{SB_{pix}}$ to the SB uncertainty per pixel. The SB uncertainty per pixel was calculated as the rms noise of out-of-source regions. For the extraction of the SB profiles, we masked the central source region, defined as a circle of radius 0.5 $\mathrm{arcsec^2}$, since this was used as a normalization region to subtract the quasar PSF and is affected by residuals. The profile is plotted up to the size of the nebula estimated using the S/N>2 isophote.

\section{Results} \label{sect:results}

\subsection{Spatially resolved and extended emission} \label{sect:res_emission}

After modeling the [\ion{O}{iii}] line emission across the FoV following our spectral fitting procedure (Sect. \ref{sect:spax_fit}), it is necessary to establish whether the observed emission is truly extended and spatially resolved, rather than being dominated by instrumental PSF or beam-smearing effects. Following the approach of \citet{kakkad2020super}, we conducted two independent tests to address this question.

\begin{figure*}[ht]
    \centering
    \setlength{\tabcolsep}{3pt} 
    \begin{tabular}{@{}ccc@{}} 
        \includegraphics[width=0.33\textwidth]{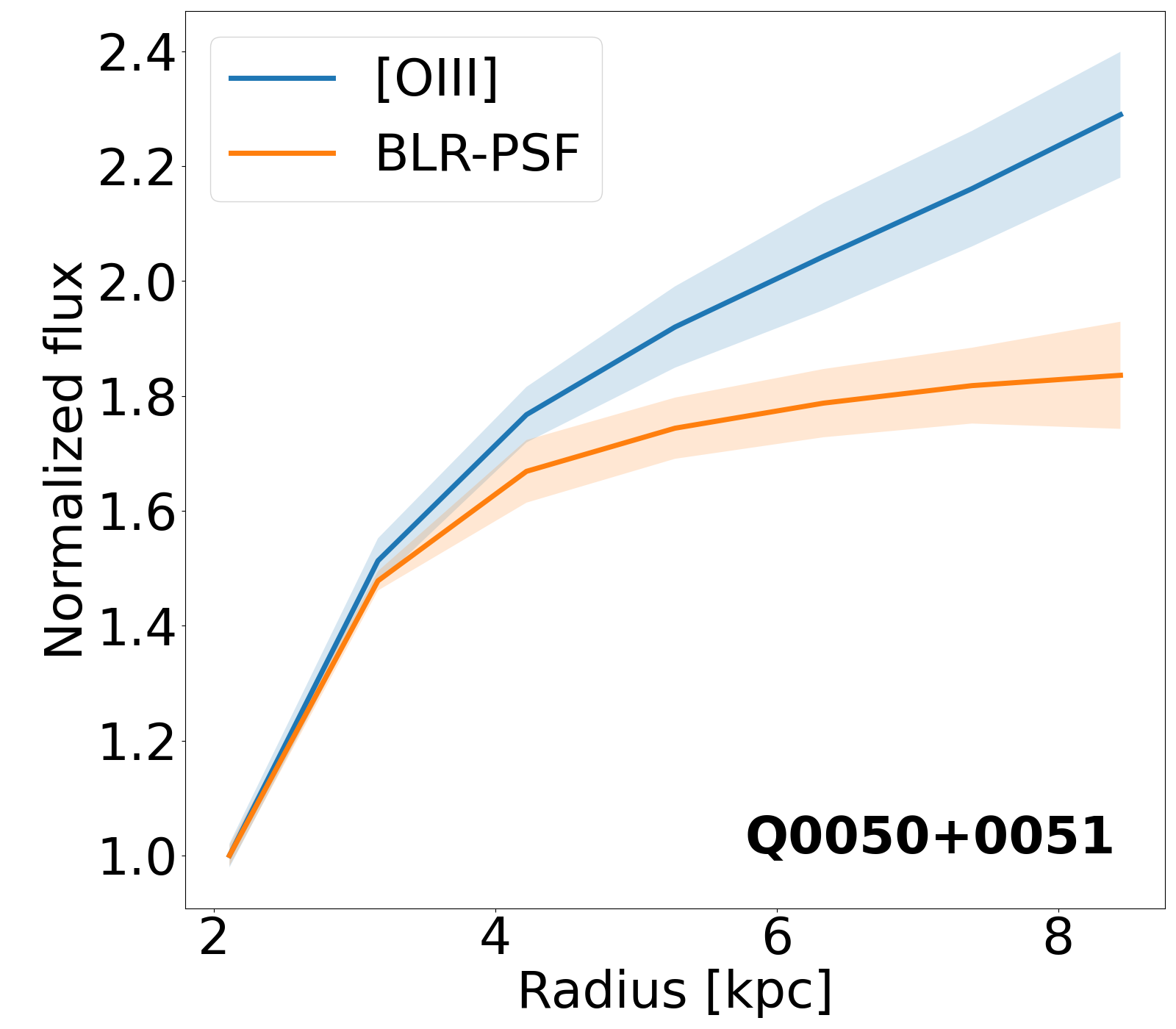} &
        \includegraphics[width=0.33\textwidth]{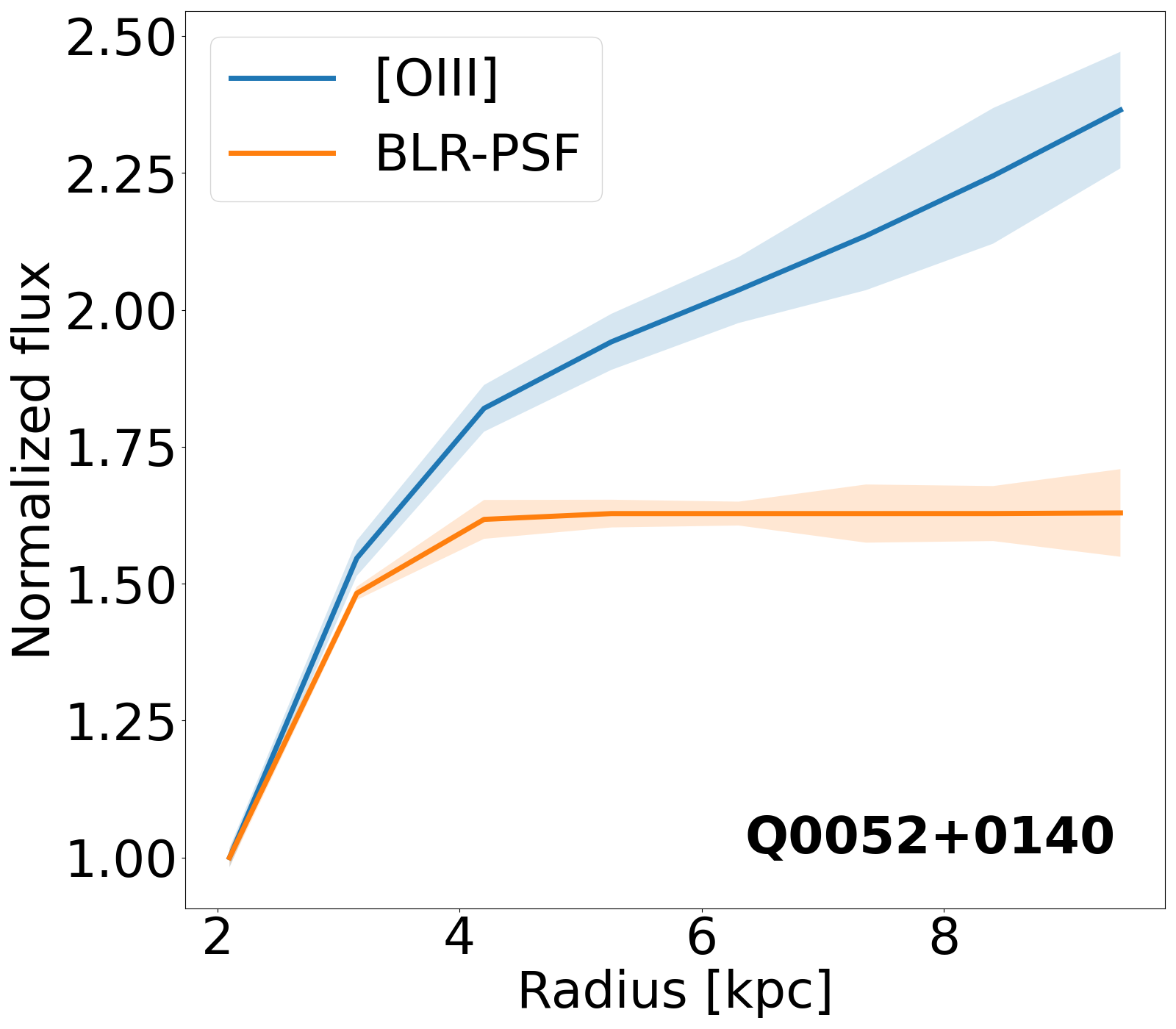} &
        \includegraphics[width=0.33\textwidth]{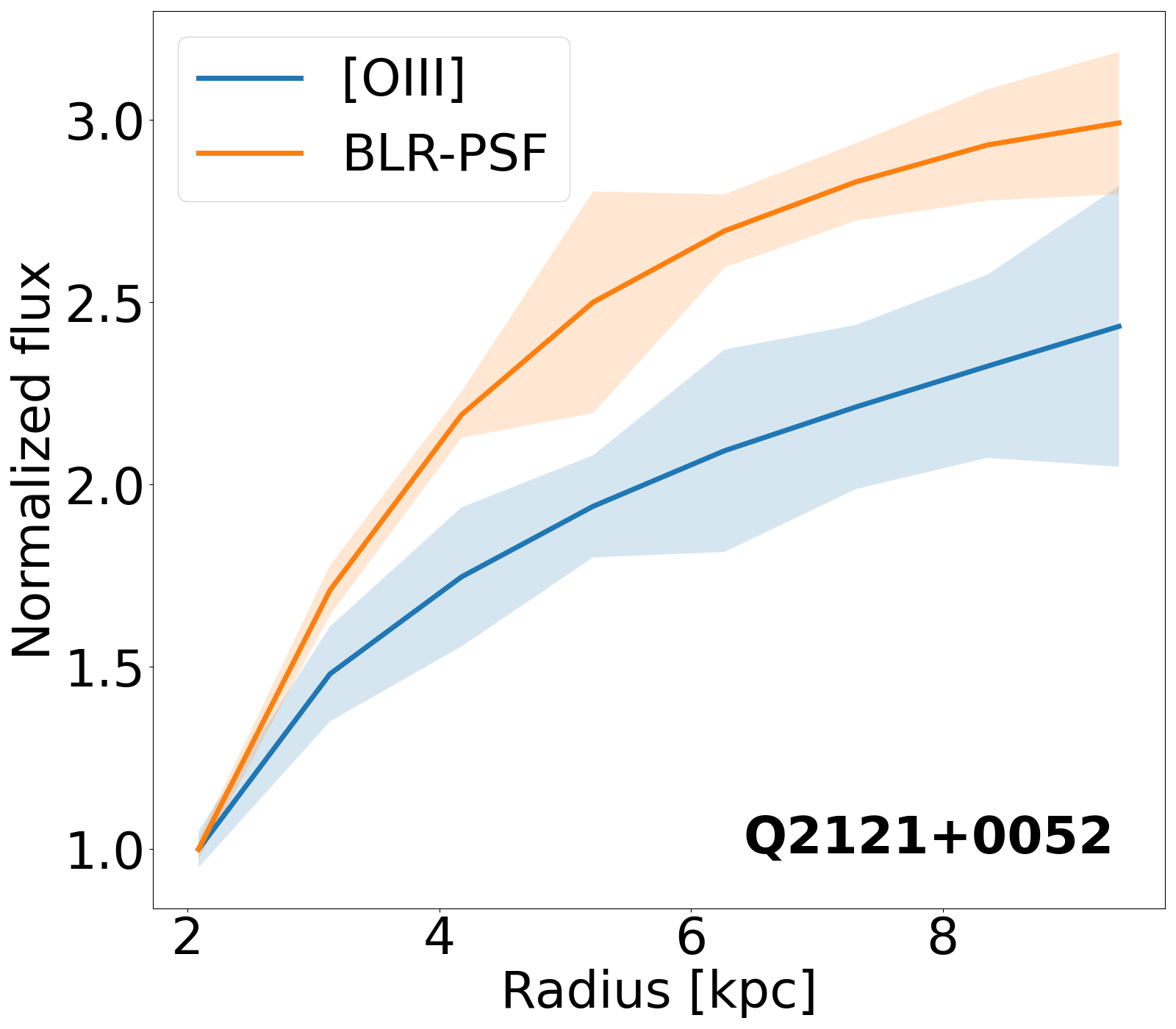} \\
        \includegraphics[width=0.33\textwidth]{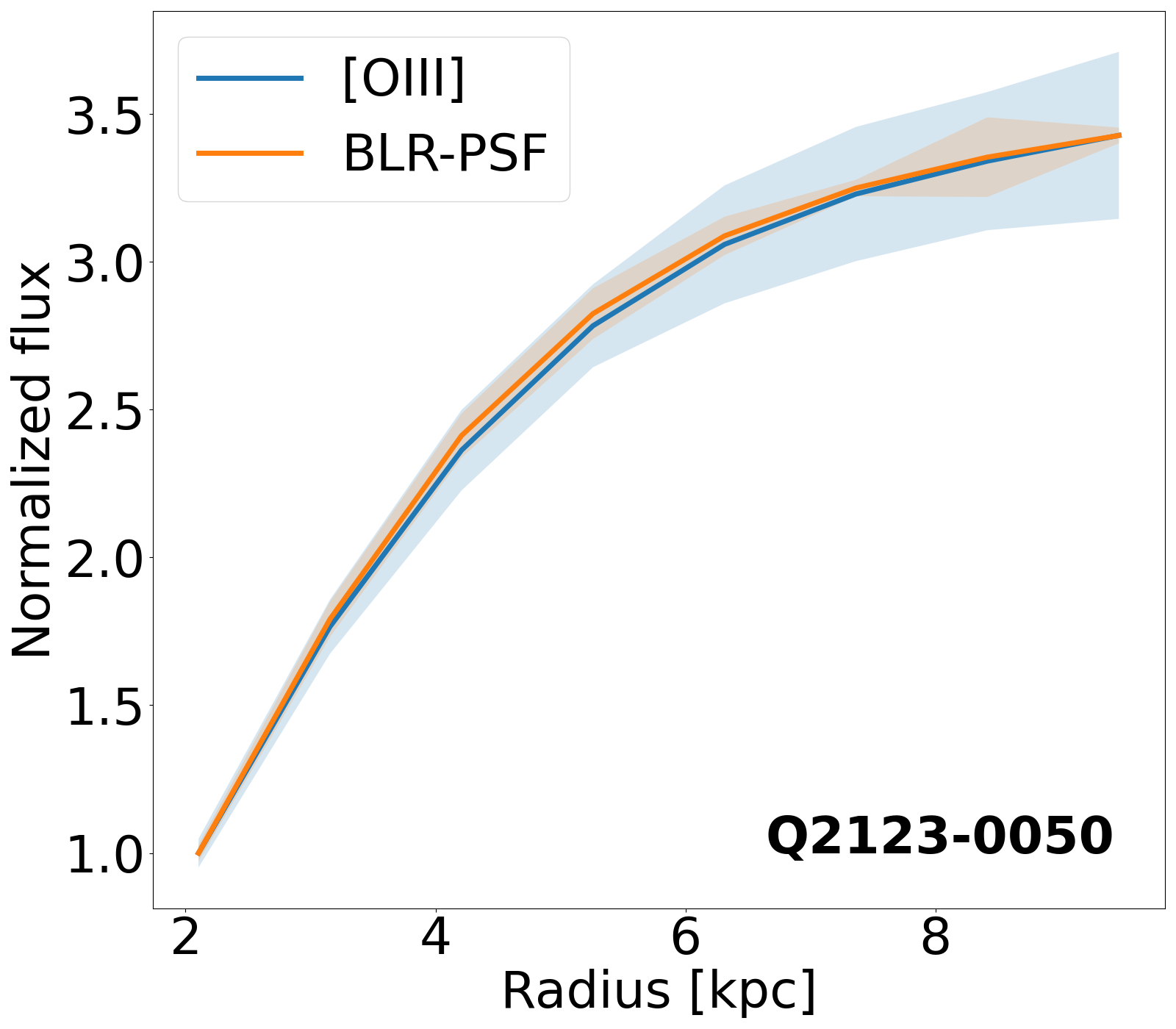} &
        \includegraphics[width=0.33\textwidth]{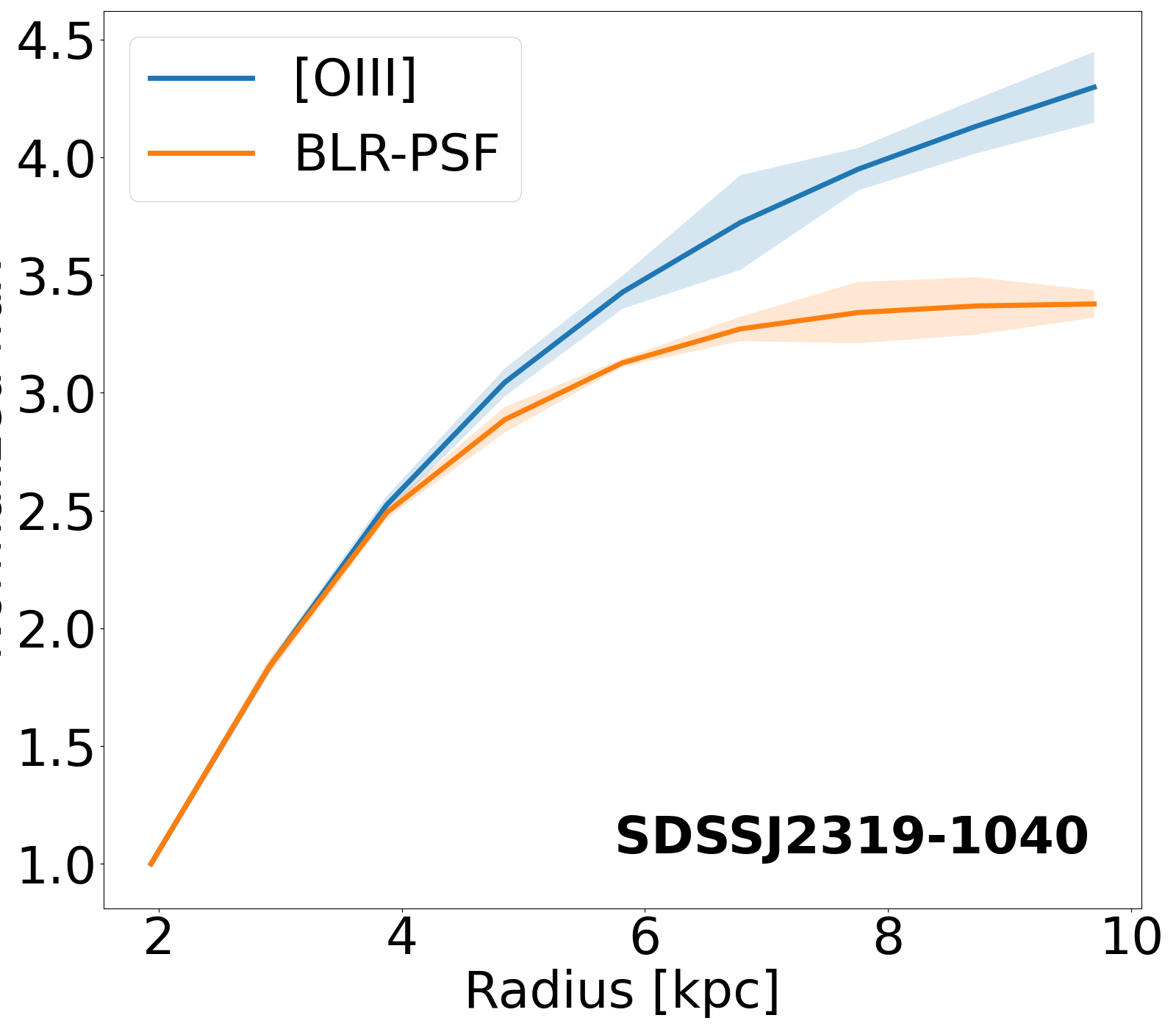} &
        \includegraphics[width=0.33\textwidth]{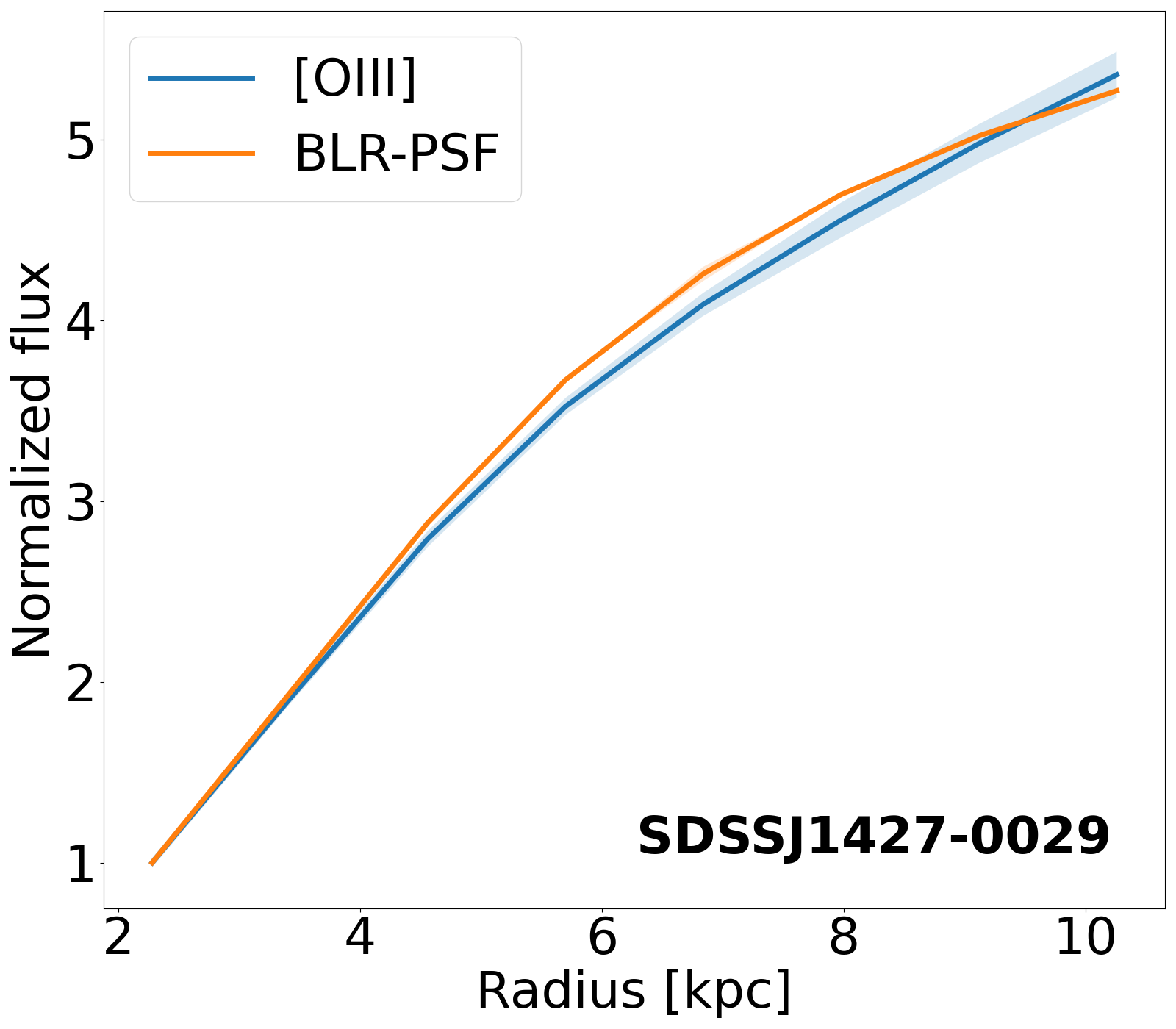} \\
    \end{tabular}
    \caption{COG of the [\ion{O}{iii}] (blue) and BLR normalized flux (red), in log scale. Targets Q0050+0051, Q0052+0140 and SDSSJ2319-1040 show evidence of spatially resolved [\ion{O}{iii}] emission.}
    \label{fig:OIII-BLR}
\end{figure*}

The first test consists of comparing the curve of growth (COG; i.e., the flux within circular apertures as a function of radius, centered on the AGN) of the [\ion{O}{iii}] emission with that of the PSF, to search for significant excess emission. As no dedicated standard-star observation is available, we used the unresolved point-like BLR emission as a proxy for the PSF. To construct the COGs of both the [\ion{O}{iii}] and BLR emission, we first created their flux maps based on the spaxel-by-spaxel modeling described in Sect. \ref{sect:spax_fit}. Using such maps, we measured the mean flux in concentric circular apertures spanning radii from two to ten kiloparsec in one-kiloparsec steps, and subsequently normalized by the flux in the innermost aperture. The resulting COGs are presented in Fig. \ref{fig:OIII-BLR}, where the [\ion{O}{iii}] emission is shown in blue and the BLR-PSF in red.

Figure \ref{fig:OIII-BLR} shows that the [\ion{O}{iii}] emission in Q0050+0051, Q0052+0140, and SDSSJ2319$-$1040 is spatially resolved, as its normalized flux increases more rapidly with the radius than that of the PSF. In contrast, the COG of the [\ion{O}{iii}] emission in Q2121+0052 remains below that of the PSF at all radii, indicating that the emission is not resolved. For Q2123$-$0050 and SDSSJ1427$-$0029, the normalized [\ion{O}{iii}] and BLR fluxes grow at a similar rate across all radii, suggesting that the emission is unresolved as well.

\begin{figure*}[h]
    \centering
    \includegraphics[width=1\linewidth]{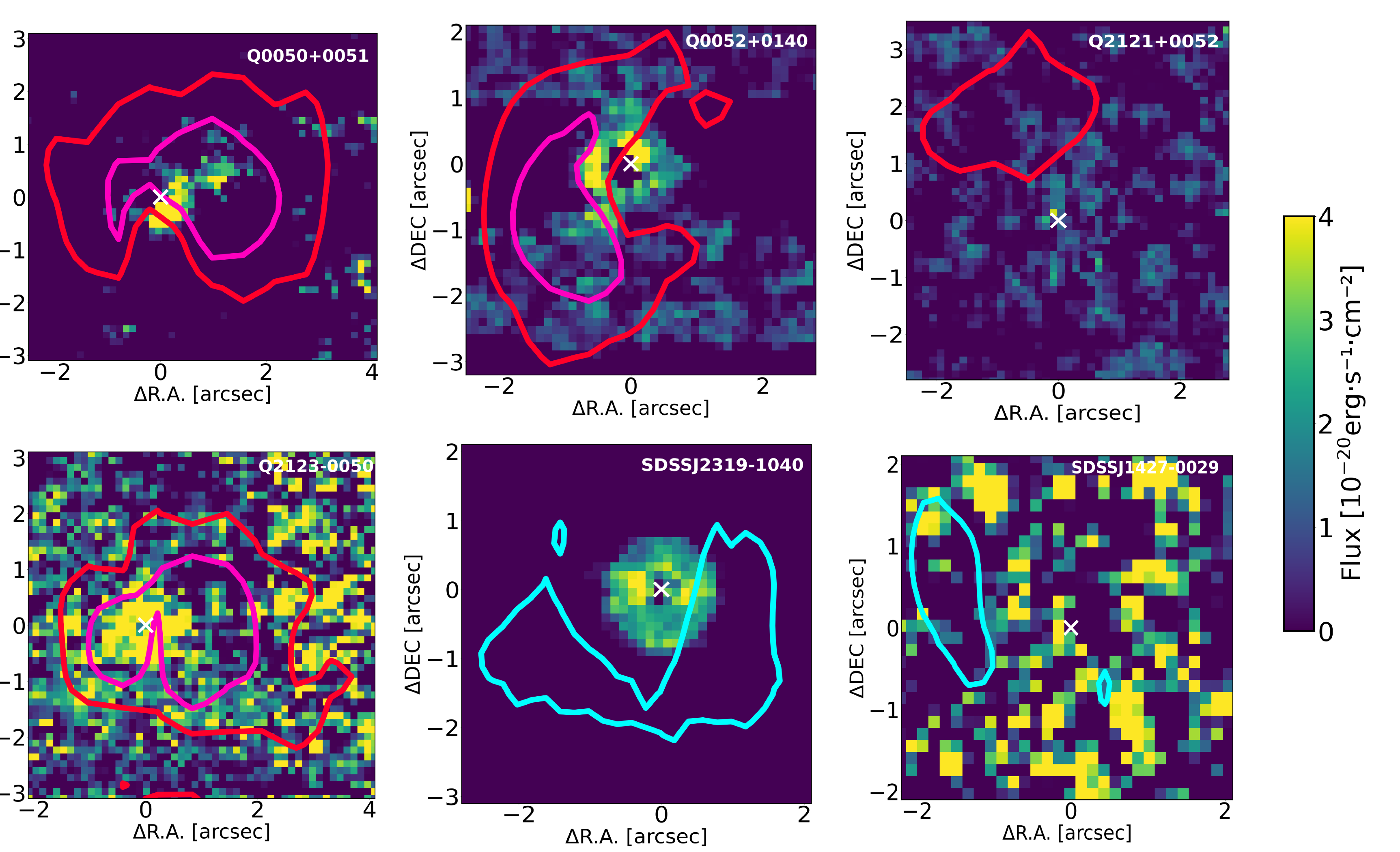}
    \caption{Median maps of the residual [\ion{O}{iii}] emission after PSF subtraction. Targets Q0050+0051, Q0052+0140, Q2123-0050 and SDSSJ2319-1040 show evidence of extended [\ion{O}{iii}] emission. The magenta, red and cyan contours represent 5$\sigma$, 10$\sigma$ and 30$\sigma$ contours of the Ly$\alpha$ SB maps, respectively. The white cross represents the AGN center. North is up and East is left.}
    \label{fig:extended-OIII}
\end{figure*}

For the second test, we follow the PSF-subtraction method described in \citet{husemann2016large} and \citet{kakkad2020super} to identify the presence of extended emission. We first extracted the nuclear spectrum from an aperture of one pixel around the quasar. This spectrum was fitted with a model $f(x)$, identical to the one described in Sect. \ref{sect:mod_BLR}. The rest of the pixels were fitted with a model $f'=a*f(x)$, where $a$ corresponds to the rescaling factor. This model $f'$, accounting for the contribution from the PSF, was subtracted from the observed spectrum in each pixel. After such subtraction, any remnant [\ion{O}{iii}] emission is considered to be extended and not BLR-dominated as a consequence of beam smearing from the PSF. To map such emission, we created narrow median maps (between 5-15 $\AA$ wide) around the center of the [\ion{O}{iii}] line. The results are shown in Fig. \ref{fig:extended-OIII}. We observe that four targets show signs of extended [\ion{O}{iii}] emission. Three of these targets, namely Q0050+0051, Q0052+0140, and SDSSJ2319$-$1040, had previously been identified to be spatially resolved in the first test carried out from Fig. \ref{fig:OIII-BLR}. Although Q2123$-$0050 had previously shown unresolved emission, Fig. \ref{fig:extended-OIII} shows the presence of extended emission above the $2\sigma$ level. We can then conclude that any [\ion{O}{iii}] emission detected by the spaxel-by-spaxel fitting from these four targets is extended and/or spatially resolved, and we can rely on the kinematic maps described in Sect. \ref{sect:spat_res} to characterize the extension and morphology of the ionized outflows. In contrast, Q2121+0052 and SDSSJ1427$-$0029 show no evidence of extended [\ion{O}{iii}] emission. As these targets fail to show resolved and/or extended emission in both tests, we refrain from performing the spatially resolved analysis for them. 

\subsection{Ionized gas kinematics} \label{sect:spat_res}

We characterize the [\ion{O}{iii}] line by measuring the velocity at the 10th, 50th, and 90th percentile ($v_{10}$, $v_{50}$ and $v_{90}$, respectively). The extreme velocities $v_{10}$ and $v_{90}$ are representative of the blue and red-shifted emission in the wings of the line, respectively. The velocity width is measured by $w_{80}$, which contains $80\%$ of the total line flux ($w_{80}=v_{90}-v_{10}$). The values of $v_{10}$, $v_{90}$, and especially $w_{80}$ are widely used in the literature to test the presence of fast outflowing gas \citep[e.g.][]{liu2013observations, harrison2016kmos, temple2019iii,kakkad2020super, tozzi2024super}.

\begin{figure*}[ht]
    \centering
    \begin{tabular}{cccc}
        [\ion{O}{iii}] Flux & \textbf{$v_{10}$} & \textbf{$v_{90}$} & \textbf{$w_{80}$} \\
        \includegraphics[width=0.22\textwidth]{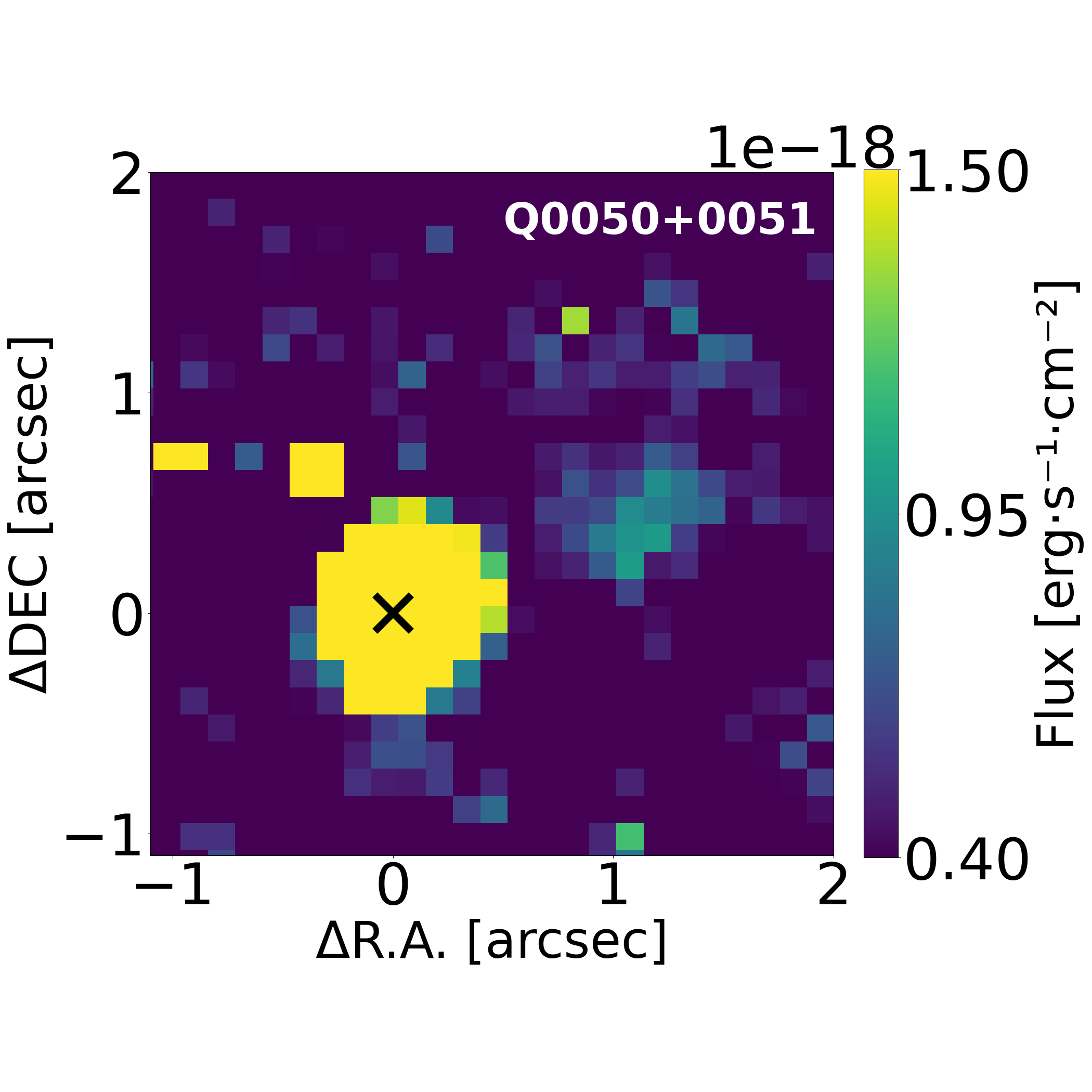} &
        \includegraphics[width=0.22\textwidth]{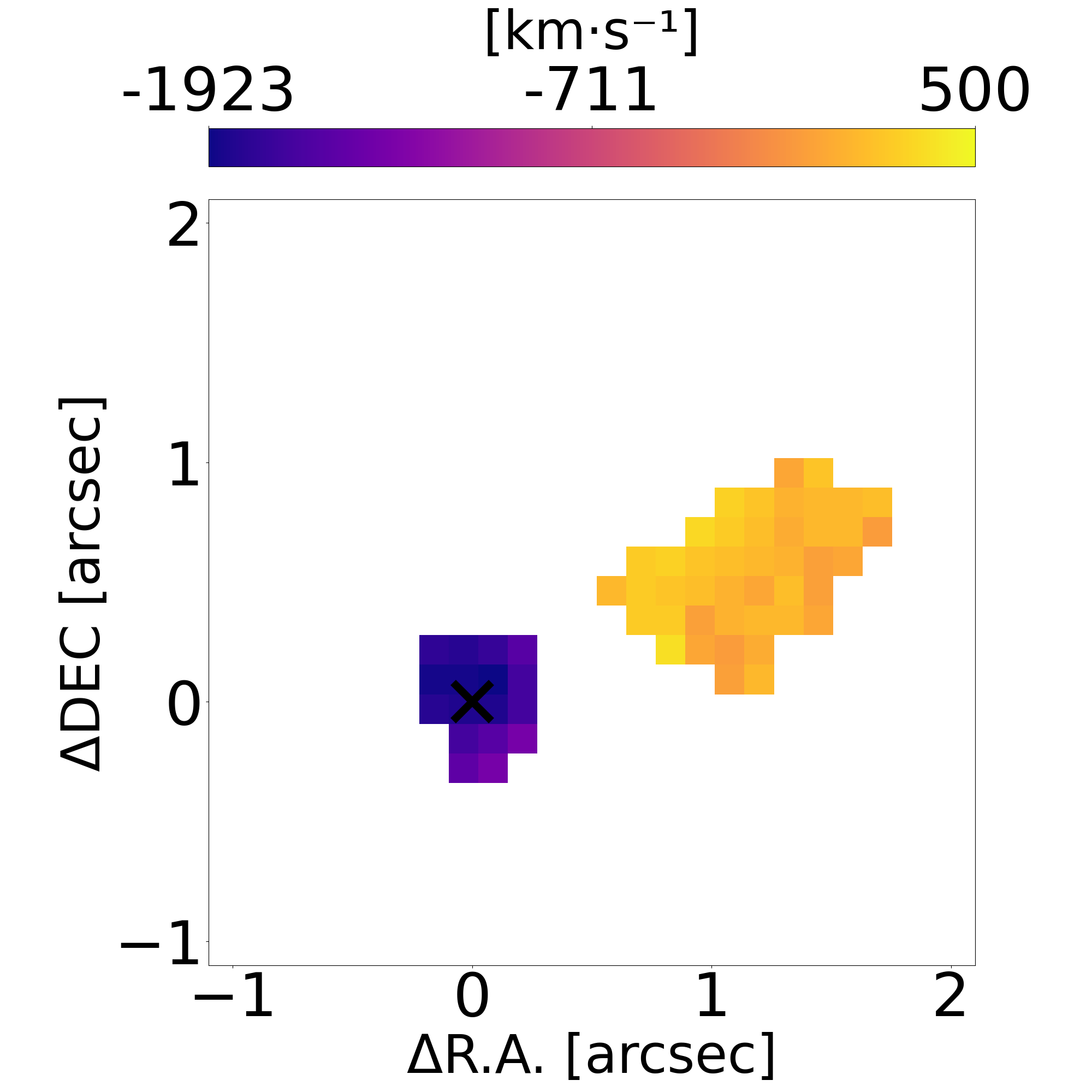} &
        \includegraphics[width=0.22\textwidth]{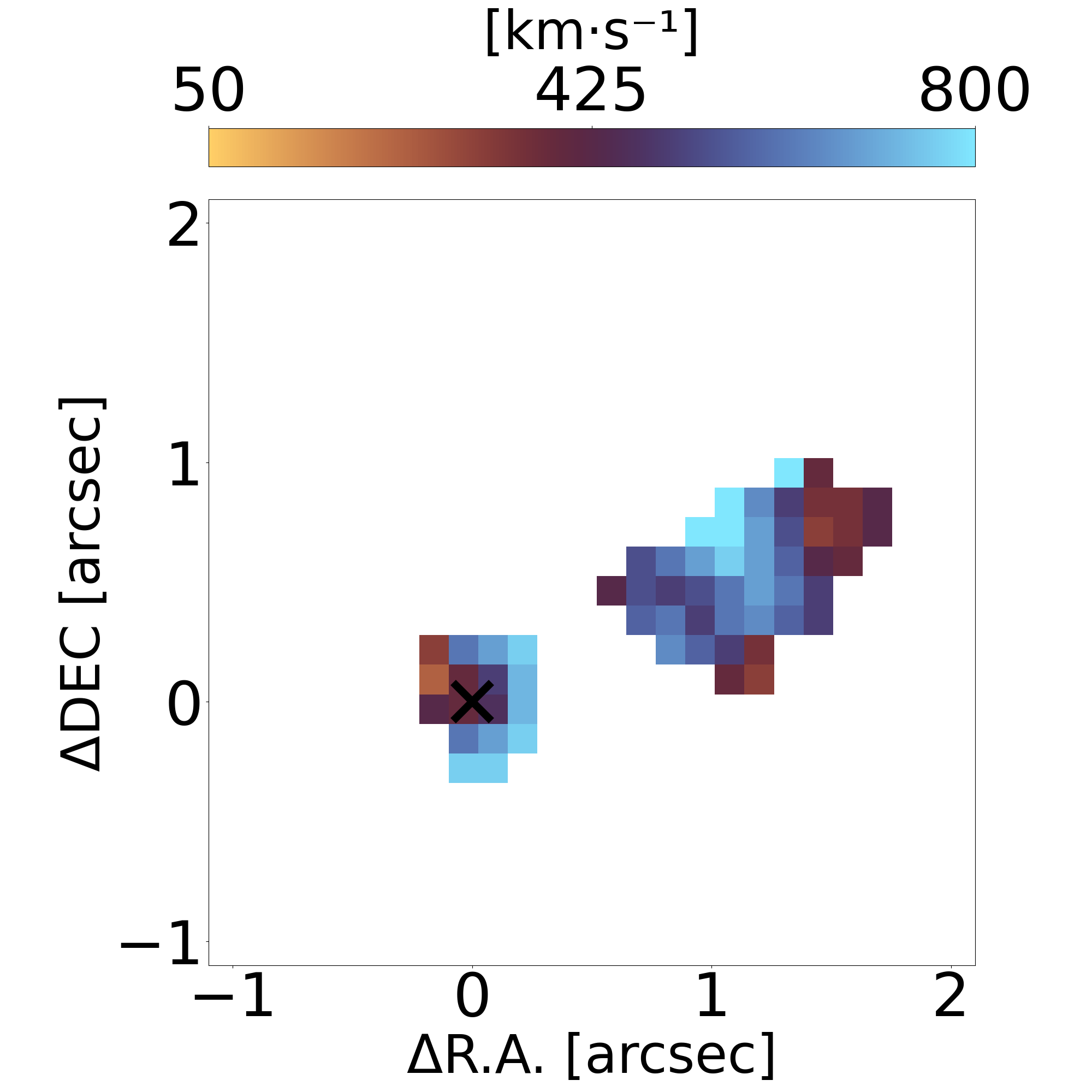} &
        \includegraphics[width=0.22\textwidth]{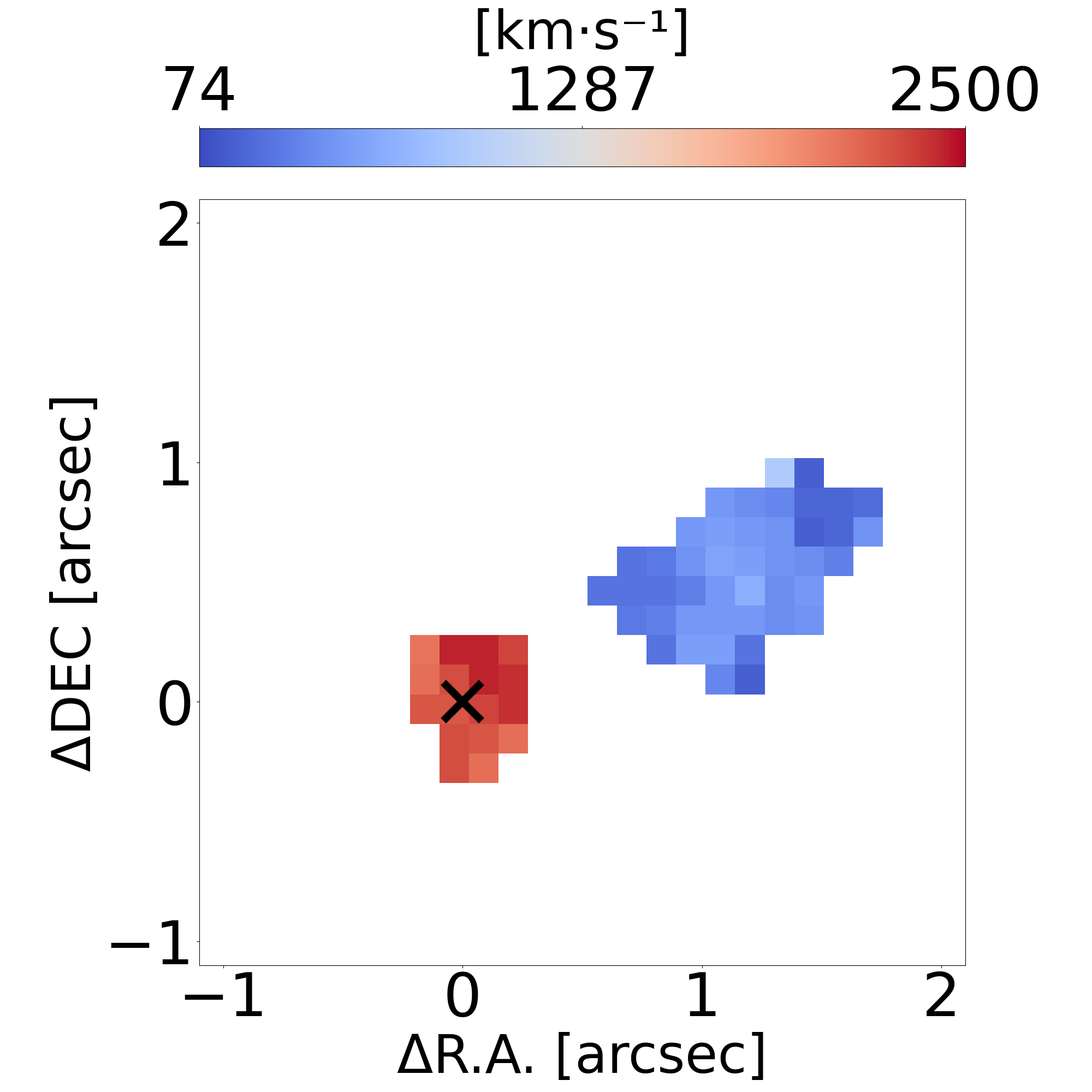} \\

        \includegraphics[width=0.22\textwidth]{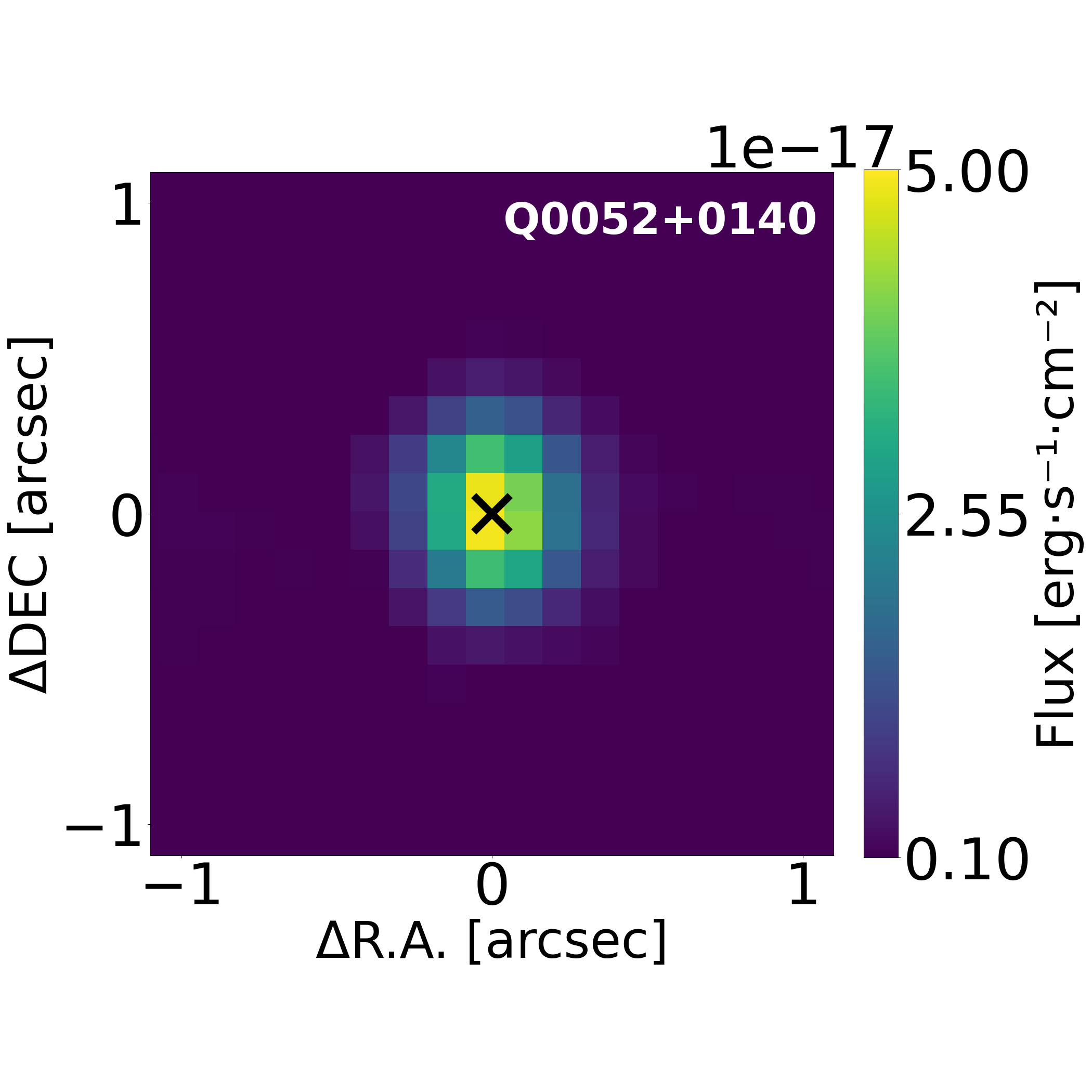} &
        \includegraphics[width=0.22\textwidth]{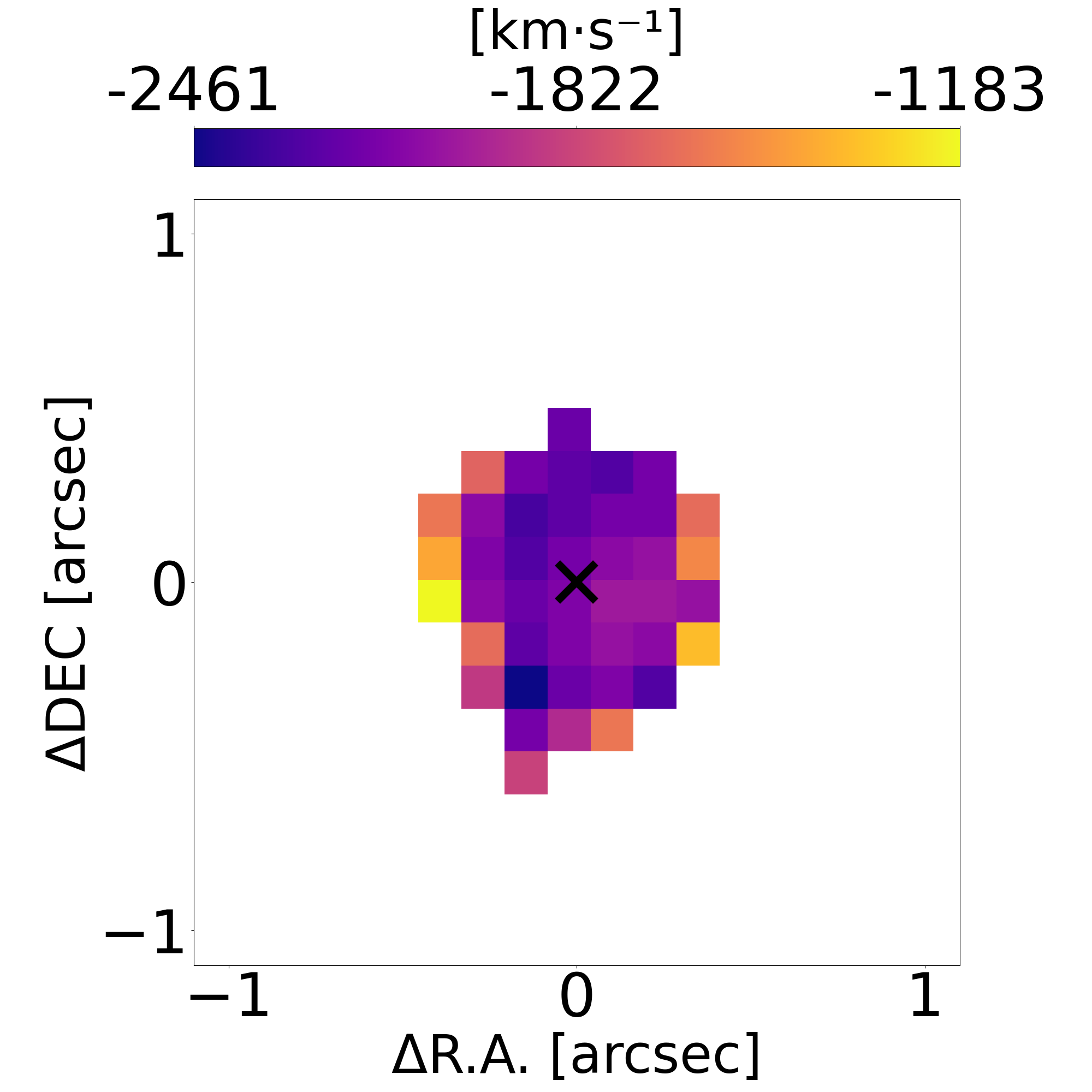} &
        \includegraphics[width=0.22\textwidth]{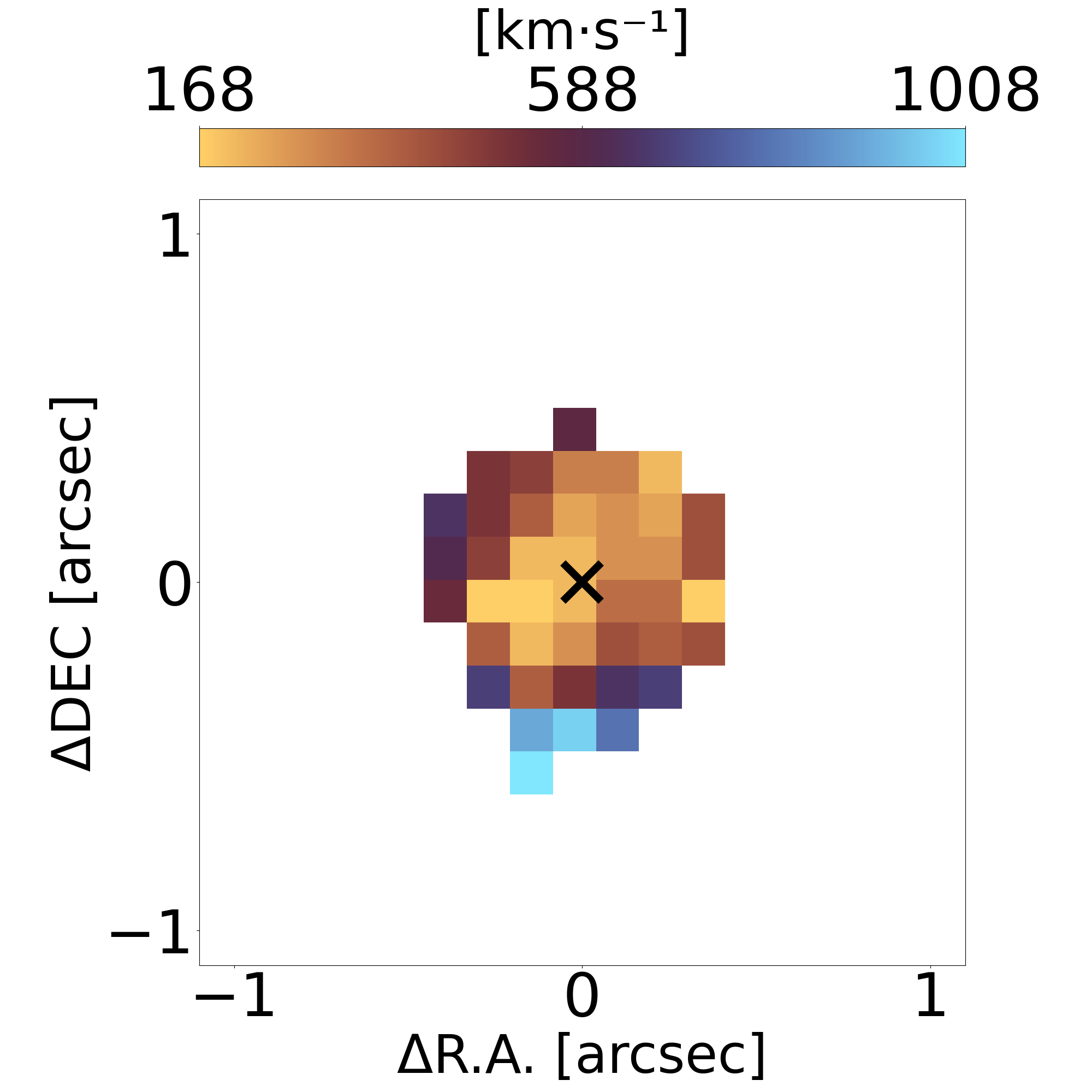} &
        \includegraphics[width=0.22\textwidth]{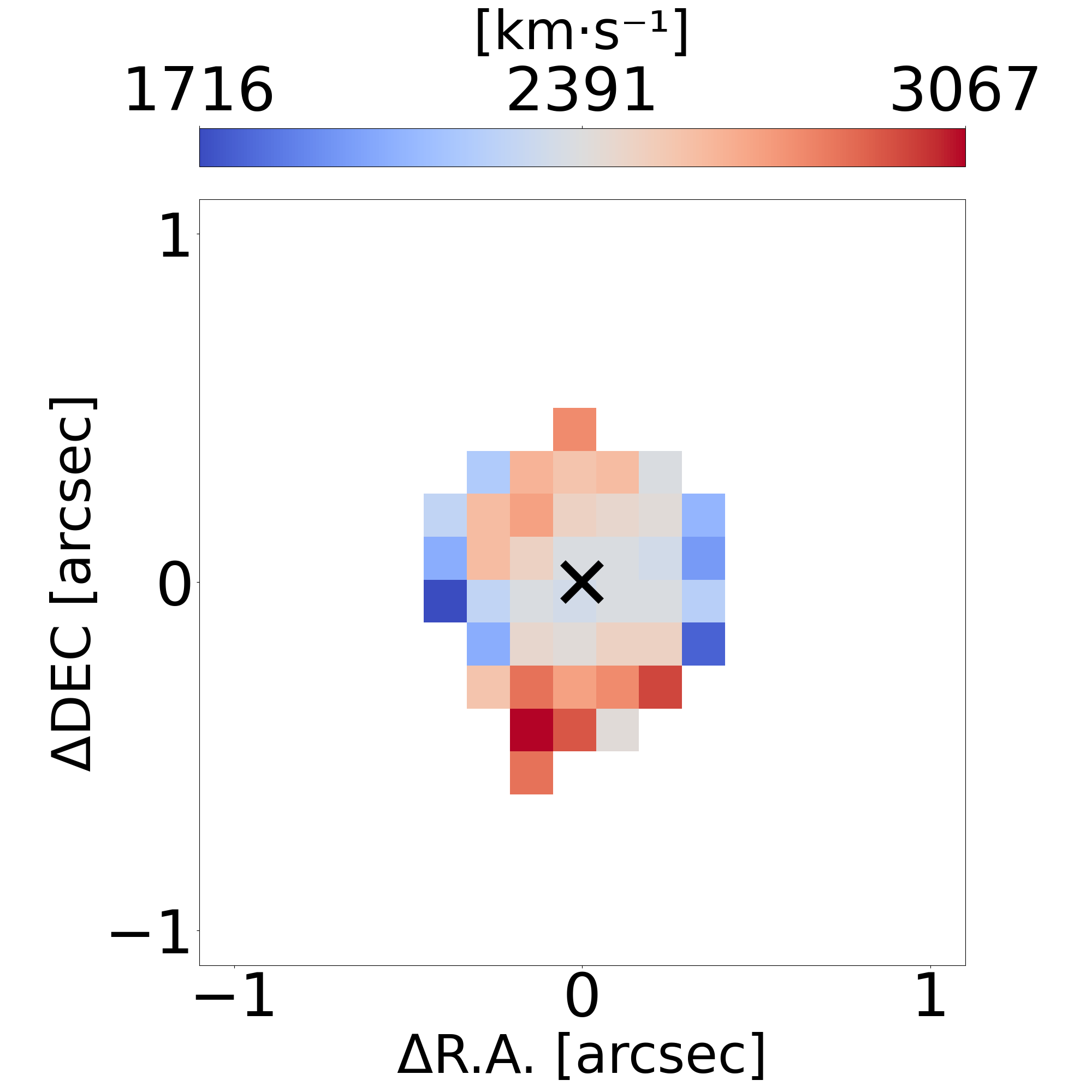} \\

        \includegraphics[width=0.22\textwidth]{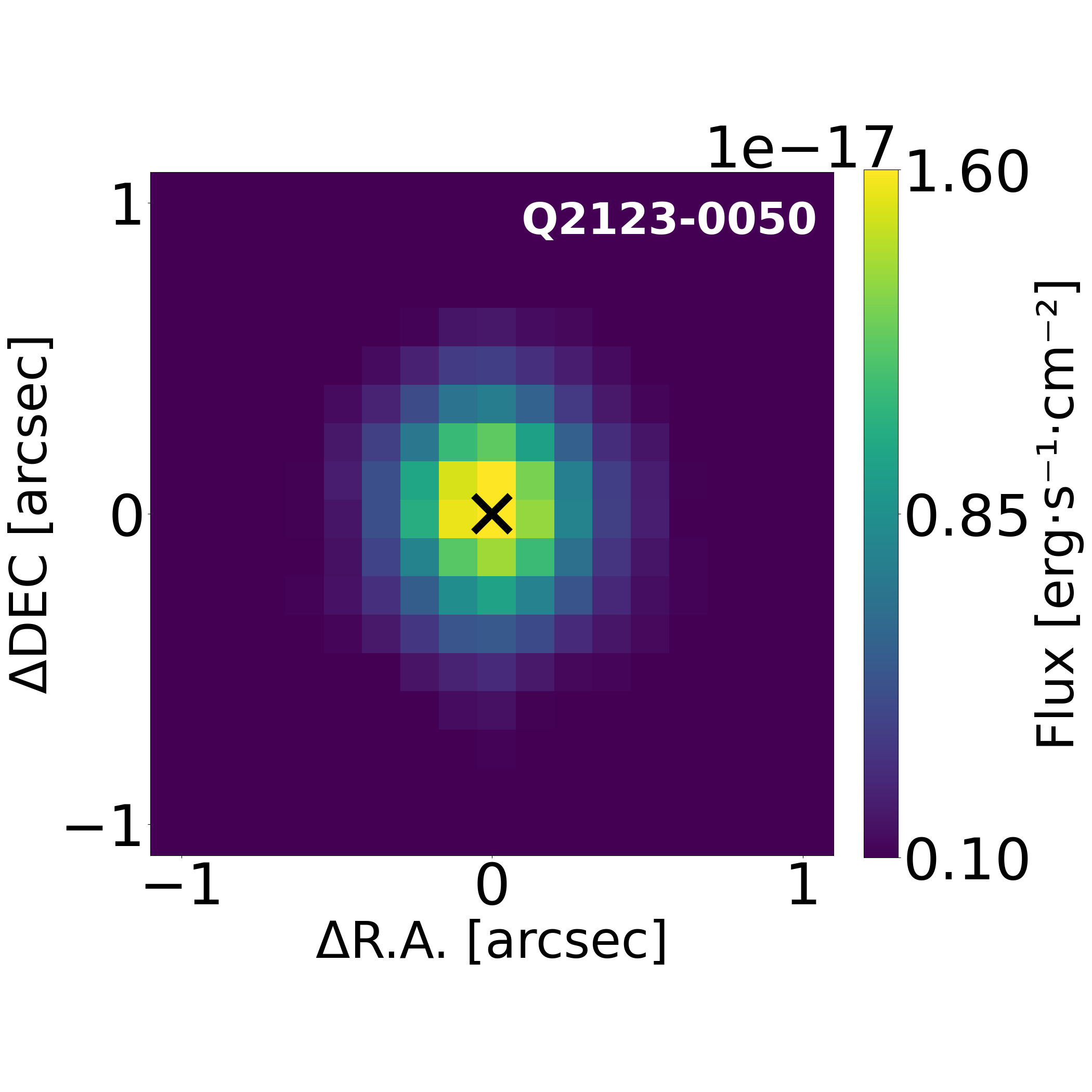} &
        \includegraphics[width=0.22\textwidth]{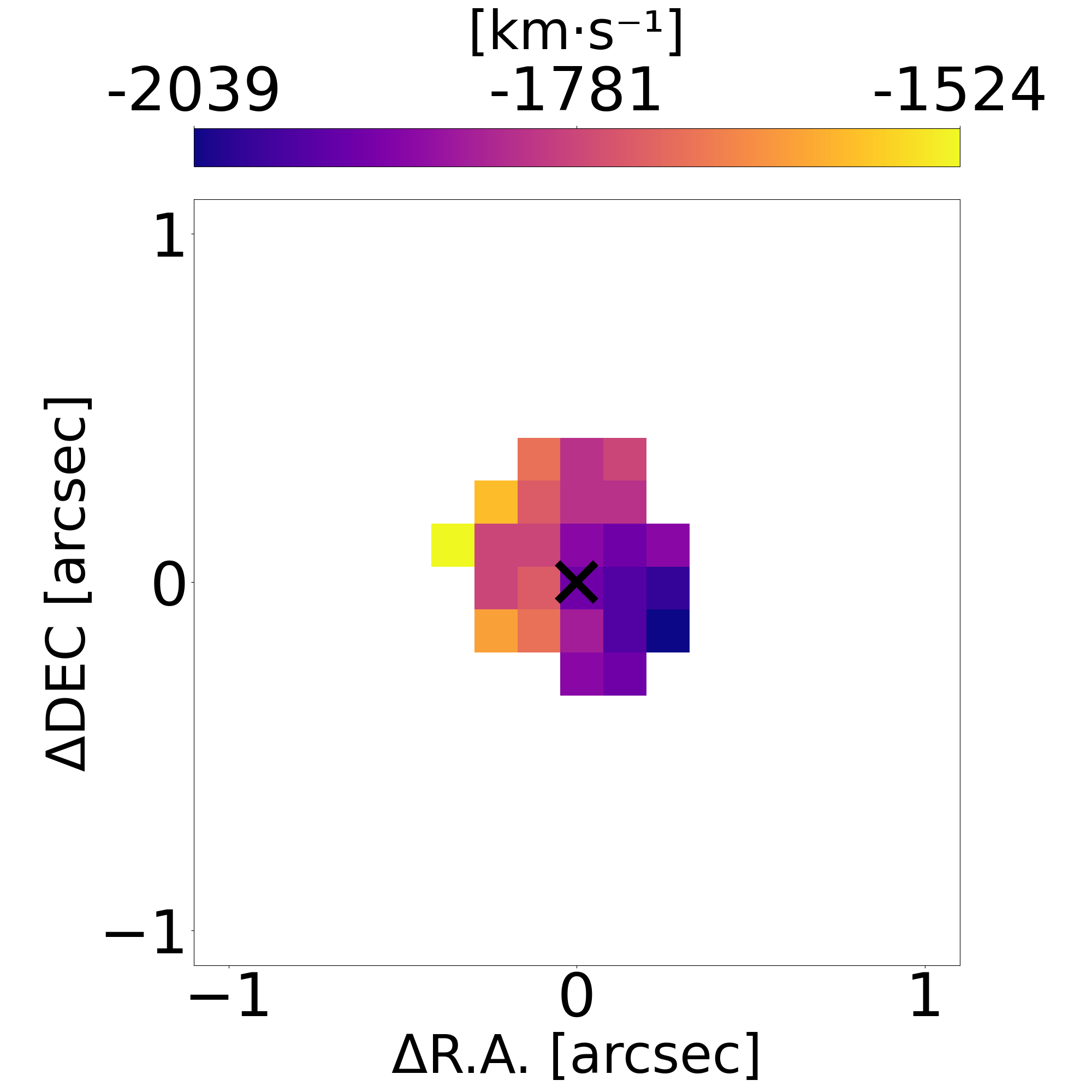} &
        \includegraphics[width=0.22\textwidth]{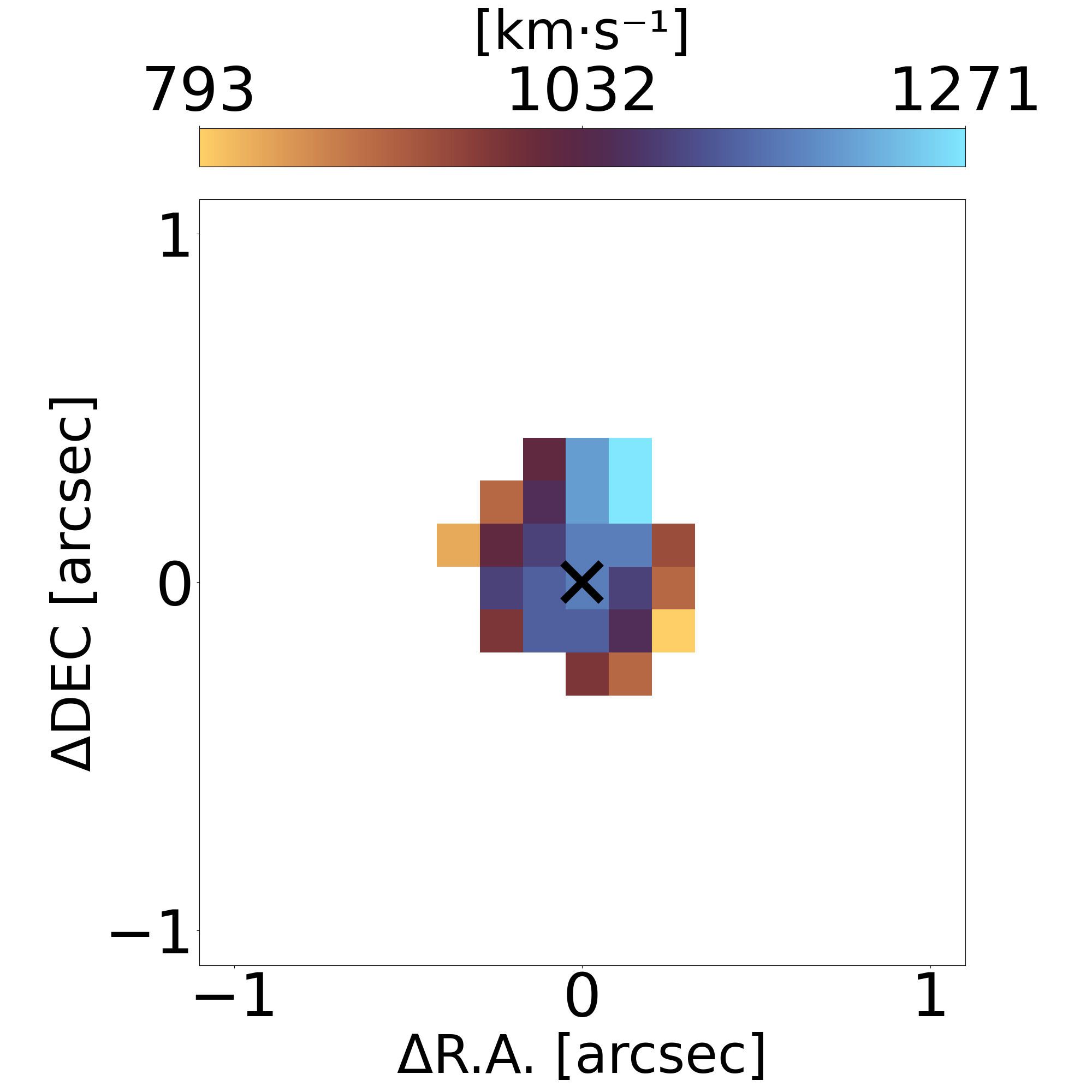} &
        \includegraphics[width=0.22\textwidth]{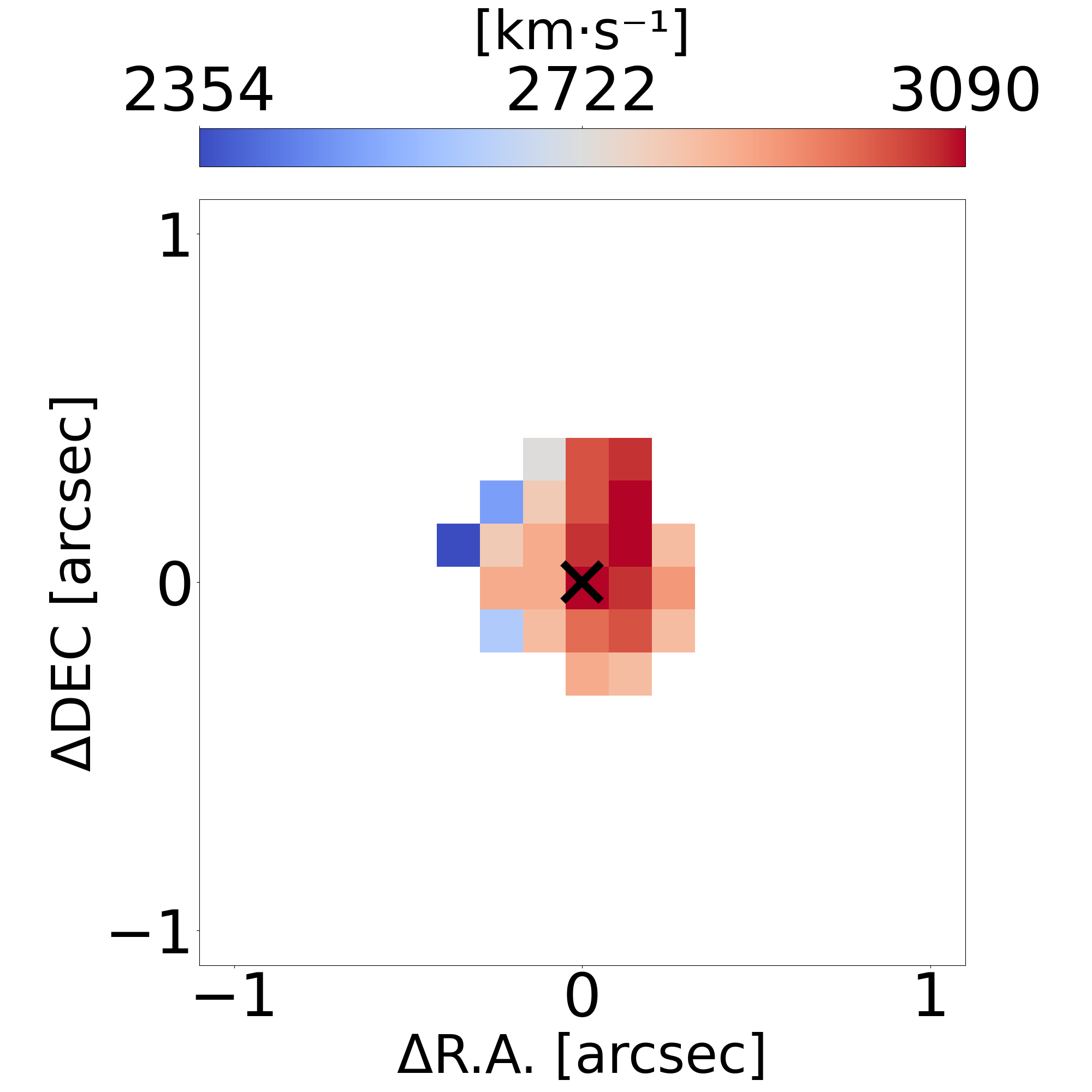} \\

        \includegraphics[width=0.22\textwidth]{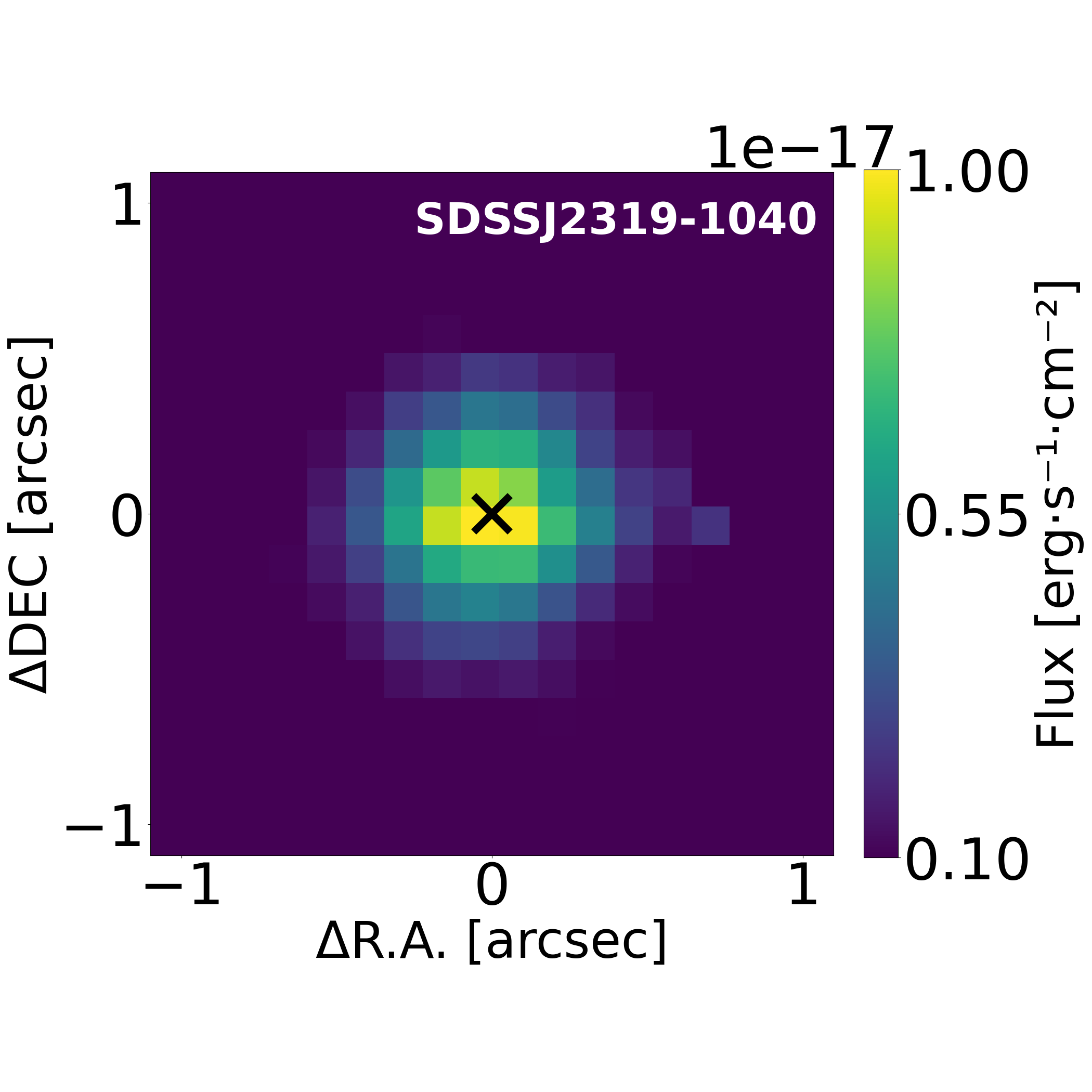} &
        \includegraphics[width=0.22\textwidth]{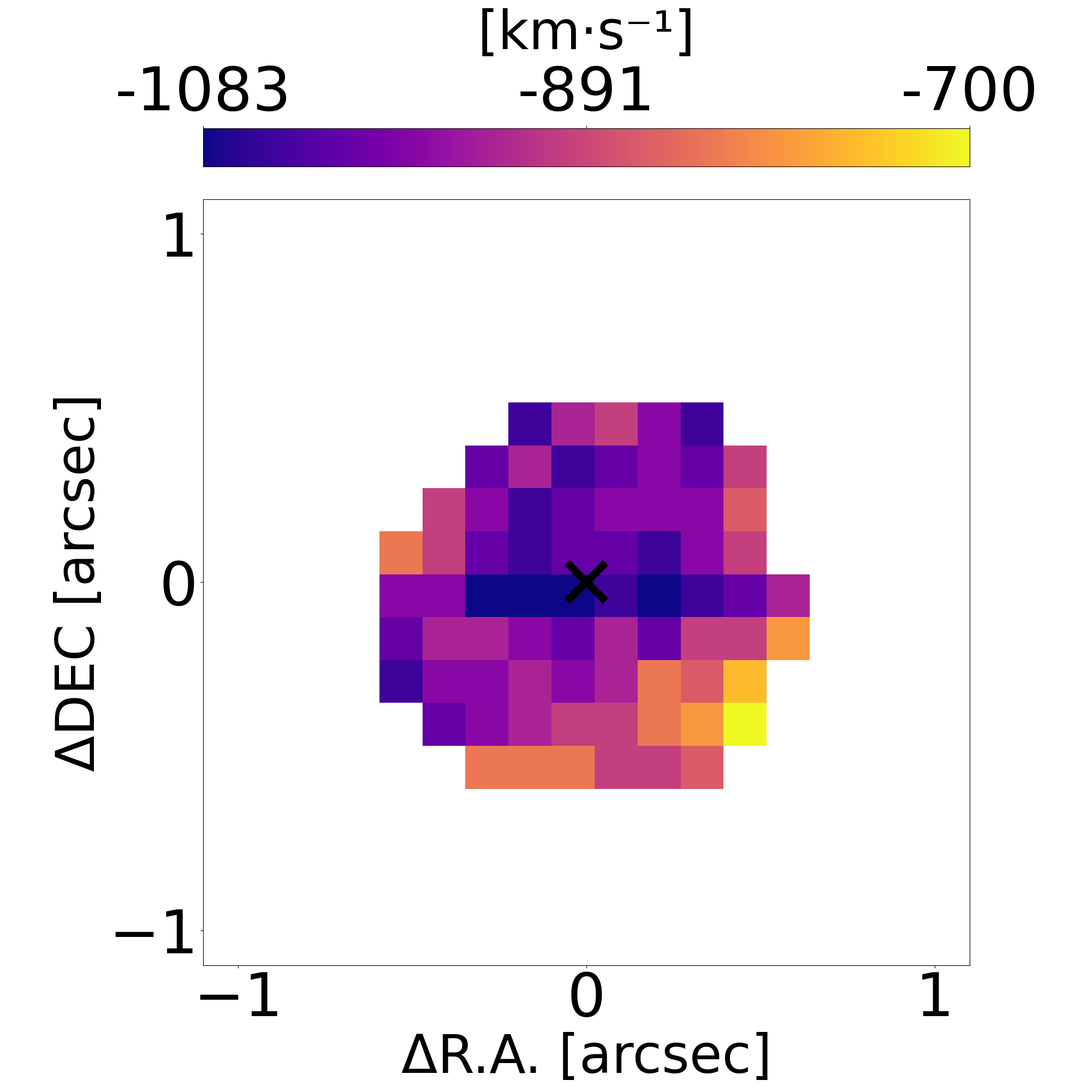} &
        \includegraphics[width=0.22\textwidth]{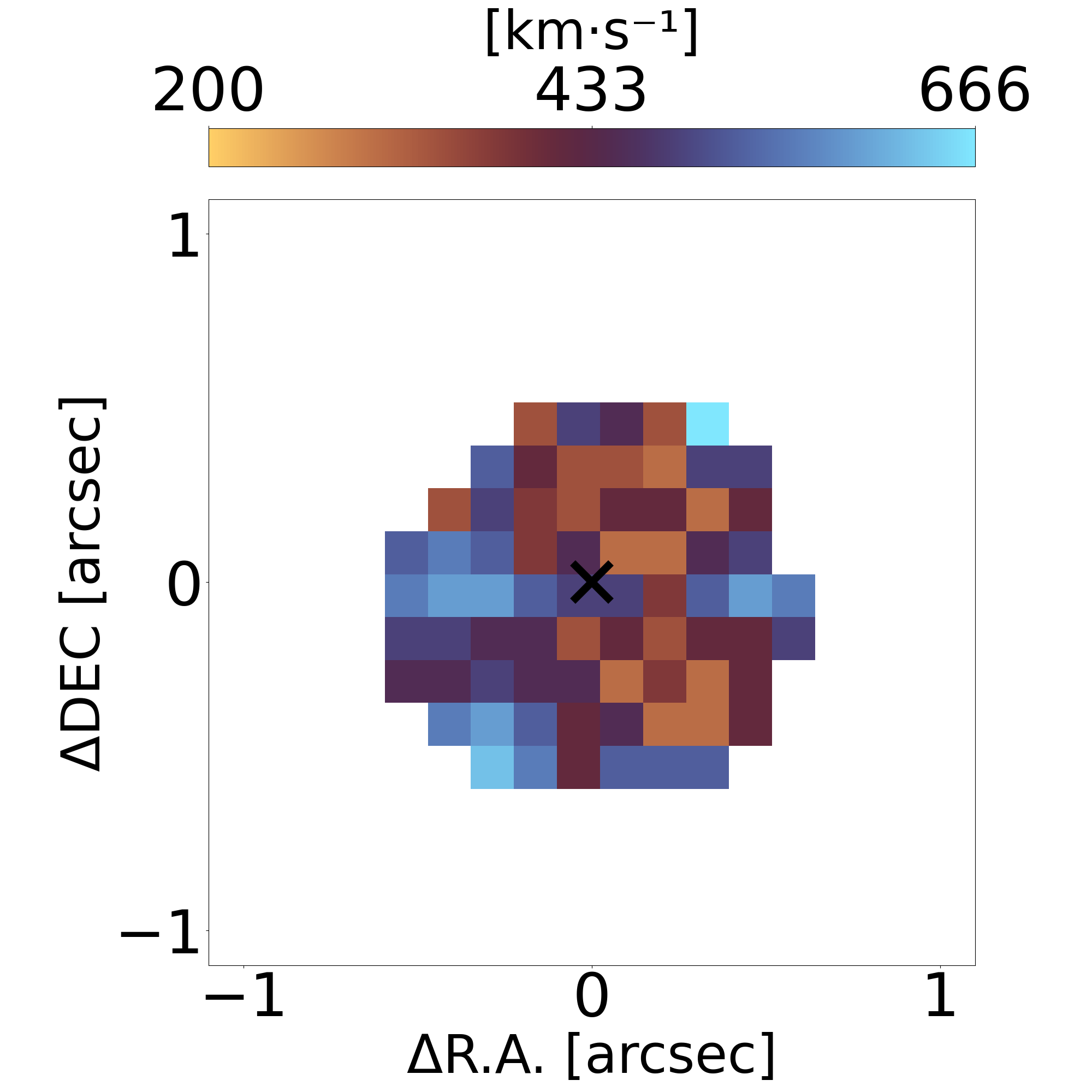} &
        \includegraphics[width=0.22\textwidth]{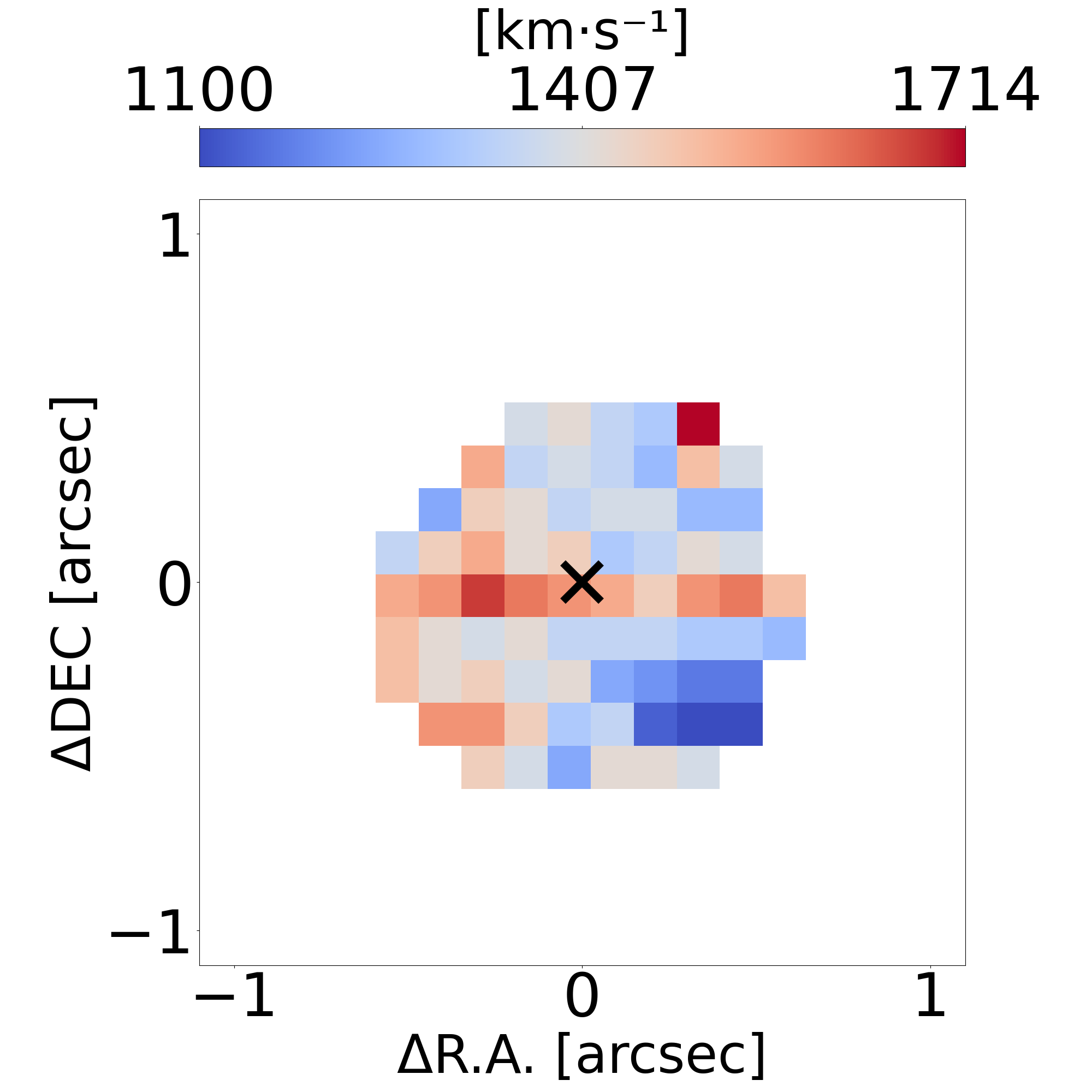} \\
        
    \end{tabular}
    \caption{Flux and kinematics maps of the [\ion{O}{iii}] emission for the targets showing spatially resolved and/or extended emission. The first column shows the flux map of the  [\ion{O}{iii}] emission. The second, third, and fourth, the $v_{10}$, $v_{90}$, and $w_{80}$ velocity maps, respectively. The black cross marks the AGN center. North is up and East is left. }
    \label{fig:kinems_target}
\end{figure*}

Fig. \ref{fig:kinems_target} shows the [\ion{O}{iii}] total flux and velocity maps for the targets in the sample with spatially resolved and/or extended emission, as demonstrated in Sect. \ref{sect:res_emission}. The [\ion{O}{iii}] total flux corresponds to the flux of the sum of the Gaussian components used to model the line. The second, third and fourth column show the velocities $v_{10}$, $v_{90}$ and the width $w_{80}$ respectively. The cross marks the position of the AGN center, which we determined by estimating the photo-centroid of the flux map at $5100~\AA$. At this wavelength, the emission is dominated only by the AGN continuum, and therefore it is a good tracer for the emission close to the supermassive BH. The kinematics maps are created by excluding pixels with $\mathrm{S/N}<2$. 

\subsection{Outflow detection} \label{sect:outflow_detec}

The kinematics of the NLR is typically traced by relatively narrow, systemic Gaussian components. In the presence of AGN-driven outflows, however, the [\ion{O}{iii}] line profile becomes asymmetric, showing extended wings that require additional broad Gaussian components to accurately reproduce the spectrum. The resulting asymmetry of the [\ion{O}{iii}] line profile is indicative of outflowing ionized gas, which translates into extreme velocity values. As discussed in Sect. \ref{sect:spat_res}, different studies have adopted various definitions of outflow velocity based on different parameters. In this work, we adopt the non-parametric line width $w_{80}$ as our tracer of the ionized gas velocity. Compared to $v_{10}$ or $v_{90}$, $w_{80}$ is less sensitive to projection effects (e.g., due to galaxy inclination; \citealt{harrison2012energetic}), although it may be more affected by dust extinction \citep{tozzi2024super}. In our case, this effect is negligible, as our sample consists of bright Type I QSOs where obscuration is expected to be minimal.

We use the $w_{80}$ maps presented in the fourth column of Fig. \ref{fig:kinems_target} to investigate the presence of AGN-driven outflows in our sample. The ionized gas in a given spaxel is classified as part of an outflow when $w_{80} \geq 600~\mathrm{km~s^{-1}}$ \citep[e.g.,][]{harrison2012energetic, harrison2016kmos, kakkad2016tracing, kakkad2020super}. This threshold is deliberately conservative: numerical experiments of AGN winds interacting with a clumpy ISM show that outflowing gas can exhibit considerably lower bulk velocities, down to $\sim100~\mathrm{km~s^{-1}}$, while still being distinguishable from the effect of the host-galaxy kinematics \citep{ward2024agn}. In observations, however, decoupling the low-velocity outflowing component from the host's contribution is not straightforward. Therefore, we limit our analysis to the gas that can be unambiguously classified as outflowing material, i.e., that with $w_{80} \geq 600~\mathrm{km~s^{-1}}$. 

Following our definition, all spatially resolved targets in our sample show evidence for AGN-driven outflows. In Q0050+0051, we identify an [\ion{O}{iii}] emission region northwest of the source; however, this structure is distinct from the main outflow, where gas velocities exceed $2000~\mathrm{km~s^{-1}}$ (see Sect. \ref{sect:spat_rel} for further discussion). For the remaining three resolved sources, we measure ionized gas velocities of at least $1100~\mathrm{km~s^{-1}}$, consistent with gas moving significantly faster than the systemic velocity of the host galaxy. Regions of high $w_{80}$ show a mild spatial gradient in Q0051+0140 and Q2123$-$0050, while they appear more spherically symmetric in Q0050+0051 and SDSSJ2319$-$1040. A comparison with the $v_{10}$ maps (first column of Fig. \ref{fig:kinems_target}) shows that the $w_{80}$ peaks coincide with the highest velocity shifts ($v_{10} < -1000~\mathrm{km~s^{-1}}$), indicating the presence of a strongly blueshifted component. Similarly, the $v_{90}$ maps (second column of Fig. \ref{fig:kinems_target}) reveal milder redshifted components (up to $\sim1270~\mathrm{km~s^{-1}}$) coincide spatially with the $w_{80}$ and $v_{10}$ peaks. This configuration is consistent with a biconical outflow geometry, where the approaching (blueshifted) side is visible, while the receding (redshifted) counterpart is likely obscured by dust in the galactic disc.

Alternative scenarios such as SF-driven outflows, inflows, mergers, or turbulence are unlikely to explain the observed high-velocity dispersions and velocity shifts, as well as the asymmetric line profiles. SF-driven outflows and inflows typically exhibit much lower velocities of $\sim500~\mathrm{km~s^{-1}}$ \citep{arribas2014ionized, schreiber2019kmos3d} and $\sim200~\mathrm{km~s^{-1}}$ \citep{bouche2013signatures, liu2025discovery}, respectively, with the latter usually detected in absorption. In contrast, merger-driven kinematics would produce symmetric double-peaked line profiles rather than the asymmetric blueshifted components observed here.

After successfully detecting outflows in the resolved targets of our sample, we estimated their extension and velocity, which are needed to determine the outflow energetics. We used the kinematics maps in Fig. \ref{fig:kinems_target} to estimate these values. The extension of the outflow $R_{\mathrm{out}}$ is defined as the largest distance between the ionized gas and the AGN center.  The velocity of the outflow $v_{\mathrm{out}}$ is defined as the maximum value observed in the $w_{80}$ map, assuming the outflow has a constant velocity and therefore any smaller velocities are produced by projection effects. 

We define the outflow luminosity $L_{\mathrm{out}}^{[\mathrm{OIII}]}$ as the luminosity estimated from the flux contained in the high-velocity channels ($v \geq 300~\mathrm{km~s^{-1}}$, \citealt{kakkad2020super}) of all spaxels with S/N $>2$. The estimated values of extension, velocity, and luminosity of the AGN-driven outflows in our sample are shown in the second, third and fourth column of Table \ref{tab:outflow_props}, respectively.

For the unresolved targets Q2121+0052 and SDSSJ1427$-$0029, we are limited by the PSF to perform any spatial analysis on the ionized gas. However, we can still detect the presence of AGN-driven outflows by analyzing the [\ion{O}{iii}] line profile extracted from an integrated spectrum centered around the quasar. The aperture of this spectrum was set to $0.25 \arcsec$ for Q2121+0052 (see Fig. \ref{fig:app_int_q2121}) and $0.75 \arcsec$ for SDSSJ1427$-$0029 (see Fig. \ref{fig:app_int_j1427}), corresponding to roughly 2 kpc and 6 kpc, respectively. These values were determined by creating a channel map around the location of the [\ion{O}{iii}] line, and determining the radius where $99.5\%$ of the flux is contained. From this integrated spectrum, we observed an asymmetric [\ion{O}{iii}] profile, which we modeled as described in Sect \ref{sect:spec_fitting}. The non-parametric analysis of the line profile indicates ionized gas $w_{80}$ velocities consistent with our definition of AGN-driven outflows. We estimate the properties of this outflow in the same way as for the spatially resolved sources, with the difference that the resulting values on the extension, velocity and luminossity of the outflow correspond only to upper limits. 

The fitting uncertainties on $v_{\mathrm{out}}$ and $L_{\mathrm{out}}^{[\mathrm{OIII}]}$ are estimated as follows. We extract an integrated spectrum containing roughly $95\%$ of the total [\ion{O}{iii}] flux of the quasar. By adding rms noise to the best fit of this spectrum, we created 100 mock spectra and fit each of them using the procedure mentioned in Sect. \ref{sect:mod_BLR}. The errors reported in Table \ref{tab:outflow_props} correspond to the $1\sigma$ uncertainties resulting from these iterations. Considering the resolving power of VLT/ERIS and GEMINI/GNIRS, the uncertainty on $w_{80}$ due to the instrumental resolution is expected to be negligible compared to the fitting uncertainty. Although we do not have an accurate estimate on the uncertainties of $R_{\mathrm{out}}$, we do not expect them to be smaller than the pixel scale of the observations with VLT/ERIS, corresponding to $0.125"/\mathrm{pix}$.

\subsection{Outflow energetics} \label{sect:out_energ}

The outflow energetics can be derived using the outflow extension, velocity, and luminosity $R_{\mathrm{out}}$, $v_{\mathrm{out}},$ and $L_{\mathrm{out}}^{[\mathrm{OIII}]}$ estimated in the previous section. 

The outflow mass $M_{\mathrm{out}}$ can be estimated as follows \citep[e.g.][]{carniani2015ionised, kakkad2020super, tozzi2024super}:

\begin{align}
    M_\mathrm{out}^{\mathrm{[OIII]}} = 0.8 \times 10^8 \mathrm{M}_\odot 
    \left ( \frac{L_{\mathrm{out}}^{[\mathrm{OIII}]}}{10^{44}~\mathrm{erg~ s^{-1}}} \right )
    & \left( \frac{n_\mathrm{e}}{500~\mathrm{cm^{-3}}} \right)^{-1} \notag \\
    & \left( \frac{1}{10^{\mathrm{[O/H]} - \mathrm{[O/H]}_\odot}} \right)
\end{align}

where all oxygen is assumed to be ionized to $\mathrm{O}^{2+}$, the electron temperature is taken to be $T \sim 10^4$~K, $n_\mathrm{e}$ denotes the electron density, and [O/H] represents the oxygen abundance, assumed to be solar. The electron density is commonly inferred from the [\ion{S}{ii}$]\lambda6716$/[\ion{S}{ii}]$\lambda6731$ flux ratio; however, our observations do not cover such emission lines. In the literature, the electron density $n_\mathrm{e}$ spans several orders of magnitude, typically ranging from $10^{2}$ to $10^{4},\mathrm{cm}^{-3}$ \citep[e.g.,][]{cano2012observational, fiore2017agn, perna2017x, baron2019discovering}. We adopt a fiducial value of $n_\mathrm{e}=500 \pm 250~\mathrm{cm}^{-3}$, consistent with previous studies and with measurements reported for AGN at $z\sim2$ \citep{harrison2014kiloparsec, cresci2023bubbles, tozzi2024super}. As demonstrated by \citet{kakkad2020super}, this uncertainty in $n_\mathrm{e}$ alone can result in variations of up to an order of magnitude in the inferred outflow mass rate. Consequently, the adopted value of $n_\mathrm{e}$ constitutes the dominant source of uncertainty in our estimates of the outflow energetics.

The outflow mass rate will depend on the outflow geometry. For an outflow with a bi-conical or spherical geometry, filled with uniformly distributed clouds of the same density and constant outflow velocity, the outflow mass rate and outflow kinetic power are independent of the outflow opening angle and the filling factor of the clouds in the cone \citep{husemann2016large}.

In this scenario, the outflow mass rate, corresponding to the instantaneous outflow mass rate of ionized gas crossing the spherical sector at a distance $R_{\mathrm{out}}$ from the AGN, can be estimated as:

\begin{equation} \label{omr_cone}
    \Dot{M}_\mathrm{out}^{\mathrm{[OIII]}} = 3 \frac{M_\mathrm{out}^{\mathrm{[OIII]}}~v_{\mathrm{out}}}{R_{\mathrm{out}}}
\end{equation}

For an outflow propagating in thin shells of thickness $\Delta R$ that shock against the ISM, the localized outflow mass rate at the shell distance from the nucleus is independent of the outflow opening angle, and can be estimated as:

\begin{equation}
    \Dot{M}_\mathrm{out}^{\mathrm{[OIII]}} = \frac{M_\mathrm{out}^{\mathrm{[OIII]}}~v_{\mathrm{out}}}{\Delta R}
\end{equation}

In this work, we adopt a simplified bi-conical or spherical outflow geometry, described by Equation \ref{omr_cone}. This choice is conservative, as a shell-like geometry yields higher estimates of the outflow mass rate \citep{maiolino2012evidence, husemann2019close, kakkad2020super}, while our data lacks the spatial resolution required to constrain the shell thickness in such a scenario. Additionally, as discussed in \citet{cicone2014massive}, a bi-conical or spherical geometry provides a physically motivated description of feedback scenarios in which the outflow is continuously replenished by clouds ejected from the galactic disk.

The outflow kinetic power is estimated as:

\begin{equation}
    \dot{E}_\mathrm{kin}^{\mathrm{[OIII]}} = \frac{3}{2} \frac{M_\mathrm{out}^{\mathrm{[OIII]}}~v_{\mathrm{out}}^3}{R_{\mathrm{out}}}
\end{equation}

The values of the ionized outflow energetics are reported in Table \ref{tab:outflow_props}. The associated uncertainties are estimated by error propagation. As discussed in Sect. \ref{sect:outflow_detec}, we adopted a conservative velocity threshold of $w_{80} \geq 600~\mathrm{km~s^{-1}}$ to identify outflowing gas, implying that only the high-velocity component is considered in the estimation of the outflow energetics. Consequently, the derived ionized outflow mass should be interpreted as a lower limit, since a significant fraction of the mass is expected to move at lower velocities \citep{costa2018quenching}. The effect of this threshold on the kinetic energy is, however, negligible, as the bulk of the power originates from the high-velocity gas \citep{ward2024agn}.

\begin{table*}[t]
    \renewcommand{\arraystretch}{1.4}
    \centering
    \caption[]{Properties of the [\ion{O}{iii}] ionized outflows.}
    \label{tab:outflow_props}
    \begin{tabular}{c c c c c c c c}
        \hline \hline
        Target & $R_{\mathrm{out}}$ & $v_{\mathrm{out}}$ & $L_{\mathrm{out}}^{[\mathrm{OIII}]}$ & $M_{\mathrm{out}}^{[\mathrm{OIII}]}$ & $ \dot{M}_\mathrm{out}^{\mathrm{[OIII]}}$ & $\dot{E}_\mathrm{kin}^{\mathrm{[OIII]}}$ & $\dot{E}_\mathrm{kin}^{\mathrm{[OIII]}}/L_{\mathrm{bol}}$  \\
        & [kpc] & [$\mathrm{km~s^{-1}}$] & $[10^{43}~\mathrm{erg~s^{-1}}]$ & [$10^6~\mathrm{M}_\odot$] & [$ \mathrm{M_\odot~yr^{-1}}$] & [$10^{43}~\mathrm{erg~s^{-1}}$] & [$10^{-4}$]\\
        \hline
        Q0050+0051 &  2.98 & 2425 $\pm$ 161 & 0.82 $\pm$ 0.07 & 6.55 $\pm$ 3.32 &  16.29 $\pm$ 8.37 & 3.02 $\pm$ 1.65 & 2.16 \\
        Q0052+0140 &  4.33 & 3068 $\pm$ 126 & 2.55 $\pm$ 0.13 & 20.36 $\pm$ 10.23 & 44.25 $\pm$ 22.36 & 13.13 $\pm$ 6.81 & 4.53 \\
        Q2121+0052 &  < 2.09 & 3408 $\pm$ 308 & < 0.32 $\pm$ 0.03 & < 2.55 $\pm$ 1.29 & 12.75 $\pm$ 6.62 & 4.66 $\pm$ 2.70 & 3.11 \\
        Q2123$-$0050 &  3.32 & 3090 $\pm$ 490 &  1.08 $\pm$ 0.17 &  8.60 $\pm$ 4.51 & 24.52 $\pm$ 13.45 & 7.38 $\pm$ 5.23 & 0.98 \\
        SDSSJ2319$-$1040 & 4.85 & 1714 $\pm$ 125 & 2.28 $\pm$ 0.16 & 18.23 $\pm$ 9.20 & 19.77 $\pm$ 10.09 & 1.83 $\pm$ 1.01 & 0.96\\
        SDSSJ1427$-$0029 & < 5.69 & 1546 $\pm$ 173  & <5.83 $\pm$ 0.47 &  46.61 $\pm$ 23.60 & 38.77 $\pm$ 20.12 & 2.92 $\pm$ 1.77 & 0.93\\
        \hline
    \end{tabular}
    
    \vspace{4mm}
    \tablefoot{
        From left to right, the columns report: the target name, extension of the ionized outflow, outflow velocity as measured by the maximum value of W80, luminosity of the outflow, outflow mass, outflow mass rate and outflow kinetic power.
    }
\end{table*}

\section{Discussion} \label{sect:discussion}

As shown in Sect. \ref{sect:outflow_detec}, for all the quasars with known Ly $\alpha$ nebulae that we have observed with VLT/ERIS and GEMINI/GNIRS, we have detected the presence of galaxy-wide AGN-driven outflows. These outflows are much more luminous and have a higher kinetic power than those detected in previous studies \citep{kakkad2020super, tozzi2024super}. While this is interesting, it is not clear proof of a causal connection between the two phenomena. To investigate how the detected outflows may be connected to the extended Ly$\alpha$ nebulae surrounding the quasar, we now focus on exploring the correlation between different physical parameters of both phenomena. 

\subsection{Outflow kinetic power vs. Ly$\alpha$ nebulae size and luminosity} \label{outflow_vs_nebula_props}

\cite{costa2022agn} find that the size and luminosity of the nebulae are set by the escape of Ly$\alpha$ and ionizing flux from the galactic nucleus. The more the gaseous reservoir in the nucleus is cleared out, the higher the escape fraction. We may thus expect a connection between the outflow kinetic power and the physical properties of the Ly$\alpha$ nebula.

\begin{figure}
    \centering
    \begin{tabular}{c}
        \includegraphics[width=0.45\textwidth]{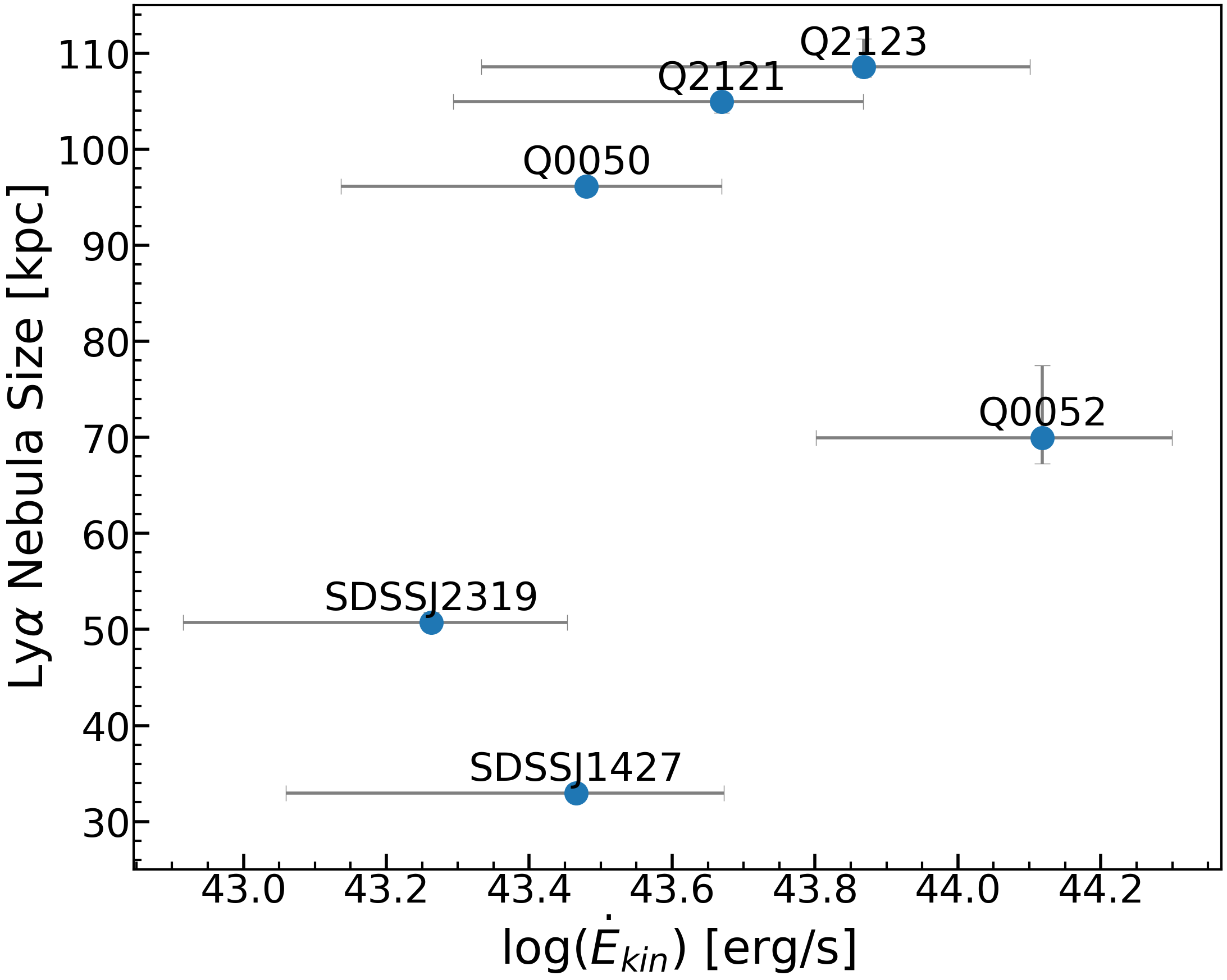} \\
    \end{tabular}
    \caption{Relation between outflow kinetic power and Ly$\alpha$ nebulae size.}
    \label{fig:power-size}
\end{figure}

\begin{figure}
    \centering
    \begin{tabular}{c}
        \includegraphics[width=0.45\textwidth]{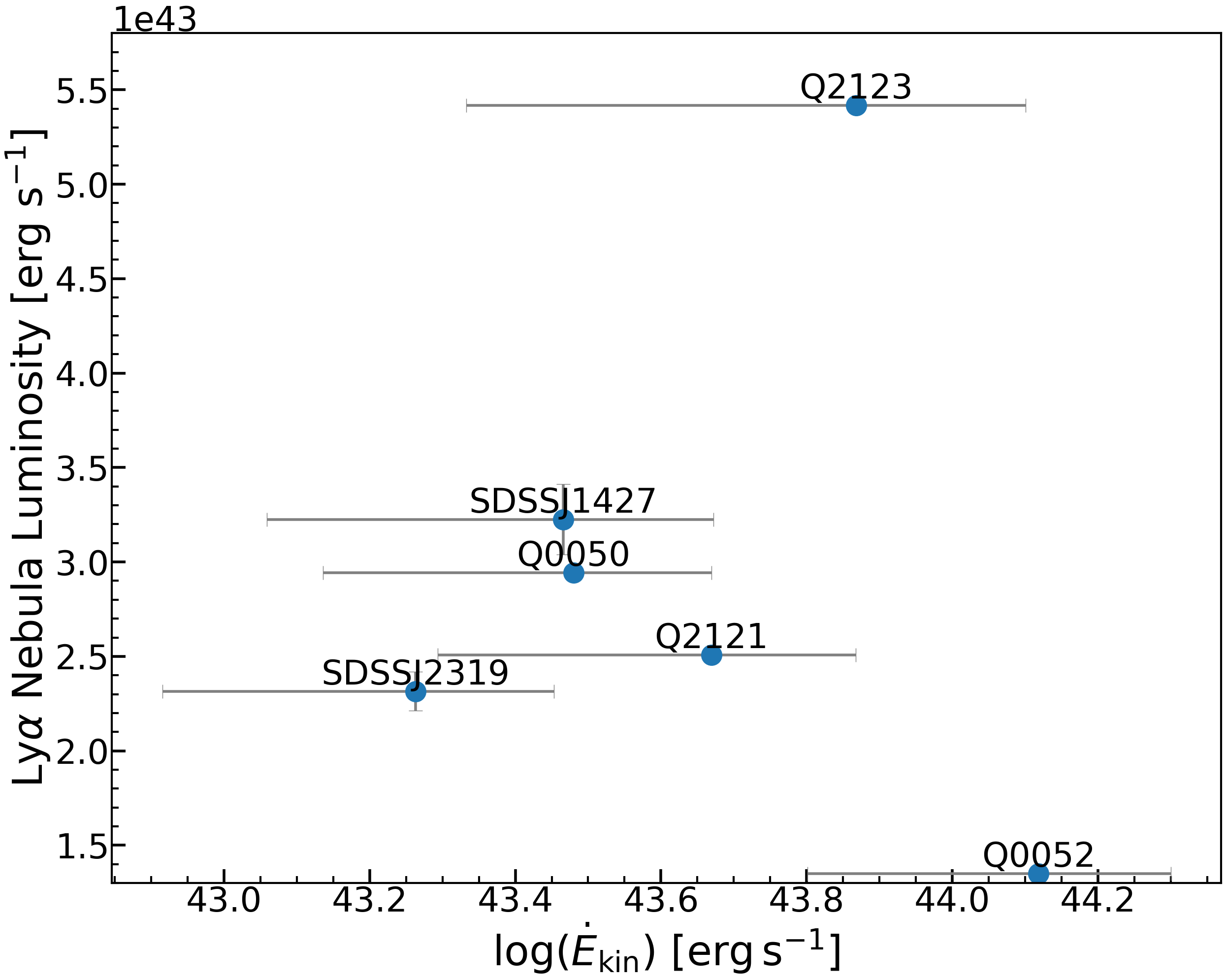} \\
    \end{tabular}
    \caption{Relation between outflow kinetic power and Ly$\alpha$ nebulae luminosity.}
    \label{fig:power-luminosity}
\end{figure}

An increasingly powerful outflow may directly (e.g., by recombination in photoionized regions) or indirectly (e.g., by opening a path of low optical depth for the Ly$\alpha$ and UV photons from the BLR to escape) contribute to further ionize the CGM and generate Ly$\alpha$ nebulae on larger scales. To test this prediction, we compare the kinetic power of the AGN-driven outflows detected in the quasars presented in this work, with the size (see Fig. \ref{fig:power-size}) and luminosity (see Fig. \ref{fig:power-luminosity}) of their corresponding Ly$\alpha$ nebulae.

Figs. \ref{fig:power-size} and \ref{fig:power-luminosity} suggest that both the size and luminosity of the Ly$\alpha$ nebulae may increase with the kinetic power of the AGN-driven outflows traced on ISM scales. To quantify the strength of these possible relations, we computed the Pearson correlation coefficients. For the relation between outflow power and Ly$\alpha$ nebula size ($\dot{E}_\mathrm{kin}^{\mathrm{[OIII]}}$–$D_{\mathrm{Ly\alpha}}$), we find $r=0.39$ with a null-hypothesis probability of $\sim43\%$. For the relation between outflow power and Ly$\alpha$ nebula luminosity ($\dot{E}_\mathrm{kin}^{\mathrm{[OIII]}}$–$L_{\mathrm{Ly\alpha}}$), we obtain $r=-0.01$ with a null-hypothesis probability of $\sim98\%$. These results indicate very weak and statistically non-significant linear correlations between the outflow power and the Ly$\alpha$ nebulae properties. However, considering that these relations may not be strictly linear, a Spearman rank correlation test would provide a more appropriate measure of any monotonic trend. The use of this non-parametric test is further motivated by the presence and effect of potential outliers. As shown in Figs. \ref{fig:power-size} and \ref{fig:power-luminosity}, Q0052+0140 deviates from the overall trends: despite exhibiting relatively high [\ion{O}{iii}] outflow kinetic power and luminosity, it is not associated with a comparably extended or luminous Ly$\alpha$ nebula. We propose two plausible explanations for this discrepancy. The first is related to orientation and scattering effects. In particular, the Ly$\alpha$ nebula in Q0052+0140 appears one-sided. As discussed by \citet{costa2022agn}, the escape of Ly$\alpha$ photons from the nuclear regions is highly anisotropic due to the combined effects of dust absorption and resonant scattering. Ly$\alpha$ photons preferentially escape along the galaxy’s rotation axis, such that lines of sight intersecting the disk plane are expected to produce fainter and more asymmetric nebulae than face-on orientations. The second explanation involves the impact of an over‑efficient AGN‑driven outflow. The outflow detected in Q0052+0140 exhibits the highest kinetic power of our sample of quasars. \cite{costa2022agn} shows that when the wind kinetic power exceeds the level required to simply open a low‑optical‑depth channel, it removes most of the neutral hydrogen from the inner halo. In this regime, Ly$\alpha$ photons escape directly from the nucleus rather than being resonantly scattered through the CGM, resulting in a compact and faint Ly$\alpha$ nebula.

To assess the impact of individual data points in the results of the test, we performed multiple iterations of the Spearman rank correlation test, each time excluding one source from the sample. The results are shown in Table \ref{tab:stat_corr}.

\begin{table}[ht]
    \renewcommand{\arraystretch}{1.4}
    \centering
    \caption[]{Statistical parameters for the Spearman rank correlation test on the scaling relations in Fig. \ref{fig:power-size} and \ref{fig:power-luminosity}.}
    \label{tab:stat_corr}
    \begin{tabular}{c c c c c}
        \hline\hline
        {Exc. source} & 
        \multicolumn{2}{c}{$\dot{E}_\mathrm{kin}^{\mathrm{[OIII]}}$–$D_{\mathrm{Ly\alpha}}$} & 
        \multicolumn{2}{c}{$\dot{E}_\mathrm{kin}^{\mathrm{[OIII]}}$–$L_{\mathrm{Ly\alpha}}$} \\
        \cline{2-5}
        & $\rho$ & $p$-value & $\rho$ & $p$-value \\
        \hline
        None & 0.60 & 0.21 & -0.08 & 0.87 \\
        Q0050+0051 & 0.60 & 0.28 & -0.09 & 0.87 \\
        Q0052+0140 & 0.89 & 0.03 & 0.60 & 0.28 \\
        Q2121+0052 & 0.60 & 0.28 & -0.09 & 0.87 \\
        Q2123-0050 & 0.60 & 0.28 & -0.39 & 0.50 \\
        SDSSJ2319-1040 & 0.39 & 0.50 & -0.39 & 0.50 \\
        SDSSJ1427-0029 & 0.39 & 0.50 & -0.09 & 0.87 \\
        \hline
    \end{tabular}
    
    \vspace{4mm}
    \tablefoot{
        From left to right, the columns report: the source excluded in each run of the test, and the test results for the scaling relation between outflow kinetic power and Ly$\alpha$ nebula size and luminosity.
    }
\end{table}

The results of the correlation analysis presented in Table \ref{tab:stat_corr} indicate that excluding Q0052+0140, identified as an outlier, yields a stronger positive monotonic relation between the outflow and the Ly$\alpha$ nebula properties, compared to the results obtained when including all sources in the correlation test. In this case, the $\dot{E}_\mathrm{kin}^{\mathrm{[OIII]}}$–$D_{\mathrm{Ly\alpha}}$ relation shows a Spearman coefficient of $\rho=0.89$ with a null-hypothesis probability of $3\%$, while the $\dot{E}_\mathrm{kin}^{\mathrm{[OIII]}}$–$L_{\mathrm{Ly\alpha}}$ relation gives $\rho=0.6$ with a null-hypothesis probability of $28\%$. Excluding any other source does not produce a significant change in the results. 

These findings suggest a strong positive correlation between the outflow kinetic power and the Ly$\alpha$ nebula size, and a weaker correlation with the Ly$\alpha$ luminosity. However, given the small sample size and the limited range in $L_{\mathrm{bol}}$ and $\dot{E}_\mathrm{kin}^{\mathrm{[OIII]}}$ (less than one order of magnitude), a larger dataset would be required to confirm these trends more robustly.

Both the outflow power \citep[e.g.,][]{fiore2017agn, liu2009host, zakamska2016discovery} and the Ly$\alpha$ nebula size and luminosity \citep[e.g.,][]{lobos2025qso, mackenzie2021revealing} are known to correlate positively with the quasar $L_{\mathrm{bol}}$. This raises the possibility that the observed relation between outflow kinetic power and Ly$\alpha$ nebula properties may be indirectly driven by their mutual dependence on $L_{\mathrm{bol}}$. To assess this, we performed a partial correlation analysis for the $\dot{E}_\mathrm{kin}^{\mathrm{[OIII]}}$–$D_{\mathrm{Ly\alpha}}$ and $\dot{E}_\mathrm{kin}^{\mathrm{[OIII]}}$–$L_{\mathrm{Ly\alpha}}$ relations while controlling for $L_{\mathrm{bol}}$. Using the simplified recursive expression for a single control variable \footnote{$\rho_{XY\cdot Z}=\frac{\rho_{XY}-\rho_{XZ}\rho_{ZY}}{\sqrt{\left(1-\rho^2_{XZ}\right)\left(1-\rho^2_{ZY}\right)}}$ where $X$ corresponds to $\dot{E}_\mathrm{kin}^{\mathrm{[OIII]}}$, $Y$ to $D_{\mathrm{Ly\alpha}}$ or $L_{\mathrm{Ly\alpha}}$, and $Z$ to $L_{\mathrm{bol}}$}, we computed the corresponding partial Spearman correlation coefficients. We find $\rho = 0.90$ for $\dot{E}_\mathrm{kin}^{\mathrm{[OIII]}}$–$D_{\mathrm{Ly\alpha}}$ and $\rho = 0.61$ for $\dot{E}_\mathrm{kin}^{\mathrm{[OIII]}}$–$L_{\mathrm{Ly\alpha}}$. These results confirm that the positive monotonic correlations between outflow kinetic power and Ly$\alpha$ nebula properties persist independently of $L_{\mathrm{bol}}$, and are therefore not driven by their shared dependence on AGN luminosity.

\subsection{Spatial relation between extended [\ion{O}{iii}] emission and Ly$\alpha$ nebulae} \label{sect:spat_rel}

If the AGN-driven outflows on ISM scales are opening a path of least resistance for the quasar's Ly$\alpha$ and UV photons to escape to the CGM, one would expect that, at such scales (inner CGM), the morphology of the Ly$\alpha$ nebulae is correlated with the orientation of the outflow. To investigate any possible spatial alignment between the ionization cone and the direction of the Ly$\alpha$ nebulae, we compare the location of the extended [\ion{O}{iii}] emission with that of the brightest regions of the nebula, represented by the highest S/N contours, S/N = 5, 10, 30.

We performed the alignment between the remnant [\ion{O}{iii}] emission and the Ly$\alpha$ nebulae contours by placing both datasets onto a common astrometric reference frame using the World Coordinate System (WCS) information from the data cubes. We converted the pixel coordinates into projected angular offsets relative to the quasar position, and both maps were recentered such that the quasar lies at position (0,0). The Ly$\alpha$ SB contours were then overplotted on the [\ion{O}{iii}] extended emission maps using these common angular coordinates. This alignment is therefore based on the morphological correspondence between the two tracers, rather than on explicit position-angle measurements. The results are shown in Fig. \ref{fig:extended-OIII}.

\begin{figure}
    \centering
    \includegraphics[width=0.7\linewidth]{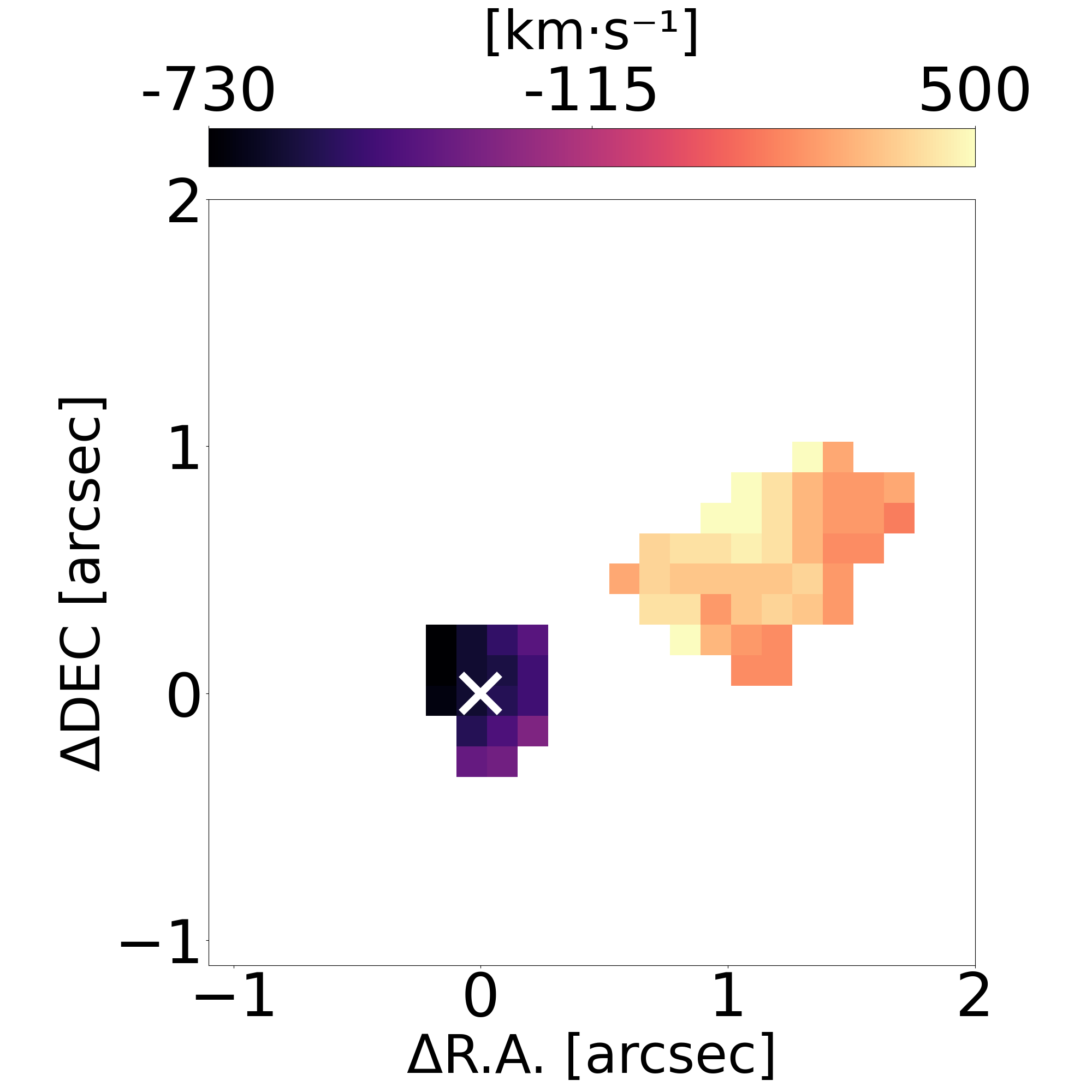}
    \caption{$v_{50}$ velocity map of Q0050+0051.}
    \label{fig:v50_0050}
\end{figure}

For Q0050+0051, we detect extended [\ion{O}{iii}] emission towards the south and northwest, with the latter spatially coinciding with the peak of the Ly$\alpha$ SB map (Fig.~\ref{fig:Lya-SB_maps}). However, the $w_{80}$ map (Fig.~\ref{fig:kinems_target}) shows that this region exhibits low velocities, inconsistent with ionized outflows. Additionally, the $v_{50}$ map (Fig.~\ref{fig:v50_0050}) shows a pronounced velocity offset ($\Delta v_{50} \sim 1200 \mathrm{km~s^{-1}}$) between the central and the extended emission. Such velocity offset exceeds the typical range expected from disk rotation in star-forming galaxies at $z\sim2$ (tens to a few hundred $\mathrm{km~s^{-1}}$) \citep[e.g.][]{yue2021alma,roman2023regular}. These results suggest that the [\ion{O}{iii}] emission cloud detected towards the northwest of Q0050+0051 likely traces dynamically quiescent gas outside the main host. Nonetheless, this interpretation does not exclude the possibility of a spatial alignment between the outflow and the nebula in this source. Given that the outflow is detected only on small scales and its direction cannot be constrained with our data, we cannot rule out the scenario in which the cloud is being ionized by the AGN. In that case, the [\ion{O}{iii}] cloud might be indicating the direction of the ionization cone, likely corresponding to the direction of the inner outflow clearing the central regions.
a

For Q0052+0140, we detect extended [\ion{O}{iii}] emission towards the west and southeast, coincident with regions of increased $w_{80}$ velocities (see Fig. \ref{fig:kinems_target}). Notably, the Ly$\alpha$ nebula is also particularly bright in these directions, suggesting a possible connection between the two phenomena. In the cases of Q2123$-$0050 and SDSSJ2319$-$1040, the [\ion{O}{iii}] emission appears symmetrically extended around the quasar, which is comparable with the symmetric morphology of their Ly$\alpha$ nebulae (see Fig. \ref{fig:Lya-SB_maps}).

Overall, the observed alignment between the extended \ion{O}{iii} emission and the regions of increased Ly$\alpha$ emission on small scales supports the scenario of a casual link between the two phenomena, with the caveat that this conclusion is based on our limited sample size.

\subsection{Alternative powering mechanisms of the Ly$\alpha$ nebulae}

The Ly$\alpha$ emission detected around high-redshift quasars is generally attributed to a combination of mechanisms, including recombination, gravitational cooling, and scattering of BLR photons. However, as demonstrated by \citet{costa2022agn}, these processes alone are insufficient to reproduce the observed bright and extended Ly$\alpha$ nebulae without the inclusion of AGN-driven outflows. In particular, considering BLR scattering as the only source of Ly$\alpha$ photons results in emission confined to the central $\sim0.5$ kpc, preventing the formation of an extended nebula. Conversely, considering only recombination and collisional excitation yields nebulae with larger spatial extents, but they are fainter, morphologically irregular, and not centered around the quasar.

Inspecting the Ly$\alpha$ SB maps of our sample (Fig. \ref{fig:Lya-SB_maps}), we find that the nebulae associated with Q2121+0052 and SDSSJ2319$-$1040 closely resemble the recombination + collisional excitation scenario. These nebulae are fainter than those of the rest of the sample, and in the case of Q2121+0052, the emission is offset from the quasar location. Notably, Q2121+0052 also hosts the least extended and least luminous outflow in the sample, with the lowest outflowing mass, although its kinetic power lies within the typical range. In contrast, SDSSJ2319$-$1040 exhibits a widely extended but low-velocity outflow, corresponding to the lowest kinetic power among the sample. In light of the positive correlation between outflow kinetic power and Ly$\alpha$ nebula luminosity identified in Section \ref{outflow_vs_nebula_props}, these results suggest that AGN-driven outflows play a reduced role in powering the extended Ly$\alpha$ emission in these two systems. Instead, consistent with the scenario proposed by \citet{costa2022agn}, their Ly$\alpha$ nebulae are likely produced by recombination with an increasing contribution from collisional excitation towards their outskirt. 

To test this scenario quantitatively, we follow the approach by \citet{hennawi2013quasars}, used in the literature \citep[e.g.,][]{battaia2015deep, farina2019requiem, lobos2025qso} to predict the SB of the Ly$\alpha$ emission originated from photoionization of cold gas by the AGN radiation, followed by recombination. For this purpose, we assume that quasars are surrounded by spherical gas clouds that are either optically thick ($N_\ion{H}{i} \gg 10^{17.2} \mathrm{cm^{-2}}$) or optically thin ($N_\ion{H}{i} \ll 10^{17.2} \mathrm{cm^{-2}}$) to the Lyman continuum photons. We derive the \ion{H}{i} column density of a single cloud as:

\begin{equation} \label{col_dens}
    \frac{\langle N_\ion{H}{i} \rangle}{10^{17.2}~\mathrm{cm^{-2}}} = 0.1\left ( \frac{L_{\mathrm{Ly}\alpha}}{10^{44}~\mathrm{erg~s^{-1}}} \right ) \left ( \frac{L_{\nu_{\mathrm{LL}}}}{10^{31}~\mathrm{erg~s^{-1}Hz^{-1}}} \right )
\end{equation}

where $L_{\mathrm{Ly}\alpha}$ represents the observed luminosity of the Ly$\alpha$ nebulae and $L_{\nu_{\mathrm{LL}}}$ the luminosity of the quasar at the Lyman edge. The latter can be estimated assuming that the spectral energy distribution (SED) of the quasar follows a power law of the form $L_\nu = L_{\nu_{\mathrm{LL}}}(\nu/\nu_{\mathrm{LL}})^{\alpha_{\mathrm{UV}}}$, where $L_\nu$ represents the ionizing luminosity, and $\alpha_{\mathrm{UV}}=-0.61$, following \citet{lusso2015first}. For SDSSJ2319$-$1040, $\nu L_\nu (1700~\AA) = 9.13 \times 10^{46}~\mathrm{erg~s^{-1}}$ \citep{allen2011strong}, therefore $L_{\nu_{\mathrm{LL}}} = L_\nu (912~\AA) = 3.52 \times 10^{31}~ \mathrm{erg~s^{-1}Hz^{-1}}$. And in the same way for Q2121+0052, $\nu L_\nu (1700~\AA) = 5.78 \times 10^{46}~\mathrm{erg~s^{-1}}$ \citep{allen2011strong} and $L_{\nu_{\mathrm{LL}}} = 2.23 \times 10^{31}~ \mathrm{erg~s^{-1}Hz^{-1}}$. Plugging the values for the quasar luminosity at the Lyman edge and those for the nebula luminosity reported in Table \ref{tab:Lya_props} in Equation \ref{col_dens}, we estimate that $\langle N_\ion{H}{i} \rangle = 8.12 \times 10^{-15.2} \mathrm{cm}^{-2}$ and $\langle N_\ion{H}{i} \rangle = 5.59 \times 10^{-15.2} \mathrm{cm}^{-2}$ for SDSSJ2319$-$1040 and Q2121+0052, respectively. These results are consistent with the optically thin scenario for which:

\begin{align}
    \frac{\mathrm{SB}_\mathrm{Ly\alpha}}{10^{-19}~\mathrm{erg~s^{-1}cm^{-2}arcsec^{-2}}} =  3.6 & \left ( \frac{1+z}{1+6.2}\right)^{-4} \left( \frac{f_\mathrm{C}}{0.5}\right )   \notag \\
    & \left ( \frac{n_\mathrm{H}}{1~\mathrm{cm^{-3}}} \right ) \left ( \frac{N_\mathrm{H}}{10^{20.5}\mathrm{cm}^{-2}}\right )
\end{align}

where we assume typical values of $f_\mathrm{C}=0.5$ for the covering fraction of the optically thin clouds, $n_\mathrm{H}=1~\mathrm{cm^{-3}}$ for the cloud's hydrogen volume, and $N_\mathrm{H}=10^{20.5}\mathrm{cm^{-2}}$ for its column density \citep[e.g.,][]{prochaska2013quasars, lau2016quasars}. In this regime, the expected SB does not depend on the luminosity of the quasar but only on the physical properties of the gas. Therefore, we estimate that the AGN photoionization alone will power a Ly$\alpha$ nebulae with $\mathrm{SB}  \sim 3.2 \times 10^{18} \mathrm{erg~s^{-1}cm^{-2}arcsec^{-2}}$ and $\mathrm{SB}  \sim 7.4 \times 10^{18} \mathrm{erg~s^{-1}cm^{-2}arcsec^{-2}}$, corresponding to nebula luminosities of $L_{\mathrm{Ly\alpha}}\sim 9\times10^{42} \mathrm{erg~s^{-1}}$ and $L_{\mathrm{Ly\alpha}}\sim 4\times10^{43} \mathrm{erg~s^{-1}}$ for SDSSJ2319$-$1040 and Q2121+0052, respectively. These results are consistent with the observed Ly$\alpha$ nebulae luminosities reported in Table \ref{tab:Lya_props}. Hence, we can conclude that the AGN radiation provides sufficient energy to photoionize the surrounding gas and power extended but faint nebulae such as the ones observed around the quasars SDSSJ2319$-$1040 and Q2121+0052, with a minor contribution from AGN-driven outflows.

In addition to assessing the role of alternative mechanisms in powering the Ly$\alpha$ nebulae, we also examine the impact of outflow timescales. We characterize the AGN-driven outflows by estimating their dynamical ages using the velocities and extents reported in Table \ref{tab:outflow_props}. The resulting outflow age estimates are presented in Table \ref{tab:outflow_age}.

\begin{table}[ht]
    \renewcommand{\arraystretch}{1.2}
    \centering
    \caption[]{Outflow ages}
    \label{tab:outflow_age}
    \begin{tabular}{c c}
        \hline \hline
        Target ID & Outflow age \\ 
                  & (Myr)       \\ 
        \hline
        Q0050+0051 & 1.14 $\pm$ 0.09 \\
        Q0052+0140 & 1.32 $\pm$ 0.07 \\
        Q2121+0052 & 0.57 $\pm$ 0.06 \\
        Q2123$-$0050 & 1.00 $\pm$ 1.16 \\
        SDSSJ2319$-$1040 & 2.86 $\pm$ 0.22 \\
        SDSSJ1427$-$0029 & 3.80 $\pm$ 0.43 \\
        \hline
    \end{tabular}

\end{table}

Assuming an AGN duty cycle of $\sim 1$ Myr \citep[e.g.,][]{clavijo2023agn,harrison2024observational}, defined as the timescale over which the supermassive BH alternates between active and inactive phases, the outflow timescales reported in Table \ref{tab:outflow_age} suggest that all sources in our sample, except Q2121+0052, were likely launched during previous AGN episodes. In contrast, Q2121+0052 hosts the youngest outflow, which allows us to consider a scenario in which the outflow was only recently triggered and has not yet had sufficient time to propagate through the host galaxy. As a result, its impact on increasing the escape fraction of Ly$\alpha$ photons from the nuclear regions may still be limited. Additionally, we find that all outflows are younger than 5 Myr, which, according to \citet{costa2020powering}, is sufficient time for them to propagate to scales below 10 kpc. This suggests that the observed Ly$\alpha$ nebulae may be the result of the cumulative effect of AGN events over the lifetime of the host galaxy, rather than due to a single episode. However, as the ionizing photons can travel faster than the outflow itself, the Ly$\alpha$ emission on large scales might be the result of photoionization followed by recombination processes.

\section{Summary and conclusions}

We presented VLT/ERIS and GEMINI/GNIRS observations of six quasars previously known to host extended Ly$\alpha$ nebulae \citep{arrigoni2019qso, cai2019evolution}, identified with VLT/MUSE and Keck/KCWI. Using near-infrared IFU data, we traced the kinematics of the ISM through the [\ion{O}{iii}] emission line (Sect. \ref{sect:spat_res}).

The main results of our work are:

\begin{itemize}
    \item We detect ionized outflows in all the targets of our sample, with velocities of $v_{\mathrm{out}}>1500~\mathrm{km~s^{-1}}$. By comparing the [\ion{O}{iii}] and PSF-BLR COG (\ref{fig:OIII-BLR}), and by performing a PSF-subtraction method (\ref{fig:extended-OIII}), we spatially resolve the outflows in four of the targets in our sample (\ref{fig:kinems_target}). These outflows have extensions between $\sim3-6$ kpc. For the remaining two spatially unresolved quasars, the emission is dominated by the PSF and beam smearing effects. Therefore, their outflow properties are constrained only through integrated spectra, resulting in upper limits on their sizes.
    
    \item  We find a positive monotonic correlation between the properties of the AGN-driven outflows and those of the Ly$\alpha$ nebulae. These correlations become statistically significant only when excluding Q0052+0140, whose nebula is comparatively compact and faint given its powerful outflow. This scenario might be a result of increased scattering in the line of sight or due to an over-efficient outflow. The correlation between outflow kinetic power and nebula size (\ref{fig:power-size}) is strong, with a Spearman coefficient of $\rho=0.89$ and a $p$-value=0.03. We find a weaker correlation between outflow power and Ly$\alpha$ nebula luminosity (\ref{fig:power-luminosity}), with $\rho=0.60$ and a $p$-value=0.28. 

    \item We find observational evidence of spatial alignment between the location of the extended [\ion{O}{iii}] emission and the inner and brightest regions of the Ly$\alpha$ nebulae. This might be an effect of the ionization cone created by the AGN-driven outflow, although constraining its direction is beyond the sensitivity of our data.
    
    \item These observational results show tentative evidence that supports the theoretical prediction by \cite{costa2022agn}. On the one hand, AGN-driven outflows might open a low-optical-depth path allowing Ly$\alpha$ photons from the nucleus to escape to CGM scales and scatter to form an extended nebula. On the other hand, such outflows might lead to the escape of ionizing flux, enhancing the effect of recombination. However, given our small sample size and the limited range in bolometric luminosity and outflow power, a larger dataset is required to test this scenario more robustly.

    \item We find evidence of alternative powering mechanisms of the Ly$\alpha$ nebula. The least powerful and younger outflows are correlated with more compact and less luminous nebulae, possibly as a consequence of different driving mechanisms such as recombination or gravitational cooling, with a minor contribution from AGN-driven outflows.
    
\end{itemize}

Although our results provide a first observational test of the theoretical prediction linking AGN-driven outflows on ISM scales to extended Ly$\alpha$ nebulae on CGM scales, further observations and studies are required to shed light on three key aspects. First, establishing the kinematic coupling between these two phenomena is essential to determine whether the energy injected by the outflow is transferred to the cold CGM, increasing the velocity dispersion of the gas in the nebula. Testing such a coupling would provide key evidence toward distinguishing a causal connection from a purely correlative relationship between AGN-driven outflows and Ly$\alpha$ nebulae. \citet{ginolfi2018extended} suggested that the high velocity dispersion observed in the central halo of a BAL QSO may be associated with outflowing material; testing whether this scenario extends to kpc-scale outflows would offer valuable insights into the physical mechanisms of AGN feedback. While our current data allows a partial exploration of this question, wide-field IFU observations with JWST are required to trace the [\ion{O}{iii}] outflows from ISM to CGM scales and to spatially match them with the Ly$\alpha$ nebulae observed with VLT/MUSE. Second, on the link between the extended Ly$\alpha$ nebulae and the sub-parsec scale nuclear winds that initiate the feedback process. Such winds are typically implemented as the seeds of AGN feedback in cosmological simulations; therefore, observational evidence of their impact on the CGM would provide crucial constraints for these models. And third, on the physical mechanism that mediates the link between AGN-driven outflows and Ly$\alpha$ nebulae. Future observations of H$\alpha$ emission on CGM scales will allow measurements of the Ly$\alpha$/H$\alpha$ ratio, helping to disentangle the relative contributions of recombination, gravitational cooling, and BLR resonant scattering to the creation of the nebula.

\begin{acknowledgements}

G.T. is funded by the European Union (ERC Advanced Grant GALPHYS, 101055023). Views and opinions expressed are, however, those of the author only and do not necessarily reflect those of the European Union or the European Research Council. Neither the European Union nor the granting authority can be held responsible for them. CMH acknowledges funding from a United Kingdom Research and Innovation grant (code: MR/V022830/1). E.P.F. is supported by the international Gemini Observatory, a program of NSF NOIRLab, which is managed by the Association of Universities for Research in Astronomy (AURA) under a cooperative agreement with the U.S. National Science Foundation, on behalf of the Gemini partnership of Argentina, Brazil, Canada, Chile, the Republic of Korea, and the United States of America.

\end{acknowledgements}


\bibliographystyle{aa}
\bibliography{bibliography}

\begin{appendix}

\section{Integrated spectra for unresolved targets}

\begin{figure}[!h]
    \centering
    \includegraphics[width=1\linewidth]{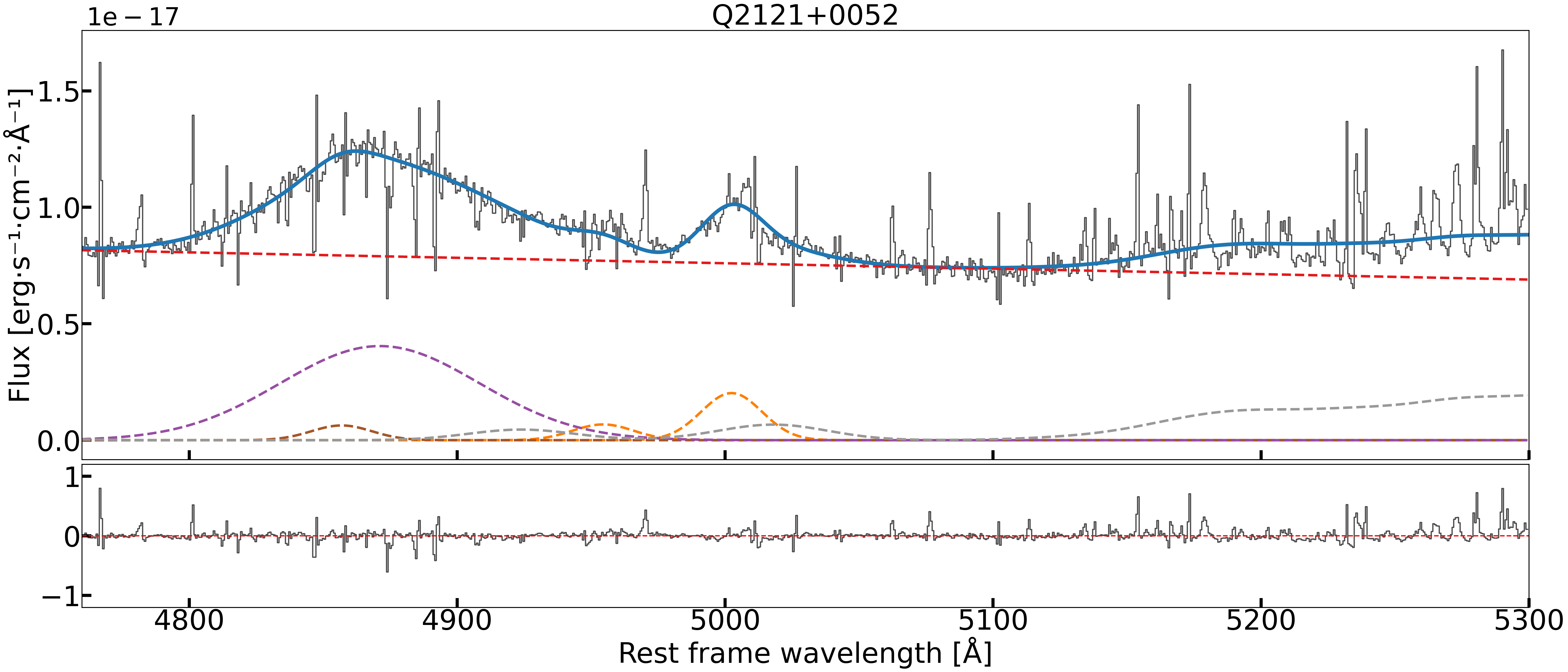}
    \caption{Q2121+0052. Same as in Fig. \ref{fig:org_fit} for an integrated spectrum with an aperture of 2 pixels.}
    \label{fig:app_int_q2121}
\end{figure}

\begin{figure}[!ht]
    \centering
    \includegraphics[width=1\linewidth]{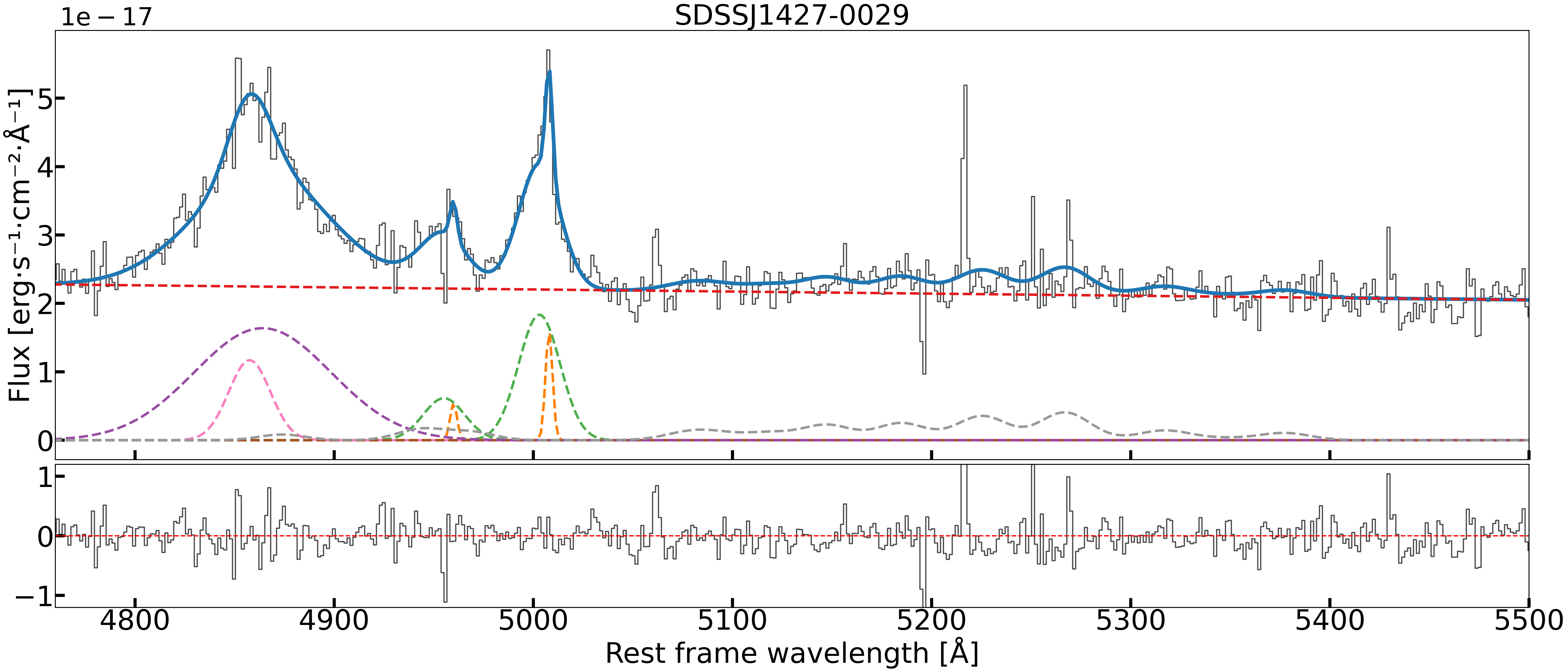}
    \caption{SDSSJ1427-0029. Same as in Fig. \ref{fig:org_fit} for an integrated spectrum with an aperture of 2 pixels.}
    \label{fig:app_int_j1427}
\end{figure}

\end{appendix}

\end{document}